\documentclass[fleqn,usenatbib]{mnras}

\usepackage{verbatim, graphicx, ifthen, pbox}
\usepackage{eso-pic}
\usepackage{CJKutf8}
\usepackage{xcolor}
\usepackage{hyperref}
\usepackage{amsmath}
\usepackage{graphicx}
\usepackage{verbatim}
\usepackage{booktabs}
\usepackage[T1]{fontenc}
\usepackage{ae,aecompl}
\usepackage{booktabs}

\usepackage{xspace}
\usepackage{amssymb}
\usepackage{bm}
\usepackage{makecell}

\newcommand{\ra}[1]{\renewcommand{\arraystretch}{#1}}

\definecolor{mgm}{rgb}{0.4, 0.1, 1.0}

\newcommand{\change}[1]{\textcolor{black} {#1}}

\newcommand{\changethree}[1]{\textcolor{black} {#1}}
\defcitealias{MacDonald2020}{M20}

\usepackage{newtxtext,newtxmath}

\title{Effects of Pebble Accretion Isolation Mass on Observable Exoplanet Properties}

\author[Brefka et al.]{Lucas Brefka, $^{1,2}$\thanks{E-mail:lfb5454@psu.edu}
Rebekah I. Dawson,$^{1,2}$
Mariah G. MacDonald,$^3$
Phoebe Sandhaus$^{1,2}$
and Eric B. Ford$^{1,2,4,5}$
\\
$^{1}$Department of Astronomy and Astrophysics, 525 Davey Laboratory, 251 Pollock Road, Penn State University, University Park, PA 16802, USA\\
$^{2}$Center for Exoplanets and Habitable Worlds, 525 Davey Laboratory, 251 Pollock Road, Penn State University, University Park, PA 16802, USA\\
$^{3}$Department of Physics, The College of New Jersey, 2000 Pennington Road, Ewing, NJ 08628, USA\\
$^{4}$Institute for Computational and Data Sciences, Penn State University, University Park, PA 16802, USA\\
$^{5}$Center for Astrostatistics, 525 Davey Laboratory, 251 Pollock Road, Penn State University, University Park, PA 16802, USA
}

\date{Accepted XXX. Received YYY; in original form ZZZ}

\pubyear{2026}

\begin{document}
\label{firstpage}
\pagerange{\pageref{firstpage}--\pageref{lastpage}}
\maketitle

\begin{abstract} 

The \emph{Kepler} Mission has discovered a plethora of planetary systems with super-Earth sized planets.
These systems exhibit many properties, from widely-spaced planets with non-negligible eccentricities and inclinations, to tightly-spaced, coplanar, and nearly circular multi-planet systems.
The observable properties of these systems, such as planet-planet spacings, multiplicity and orbital morphology, can be strongly influenced by the initial conditions of formation. 
These conditions affect the early growth of planetary embryos in the gas disk phase through pebble and/or planetesimal accretion, which then affects the planets’ final growth during the giant impact stage.
In this work, we investigate how assumptions of different limiting embryo isolation masses during early stages of planet formation affect the final properties of super-Earth planets within the inner disk, comparing our mock-observed results to each other, as well as to the \emph{Kepler} sample. 
We test several models of pebble accretion isolation mass, including pebble isolation, flow isolation, and migration feedback isolation and otherwise adopt the same parameters for the gas disk.
We find that while each model can match at least one distribution of observables in the \emph{Kepler} catalog, they fall short of matching all distributions simultaneously, even with extreme reweighting. \changethree{Our inability to match all observations} suggests that \changethree{the initial conditions and/or modeled effects in our simulations that we held fixed should} be investigated. 
This exploration sheds light on how planetary systems evolve and the processes that influence the wide range of system parameters we observe today, helping place our own Solar System in context.

\end{abstract} 


\section{Introduction}\label{sec:intro}
The \emph{Kepler} Mission has revealed thousands of planetary systems with architectures that differ vastly from our own Solar System. 
Chief among this variation in system geometries are closely-orbiting planets with radii between that of Earth and the ice giants known as ``super-Earths'' or ``sub-Neptunes'', which have no Solar System analog.
These planets are quite common in the innermost 1 au of their systems \citep{Batalha2013} and exhibit a range of variations in orbital properties.
They can appear as ``dynamically cold'' systems that are tightly-spaced and exhibit ``peas-in-a-pod''-like features \citep{Millholland2017,Weiss2018}.
However, some systems can be ``dynamically hot'', with planets on more eccentric orbits with moderate mutual inclinations\change{\footnote{We define the mutual inclination of adjacent planets as the difference in their inclinations}} and wider spacings, like Kepler-69 \citep{Barclay2013}.
This considerable diversity of system architectures could be the result of any number of influencing factors, such as a perturbing outer gas giant \citep[e.g.,][]{Huang2017} or stellar obliquity causing a misalignment in the mutual inclinations of planets \citep{SpaldingBatygin2016}.



Previous work has shown that, if close-in super-Earths and sub-Neptunes form in situ \citep[e.g.,][]{ChiangLaughlin2013, HansenMurray2013}, conditions during the giant impact stage -- including the amount of solids in the formation region \citep{Dawson2015,MacDonald2020} and the amount of gas present at late stages \citep{Dawson2016} -- can affect the final orbital architecture of the system, as well as the properties of the planets within. 
\change{\citet{Dawson2015} found that whether a planetary embryo will go on to form a super-Earth or a sub-Neptune depends primarily on the solid surface density of the disk, with massive embryos forming more rapidly in higher surface density disks and accreting significant gas envelopes.
\citet{Dawson2016} explored how initial eccentricities, inclinations, spacings, and dynamical friction due to gas damping affect the final orbital properties and compositions of planets.
\citet{MacDonald2020} (henceforth referred to as \citetalias{MacDonald2020}) varied the slope of the distribution of solids, amount of gas damping, and solid surface density, attempting to disentangle the effects of the initial conditions.
They found that the distributions of most of the \emph{Kepler} observables, namely adjacent planet period ratio, mutual Hill spacing, transit duration ratio, and system transit multiplicity, can be reproduced by considering a distribution of initial disk solid surface densities near the end of the gas disk phase. 
Using the best fit simulations from \citetalias{MacDonald2020}, \citet{Morrison2020} found that they reproduced the correct fraction of resonant systems, finding that $\sim5\%$ of systems harbored at least one resonant pair. 
They also reproduced both dynamically cold systems with low eccentricities and small mutual inclinations, as well as dynamically hot systems with higher eccentricities and larger mutual inclinations; they did not however match the extended tail of the eccentricity distribution at $e>0.3$.
\citet{Sandhaus2025} resolved this issue by incorporating giant planets into the initial systems of \citetalias{MacDonald2020}, finding that a mixture of up to $10\%$ of systems dynamically interacting with exterior giants could maintain their successful output whilst also accounting for the dynamically hottest systems.
}


These studies assumed that the solids incorporated into planetary embryos were radially distributed according to an \emph{ad hoc} distribution and assumed a classic isolation mass, 
\change{which we reproduce below in Equation~\ref{eq:classicmass},} 
to build their initial embryos. 

The classic core accretion model \citep{Pollack1996}, where dust particles collide and merge to form cores of several $\mathrm{M_\oplus}$, is sufficient to explain the formation of large planets on large timescales, but cannot \change{alone} explain the formation of super-Earths in close-in orbits.
Observations suggest gas disk lifetimes of a few Myr \citep{Haisch2001,Kraus2012}.
\change{While core accretion can form super-Earth systems in situ within this timescale, the resulting planets are much more tightly packed than the observed population, tending to pile up at the inner edge of the disk \citep{Ogihara2015} if one does not invoke changes to the solid surface density, gas density, or the effects of giant planets within the system.}

The pebble accretion formation model, in which small solid particles are captured via gas drag and accreted by a larger body \citep[e.g.,][]{JohansenLacerda2010, OrmelKlahr2010}, can account for both short formation timescales and wider planet spacings. 
When solid particles reach sizes on the order of mm-cm, growth by particle-particle collisions halts and they settle toward the midplane of the protoplanetary disk, with gas drag causing rapid inward radial drift on timescales of 100-1000 years \citep{Weidenschilling1977}. 
\citet{LambrechtsJohansen2012} show that under the assumption of pebble accretion, planetesimals can accrete pebbles rapidly enough to form embryos well within the lifetime of the gas disk.
Here we investigate isolation masses arising from pebble accretion and how different assumptions about the limiting mass and radial distribution of planetary embryos affect the final orbital properties of simulated planetary systems.

\change{We maintain the assumption of \citet{Dawson2015, Dawson2016}, \citetalias{MacDonald2020} and \citet{Sandhaus2025} that the initial planetary embryos in our simulations form in situ, in that we do not artificially change their semi-major axes. Our assumption of in situ formation requires that the inner disk have sufficiently high \textit{local} densities to form the embryos. Such regions of high density can be achieved via pile-ups \citep[e.g., ][]{Drazkowska2016}, high particle drift rates \citep[e.g., ][]{Powell2019}, and even stellar fly-bys \citep[e.g., ][]{Su2026}.
\citetalias{MacDonald2020} and \citet{Sandhaus2025} were able to match observations under this assumption.
Other \emph{formation} models, such as the ``breaking the chains'' model \citep{Izidoro2021, Izidoro2022}, posit that large scale planetary migration is necessary to produce systems that match observations.
While these works were able to match observations like the distribution of period ratios and the intra-system similarity phenomenon \citep[``peas-in-a-pod'', e.g., ][]{Weiss2018}, they still produced an order of magnitude more resonant systems than observed.
Thus, we cannot consider only one assumption of formation to be valid and should continue to attempt a myriad of different models and assumptions to explain observations.}

\begin{figure}
    \centering
    \includegraphics[width = 0.45\textwidth]{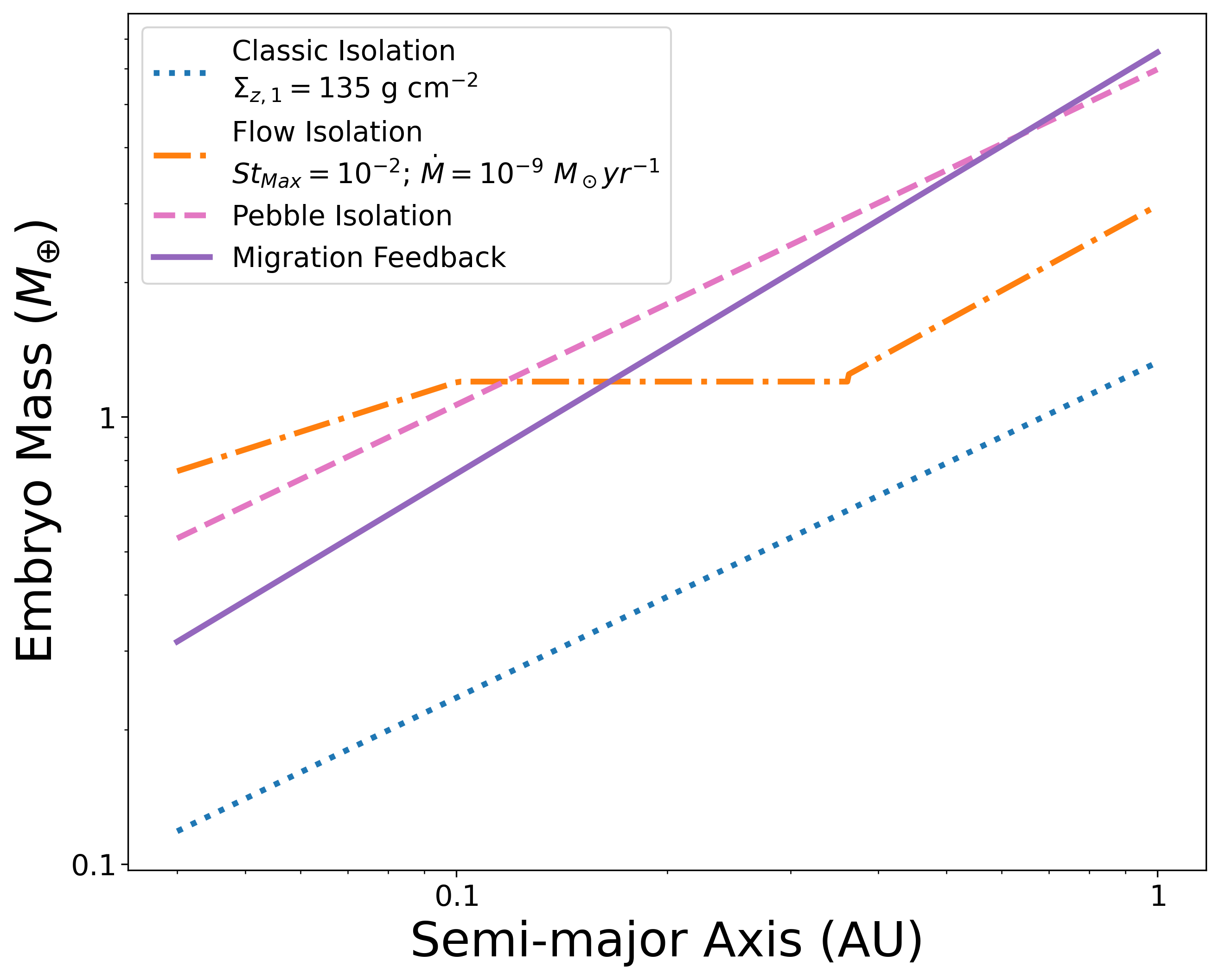}
    \caption{Embryo mass versus semi-major axis for the classic isolation (blue, dotted), flow isolation (orange, dot-dashed), migration feedback (purple, solid), and pebble isolation (pink, dashed) models.
    We plot and label the various models using typical values for their parameters, exhibiting the differences in mass distribution between each.}
    \label{fig:embryomass}
\end{figure}

The isolation mass is defined as the mass at which a planetary embryo ceases to accrete solids, halting further growth and isolating it from interactions with other embryos until giant impacts can commence. 
Many processes can govern how embryos isolate from the gas disk; we explore three prescriptions of the pebble isolation mass, comparing them to a baseline model without pebble accretion:  
(1) In Section~\ref{subsubsec:classicisomass}, we consider the case of the classic isolation mass \citep{LinPapaloizou1993,MacDonald2020} which is reached by consuming all of the material within the embryo's feeding zone, without necessitating pebble accretion. 
We use this model as a point-of-reference for the \change{other models}. 
(2) The pebble isolation mass (Section~\ref{subsubsec:pebbleiso}) is a product of density perturbations in the local disk surrounding the planetary embryo. 
These perturbations reverse the gas flow and halt pebble accretion by the embryo \citep{LJM2014}. 
(3) When an embryo's atmosphere grows large enough to overtake the impact parameter for pebble accretion, the pebbles will instead flow with the gas around the planet, halting accretion in a process called flow isolation \citep{RMC2020}, which we discuss in further detail in Section~\ref{subsubsec:flowiso}. 
(4) The final model we consider in Section~\ref{subsubsec:migrationfeedback} is the migration feedback isolation mass \citep{FungLee2018}. In this model, torques from the gas disk cause the embryo to drift inward, resulting in a pile-up of gas interior to the embryo's orbit. This gas exerts a feedback torque on the embryo, halting the migration and building a pressure bump which blocks the flow of pebbles onto the planet. 
In Figure~\ref{fig:embryomass}, we plot typical radial mass distributions of these four isolation mass prescriptions, showing how they vary from model to model \change{and comparing them to the Classic Isolation model.}

This paper is organized as follows.  
In Section~\ref{sec:isolationmasses}, we give context for the isolation mass models we use throughout this analysis. 
We also establish the assumptions and theoretical framework of our protoplanetary disk.
In Section~\ref{sec:methods}, we describe the properties of our $N$-body simulations, the initial conditions of each model, our method of mock-detecting final planetary systems, and the subset of \emph{Kepler} data we compare our results to.
Sections~\ref{sec:flowisoplanetprop},~\ref{sec:migrationfeedback_planet_props} and~\ref{sec:pebbleiso_planetprops} investigate the results of our $N$-body simulations. 
Sections~\ref{sec:flowiso_implications},~\ref{sec:feedback_implications} and~\ref{subsec:pebbleiso_implications}  discuss these findings more broadly and their implications for the diversity of planet formation. 
In Section~\ref{sec:conclusion}, we summarize our findings and detail future work.

\section{Isolation Masses} \label{sec:isolationmasses}
An embryo reaches the classic isolation (henceforth shortened to CI) mass when it accretes all of the material in its feeding zone. However, when the feeding zone is  being replenished by drifting material, this limit can be quite high. Growth due to pebble accretion can occur on timescales much shorter than the gas disk lifetime \citep{LambrechtsJohansen2012} if the feeding zone is continually replenished, which is one pathway that allows for embryos to grow to several Earth masses before dissipation of the gas disk. However, without a halting mechanism, these embryos can continue to grow beyond the terrestrial masses we observe in close-in systems \citep[e.g., ][]{Lin2018}. \citet{LJM2014} propose that growth halts when a pressure gradient is formed in the surrounding gas disk, trapping pebbles exterior to the planet's orbit. Here, we summarize the CI mass model and three formulations for the pebble accretion isolation mass.

\subsection{Disk Conditions} \label{subsec:disk}
 We assume planetary embryos are formed in situ from a reservoir of constantly replenished solid material within the inner disk while the gas disk is still present, accreting solid material until they reach a particular isolation mass.
 We assume the disk gas surface density ($\Sigma_g$) declines over time, eventually reaching a critical value of $\Sigma_{dis}$ \citep[equation 17 of][]{Owen2011} and dissipating entirely soon after, at some age $\tau_{dis}$ \citep{Owen2011, Owen2012}.
 Once the gas is fully dissipated, the planets undergo gravitational interactions unhindered by gas damping.
We \changethree{fully} describe this gas damping \changethree{of eccentricities and inclinations later} in Section~\ref{subsec:numsim}.
 We assume that planetary embryos grow in the presence of the gas disk and that the giant impact phase starts while gas is still present but partially depleted.

For the CI mass, we assume the surface density in solids $\Sigma_z$ varies as
\begin{equation}
    \label{eq:solidsurfdens}
    \Sigma_z = \Sigma_{z,1} \left(\frac{a}{\mathrm{au}}\right)^{-1.5},
\end{equation}
 where $\Sigma_{z,1}$ is the solid surface density at 1 au and $a$ is semi-major axis. \change{\citetalias{MacDonald2020} explored different slopes for this distribution and found that a slope of 1.5 produces systems that best match the distribution of period ratios, transit duration ratios, mutual Hill spacings, and transit multiplicities in systems observed by \emph{Kepler}.
 Therefore, we adopt a slope of 1.5 for the solid surface density, so as to isolate the effects of varying the isolation mass model.}
 We \change{model} the radial gas surface density and its evolution over time as
\begin{equation}
    \label{eq:gassurfdens}
    \Sigma_{g} = \begin{cases}
        1700 ~d^{-1} \left(\frac{a}{\mathrm{au}} \right)^{-1.5} ~\text{g cm}^{-2}, \hfill ~t< \tau_{dis} \\
        0, \hfill ~t > \tau_{dis}
    \end{cases}.
\end{equation}
Here $d$ represents the gas depletion factor of the disk, with $d = 1$ corresponding to the undepleted MMSN. 
We adopt $d = 100$ to represent a late stage, depleted gas disk. 
\change{\citet{Dawson2016} argued that the diversity of systems we observe could be explained through in situ formation if the giant impact stage occurred in the presence of the gas disk, but that large amounts of gas were not necessary. \citetalias{MacDonald2020} explored various levels of gas density, finding that this depletion factor of $d=100$ led to a population of planets that best matched observations. This depletion factor describes the amount of gas remaining once the embryos are finished forming, not the total amount of gas in the system after the initial formation of the disk. $d$ primarily represents a comparison of timescales, relating the time it takes to form embryos to the time it takes for the host star to remove gas from the disk. Here, we model Sun-like stars as the system hosts, which have gas disk lifetimes of roughly 3-5~Myr \citep[e.g.,][]{Kennedy2009}; a depletion factor in the disk then assumes that it also takes 3-5~Myr, but less than the dissipation timescale of the gas disk, for the embryos to fully form. }
We represent the dissipation of the gas disk as a step function, as we are beginning our simulations \change{near} the end of the disk lifetime, making the assumption that the disk depleted by a factor of $d = 100$ has reached the critical density $\Sigma_{dis}$ and will soon dissipate entirely.

\subsection{Classic Isolation Mass} \label{subsubsec:classicisomass}
The CI mass is the mass limit at which an accreting embryo consumes all solid material available in its local feeding zone. The final embryo mass is
\begin{equation}
    \label{eq:classicmass}
    M_{emb} = 2 \pi a(\Delta a)\Sigma_z,
\end{equation}
where $a$ is the semi-major axis, $\Sigma_z$ is the \change{local} solid surface density, and $\Delta a$ is the width of the \change{embryo's} feeding zone in the disk. If we substitute Equation~\ref{eq:solidsurfdens} for $\Sigma_z$ and assume $\Delta a$ is proportional to the Hill radius (Equation~\ref{eq:Hillrad} \change{below}), we can rewrite Equation~\ref{eq:classicmass} to match Equation 5 of \citetalias{MacDonald2020} and Equation 20 of \citet{Dawson2015}:
    \label{eq:classicmassmacdonald}
\begin{equation}
    \begin{split}
        \mathrm{M_p} = 0.16 ~\mathrm{M_\oplus} \left(\frac{\Delta_0}{3} \right)^{3/2}~ \left(\frac{\Sigma_{z,1}}{33 ~\mathrm{g ~cm^{-2}}} \right)^{3/2} \times \\ \left(\frac{a}{\mathrm{au}} \right)^{\frac{3}{2} \left(2 - \alpha \right)} ~\left(\frac{\mathrm{M_*}}{\mathrm{M_\odot}} \right)^{-1/2},
    \end{split}
\end{equation}
where $\Delta_0$ is the initial mutual Hill spacing (the number of Hill radii that separates each embryo in the disk), $M_*$ is the mass of the host star in solar units, and $\alpha$ represents the radial slope of the disk, typically $\alpha = 1.5$.

\subsection{Pebble Isolation Mass} \label{subsubsec:pebbleiso}
The pebble isolation mass (referred to as PI mass throughout the rest of the paper), \change{introduced} in \citet{LJM2014}, is determined primarily through numerical simulations of gravitational perturbations of the planetary embryo on the surrounding gas disk. 
If the embryo is massive enough, it can perturb the local disk surface density, causing an outward radial flare and \change{effectively} pushing pebbles away from the embryo \citep{PaardekooperMellema2006, MorbidelliNesvorny2012}. This isolation mass is given as Equation 12 of \citet{LJM2014}:
\begin{equation}
    \label{eq:pebbleiso_mmsn}
    \mathrm{M_{iso}} \approx 20 \left(\frac{a}{5~\mathrm{au}} \right)^{3/4}~\mathrm{M_\oplus}.
\end{equation}
The proportionality $M_{iso} \propto a^{3/4}$ is determined from the scale height of the disk, which  affects the amount of available material at a given location. 
The relationship between isolation mass and disk aspect ratio is expressed as $M_{iso} \propto \left(\frac{H}{a} \right)^3$, where $H$ is the disk scale height.
Here we assume a disk flaring like the minimum mass solar nebula (MMSN, \citealt{Hayashi1981}), $H/a = 0.05~ \left(\frac{a}{5~\textrm{au}} \right)^{1/4}$.

\change{To expand on previous models of the PI mass by more accurately accounting for the hydrodynamics of the protoplanetary disk, \citet{Bitsch2018} take into account disk viscosity, local pressure gradients, and pebbles traversing gaps opened within the disk gas when calculating the isolation mass.}

\change{While this model is more complex, we can replicate the initial conditions of the \citet{LJM2014} PI model with the \citet{Bitsch2018} isolation mass by assuming a local pressure gradient $\frac{d\ln{P}}{d \ln{r}} = -2.5$, a disk viscosity of $\alpha_{\mathrm{visc}} = 10^{-4}$, and a disk flaring like the MMSN.}
\change{Adopting the model from \citet{Bitsch2018} would require additional assumptions of local disk structure, primarily in exploring ranges and distributions of the local pressure gradients and the disk viscosity, in order to properly simulate and test it. We therefore adopt the simplified model of \citet{LJM2014}.
It is worthwhile to investigate the effects of these parameter ranges on planet properties within this model, but is beyond the scope of this work.}


\subsection{Migration Feedback Isolation Mass} \label{subsubsec:migrationfeedback}

\citet{FungLee2018} propose a related pebble isolation mass model called the migration feedback isolation (MFI) mass, in  \change{which a migrating embryo perturbs an inviscid gas disk, effectively halting} the flow of pebbles onto the embryo. \citet{Rafikov2002} determined the mass at which feedback from the perturbation halts an embryo's migration, parameterized in Equation 1 of \citealt{FungLee2018}. 
\change{We refactor their Equation 1 assuming a disk scale height $H$ that is consistent with the MMSN:}
\begin{equation}
    \label{eq:feedbackmass}
    \mathrm{M_{fb}} \approx 11.66 ~\mathrm{M_\oplus} ~ \left(\frac{a}{5~\mathrm{au}} \right)^{3/4} \left(\frac{\Sigma_g a^2/\mathrm{M_*}}{10^{-3}} \right)^{5/13},
\end{equation}
where $\Sigma_g$ is the \change{local} gas surface density, $a$ is the semimajor axis, and $M_*$ is the mass of the host star. 
As the gas is compressed by the migrating body, it forms a pebble trap \citep{LJM2014} \change{which effectively halts} the accretion of solid material \change{onto the embryo}. 
\citet{FungLee2018} \change{find that the resulting isolation mass is of order of $M_{fb}$ and} that the embryo will continue to accrete some material as it settles into its final orbit, \change{$M_{iso} \sim 1.4M_{\mathrm{fb}}$.} 
\change{\citet{FungLee2018} also find that the embryo experiences a slight migration once it reaches the feedback mass, which we account for in our initialization of embryos.} 

\subsection{Flow Isolation Mass} 
\label{subsubsec:flowiso}

In the flow isolation (FI) mass model \citep{RMC2018, RMC2020}, \change{gas and pebbles flow around the embryo instead of accreting onto it after the embryo reaches its FI mass. This mass is reached once the}  embryo's atmosphere grows beyond \change{the embryo's} accretion impact parameter.

\change{In the presence of a gas disk, the} size of the atmosphere is approximately the Bondi radius, the radius at which the escape velocity from the planet is equal to $c_s$,
\begin{equation}
    \label{eq:Bondi}
    R_B = \frac{G M_p}{c_s^2},
\end{equation}
where $M_p$ is the embryo mass and $c_s = \sqrt{\frac{\gamma k T}{\mu}}$ is the local gas sound speed.
The pebble accretion impact parameter 
\change{is defined as the radius at which} the gravitational force of the embryo equals the gas drag force $F_D$ \citep{RMC2020}:
\begin{equation}
    \label{eq: R_stab}
    R_{stab} = \sqrt{\frac{G M_p m}{F_D}}, 
\end{equation}
where $m$ is the mass of the pebble being accreted.\change{When $R_B > R_{stab}$, pebbles flow around the embryo instead of accreting onto it. The mass at which this occurs is what we define as the FI mass.} 
$R_{stab}$ \change{ultimately depends on the maximum particle size in the disk, because smaller particles experience stronger drag, and so \citet{RMC2020} use the Stokes number $St$ as a proxy for particle size, parameterizing the FI mass according to the maximum Stokes number.} 
\change{See Section 3.2 of \citet{RMC2020} for the explicit algorithm used to calculate the FI mass as a function of orbital distance.}

The FI mass will increase as both the maximum $St$ and the accretion rate increase. \change{As $St$ increases, or the maximum pebble size increases, $F_D$ significantly increases, effectively decreasing $R_{stab}$ without altering $R_B$. As such, the time required for $R_B$ to exceed $R_{stab}$ will increase.} 
Increasing the accretion rate will increase the temperature in the inner disk, \change{as $T\propto \dot{M}^{1/4}$}, which will in turn increase the local gas sound speed. \change{This increase in $c_s$ will in turn significantly decrease $R_B\propto c_s^{-2}$ while $R_{stab}$ will decrease as $R_{stab}\propto c_s^{-1}$, leading a longer accretion timescale until $R_B>R_{stab}.$}
The diversity in the radial mass distribution for any singular $St$ or  accretion rate is significant; \change{we show the effect of varying $St$ on the range of resulting masses in} Figure~\ref{fig:stokesrange} 
\change{and the effect on the resulting radial mass distribution in}
Figure~\ref{fig:mass_distributions}. 
\change{Because both the distribution of embryo masses and the radial distribution of masses vary, we} explore a range of these parameters for the initial conditions of our simulations. 

\section{Methodology} \label{sec:methods}
To study how the initial disk conditions and embryo distributions affect the final orbital properties of planets, we perform \emph{N}-body simulations of protoplanetary systems \change{near} the end of a typical disk lifetime using \texttt{REBOUNDX} \citep{ReinLiu2012, Tamayo2020}. \change{We employ} a custom force to model damping from a gas disk \citep{Sandhaus2025}, \change{which we detail below}.

Each model for the pebble accretion isolation mass that we explore exhibits traits which differentiate it from the CI mass model and other pebble accretion models.
The PI distribution follows the same same slope as the CI distribution, but produces embryos over three times more massive at all locations in the disk (Figure~\ref{fig:embryomass}).
Therefore, we expect the final planets in the PI model to be massive and dynamically hot.
In the MFI model, the isolation mass increases with $a$ more rapidly than other models, so we expect a more diverse population of final planets.
The FI mass remains quite flat as a function of $a$, out to a characteristic radius $r_{vis-irr}\sim0.3$ where the primary heating source for the disk transitions from viscous accretion to irradiation from the host star.
Beyond this value, the embryo mass will increase with $a$.
Where the mass distribution is flat, we expect FI planetary systems to generally be dynamically cold with low mutual inclinations.

\subsection{Numerical Simulations \label{subsec:numsim}}
To begin the simulations of late stage planet formation, we first create radial distributions of planetary embryos whose masses are determined by the three pebble isolation mass models described in Section~\ref{sec:isolationmasses}. 
The first embryo of a given simulation is placed at a semi-major axis $a_{inner}$ that is randomly drawn from a uniform distribution with a range of $0.04$ au to $0.06$ au, following the work of \citet{Dawson2016} and earlier \citet{HansenMurray2012, HansenMurray2013}.

Subsequent embryos are then placed out to a maximum of $1$ au, each separated by their mutual Hill spacing
\begin{equation}
    \label{eq:mutualhill}
    \Delta = \frac{a_2 - a_1}{R_H},
\end{equation}
where $a_1$ and $a_2$ are the respective semi-major axes of the inner and outer adjacent embryos, and $R_H$ is the Hill radius, defined as
\begin{equation}
    \label{eq:Hillrad}
    R_H = \frac{a_1 + a_2}{2} \left(\frac{M_{p,1} + M_{p,2}}{3 M_*} \right)^{\frac{1}{3}}.
\end{equation}
$M_{p,1}$ and $M_{p,2}$ represent the masses of the inner and outer adjacent embryos, respectively. 
Since we are building our initial population of embryos in order of increasing $a$, we have no prior knowledge of the mass or location of the outer embryo of a given pair. 
As a result, we calculate the spacing, and thus the new embryo's mass and location in the inner disk, using the Hill radius of the previous embryo: $R_H \simeq a\left( \frac{2M_{emb}}{3M_*}\right)^{1/3}$.
Due to the damping of the gas disk during embryo formation, initial inclinations and eccentricities are assumed to be quite small.
Each embryo's inclination is drawn from a uniform distribution between $0-0.01^\circ$, and the initial eccentricity is set to $0.0$.\footnote[1]{With \texttt{REBOUND}, if the initial inclination of every body in the system is set to (or otherwise reaches) exactly zero, then they will fall into an energy minimum and no longer be able to excite each other to larger inclinations.} \change{\citet{Dawson2016} performed integrations throughout the entire gas disk phase, and found that the final planet population of simulations that were integrated through the full disk phase were consistent with simulations that began at the depleted disk stage with equivalently small inclinations and zero eccentricity.}
The longitude of ascending node, argument of periastron, and mean anomaly are each drawn from independent uniform distributions spanning a full 360 degrees. 

Unless specified otherwise, initial mutual spacing of the embryos is $\Delta_0 = 3$. 
We choose this initial mutual Hill spacing, much tighter than $\Delta_0 \sim 10$ \citep{KokuboIda1998}, to allow the embryos to reach a self-consistent isolation mass with self-consistent orbital properties during the depleted gas stage.
The initial conditions of our simulations\changethree{---the location of the innermost embryo, the eccentricities, the inclinations,  and the spacings---are self-consistent and }mirror the properties of the embryos had we modeled the full gas disk stage, rather than the last 1 Myr \citep{Dawson2016}. They also found that among their ensembles of simulations, the embryos that had smaller initial spacings (specifically, $\Delta_0=3$) ended with wider final spacings, along with larger inclinations and eccentricities. 
At values of $\Delta_0 \gtrsim10$ embryos experience little to no significant instabilities throughout the duration of the simulation, leading to systems that do not resemble those that we observe \citep{BrefkaNote2026}.

Each suite of $N$-body simulations is integrated using \texttt{REBOUND}'s hybrid symplectic integrator \texttt{MERCURIUS}. \texttt{MERCURIUS} typically uses the symplectic integrator \texttt{WHFAST}, but when two bodies experience a close encounter at a distance of three Hill radii, it switches to using the high-order non-symplectic integrator \texttt{IAS15}. We use an initial timestep of 0.5 days for the \texttt{WHFAST} integrator and an accuracy of $10^{-9}$ for the \texttt{IAS15} integrator.


The three regimes of the gas damping timescale, as described in \citet{PapaloizouLarwood2000}, \citet{KominamiIda2002}, \citet{FordChiang2007}, and \citet{Rein2012PlanetDisk}, are given as:
\begin{equation}
    \label{eq:gasdamptimescale}
    \begin{split}   
        \tau = 0.003 d \left(\frac{a}{\mathrm{au}} \right)^2 \left(\frac{M_\odot}{M_p} \right) \text{yr} ~\times \\  
        \begin{cases}
            1, ~ \hfill v \leq c_s \\
            \left(v/c_s \right)^3, ~\hfill v > c_s,~ i < c_s/v_K \\
            \left(v/c_s \right)^4,~ \hfill i > c_s/v_K
        \end{cases}.
    \end{split}
\end{equation}
where \change{$d$ is the gas depletion factor of the disk}, $M_p$ is the mass of the body embedded in the disk in solar masses, $v_K = na$ is the Keplerian velocity (where $n$ is the mean motion of the body), and $v = \sqrt{e^2 + i^2}~v_K$, where $e$ and $i$ are the eccentricity and inclination respectively. \change{Here, we assume a sound speed of $c_s$=1.29~km~s$^{-1}~(a/$au)$^{-1/4}$, dependent only on the semimajor axis.}
The eccentricities and inclinations of bodies within the disk are damped according to $\dot{e}/e = -1/\tau$ and $\dot{i}/i = -2/\tau$, respectively. \changethree{We do not include Type I migration, as \citet{Dawson2016} and \citet{MacDonald2020} found that its inclusion does not alter the resulting population-level properties. Because the gas disk is depleted and the embryos are not massive, the inclusion of Type I migration serves only to increase the computational load.} 
See the appendix of \citet{Sandhaus2025} for further detail on the damping force. 

\change{We initialize our simulations with $d=100$, which physically relates to assuming that the embryos have reached their isolation masses after most, but not all, of the gas has dissipated.}
After 1 Myr \change{of integration within this depleted disk}, the gas disk fully dissipates, and we simulate the system for an additional 29 Myr. 
We assume particle collisions are perfectly inelastic and do not cause fragmentation, with a collisional radius corresponding to a typical density of $\rho = 1$ \text{g cm$^{-3}$}.

\subsubsection{Flow Isolation Initial Conditions} \label{sec:flowiso_init_conds}

For our suite of simulations which assumes the FI mass (Section~\ref{sec:flowisoplanetprop}), we follow Section 3.2 of \citet{RMC2020} to generate mass as a function of semi-major axis.
We choose to perform two suites of 100 simulations each, with different accretion rates of $\dot{M} = 10^{-9} ~\mathrm{M_\odot ~yr^{-1}}$ (referred to as FI1) and $\dot{M} = 4 \times 10^{-8} ~\mathrm{M_\odot ~yr^{-1}}$ (referred to as FI2), holding these values constant while we explore a range of Stokes numbers. 
We can thus more finely explore the effect of $St$ \change{on the resulting population} for a particular accretion rate, and coarsely explore how accretion rate affects planet formation for a given maximum pebble size.
The initial spacing of the FI embryos is held fixed at $\Delta_0=3$ to ensure that $St$ and accretion rate are the only two parameters which affect total embryo mass in the inner 1~au of the disk.

\begin{figure}
    \centering
    \includegraphics[width = 0.45\textwidth]{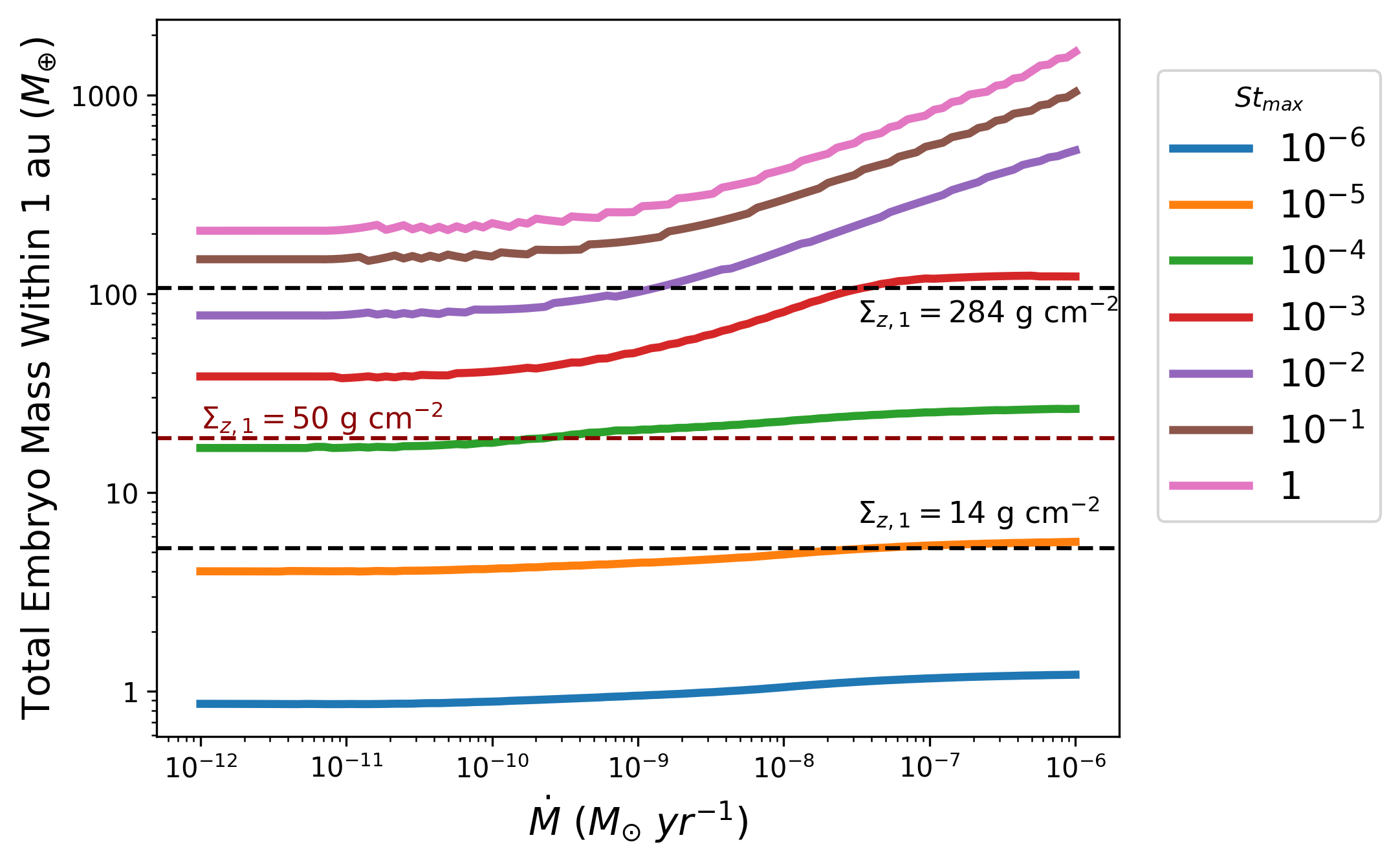}
    \caption{Total solid mass within inner 1 au of disk for Stokes numbers ranging from $10^{-6}$ to $1$ as a function of stellar accretion rate, assuming an initial mutual Hill spacing of $\Delta_0 = 3$. 
    These masses were determined by summing the individual embryos' FI masses \citep{RMC2020}.
    The dashed lines represent the total mass within 1 au given the radial profile from Equation~\ref{eq:solidsurfdens} and the listed solid surface density normalization for the CI masses. 
    Black dashed lines encompass the range explored in \citetalias{MacDonald2020} for the CI mass. $\Sigma_{z,1} = 50 ~\mathrm{g~cm^{-2}}$ represents the average solid surface density among transiting systems in their mock observed sample.}
    \label{fig:stokesrange}
\end{figure}

In Figure~\ref{fig:stokesrange}, we plot total mass within the inner disk as a function of solid accretion rate onto the host star for a range of discrete Stokes numbers. 
To motivate our parameter space, we look to the solid surface density range explored by \citetalias{MacDonald2020} and find the range of $St$ at each accretion rate which best matches the total solid mass, given the corresponding total mass calculated from solid surface density.
For both \change{simulation suites, }
we choose a lower bound of $St\geq10^{-5}$ to match the lower limit of solid surface density used by \citetalias{MacDonald2020}.
For the low accretion rate \change{suite}, we set an upper limit of $\mathrm{St_{upper}} = 10^{-2}$, and for the high accretion rate \change{suite,} we set $\mathrm{St_{upper}} = 10^{-3}$, consistent with the upper solid surface density limit. 
These suites are trimmed from 100 simulations down to 86 and 78 for FI1 and FI2 respectively to properly match the total mass range of \citetalias{MacDonald2020}.

These ranges of $St$ are motivated by Figure~\ref{fig:stokesrange}, wherein we seek to match the range of total embryo mass within the inner disk presented by \citetalias{MacDonald2020} \change{whose simulations qualitatively matched} \emph{Kepler} observable properties.

\subsubsection{Migration Feedback Isolation Initial Conditions} \label{sec:fb_init_conds}

At a given semi-major axis, the MFI mass (Equation~\ref{eq:feedbackmass}, \citealt{FungLee2018}) depends only on the gas scale height and gas surface density.
We expect embryos to form during the undepleted gas stage, but under some scenarios, their formation could be delayed to the depleted gas disk stage --- for example, if a low gas surface density is needed for large planetesimals to form via a mechanism like the streaming instability \citep[e.g.,][]{Youdin2005} and kickstart the pebble accretion process. 
To explore the effect of various initial conditions, we perform two suites of simulations: 150 simulations in which the depletion factor \change{of the gas disk }is set to $d = 1$ (labeled MFI1) and another 100 simulations where $d = 100$ (labeled MFI2), representing initial embryo formation in the undepleted and depleted disks, respectively. 
Once again we determine our parameter space by implicitly varying total embryo mass to match the \citetalias{MacDonald2020} solid surface density range.

In Figure~\ref{fig:hillrange}, \change{we compare} the total mass as a function of initial Hill spacing $\Delta_0$ for the MFI1, MFI2, and PI models.
As $\Delta_0$ increases, the total mass in the inner disk will decrease, as there will be fewer embryos in the initial simulation.
Rather than fixing $\Delta_0 = 3$, we draw $\Delta_0$ from a uniform distribution $\mathcal{U} \left(1,5 \right)$ for the depleted disk case and $\mathcal{U} \left(1,18 \right)$ for the undepleted disk. 
These ranges allow us to vary the total disk mass within 1 au and to account for the uncertainty in how disk structure affects the initial spacing of embryos, while staying within the total embryo mass range presented in \citetalias{MacDonald2020}. 

\begin{figure}
    \centering
    \includegraphics[width = 3.3in]{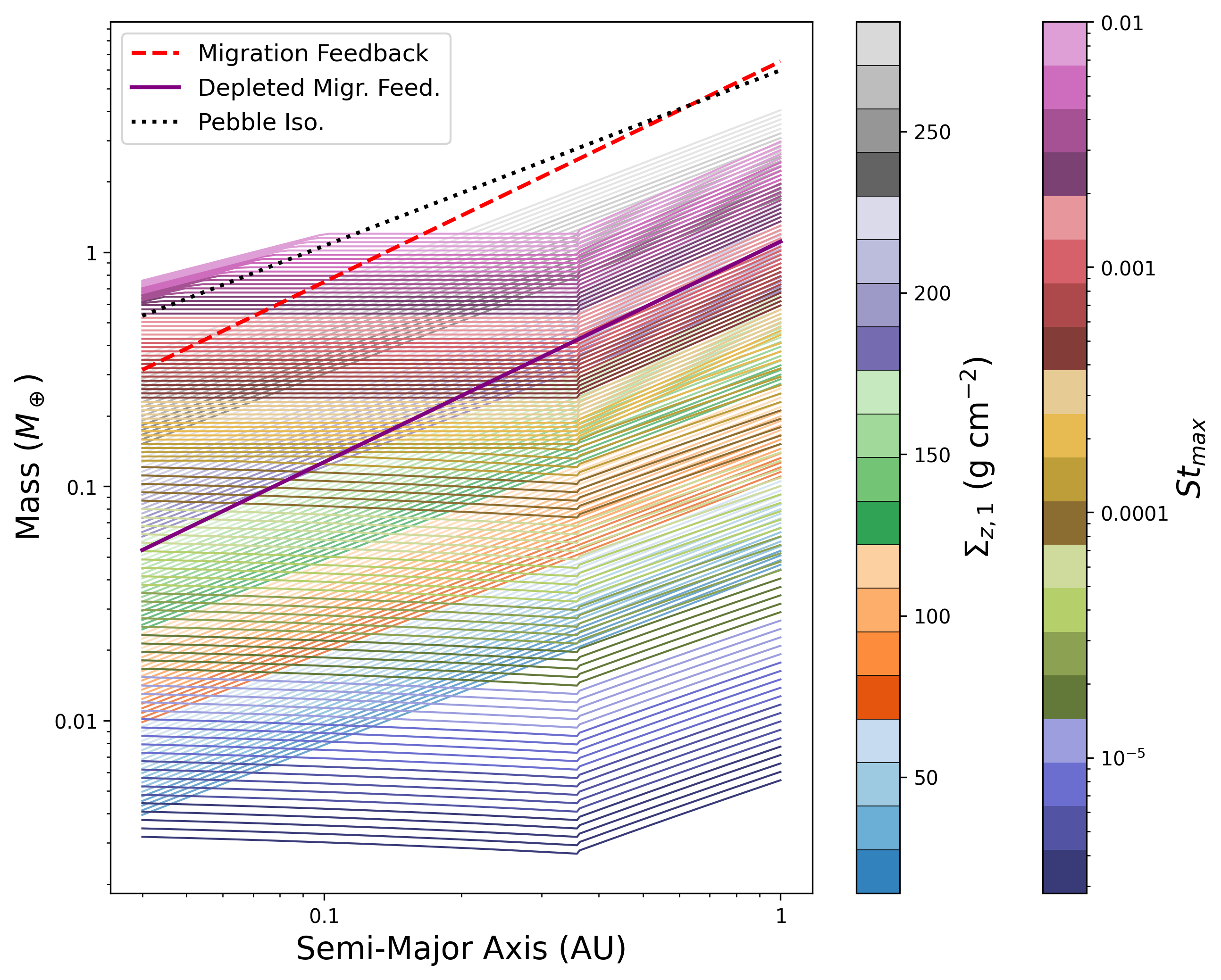}
    \caption{CI mass as a function of radial distance from the host star, for a continuum of solid surface densities.
    Over-plotted is the FI1 mass ($\dot{M}=10^{-9}~\mathrm{M_\odot~yr^{-1}}$) for a continuum of maximum Stokes numbers, as well as the MFI1 (opaque purple) and MFI2 (dashed red) masses and the PI mass (dotted black).
    The last three show no continuum of mass distributions, since we vary their initial mutual Hill spacing within the simulation. 
    Thus, all initial simulations begin from the same mass distribution model, albeit with a different number of embryos at different spacings.}
    \label{fig:mass_distributions}
\end{figure}

\subsubsection{Pebble Isolation Mass Initial Conditions} \label{sec:pebbleiso_init_conds}
\begin{figure}
    \centering
    \includegraphics[width = 0.45\textwidth]{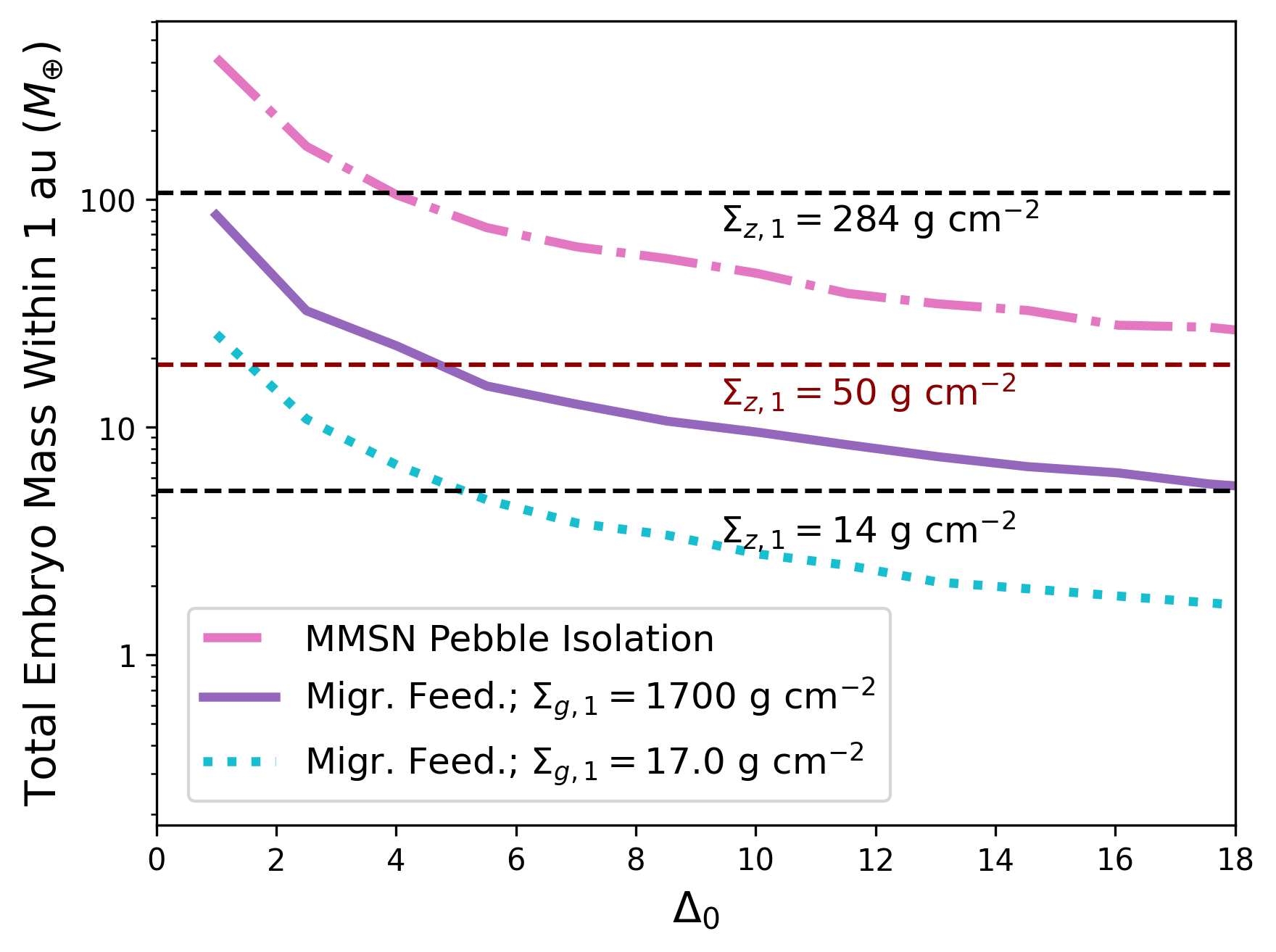}
    \caption{MFI mass \citep{FungLee2018} and PI mass \citep{LJM2014}: total mass of embryos  within inner 1 au of disk for mutual Hill spacings ranging from $\Delta_0 = 1$ to $18$. 
    The solid purple curve corresponds to an undepleted gas disk with density commensurate with the MMSN and the dotted cyan curve to a depleted disk with $d = 100$ (Eqn.~\ref{eq:gassurfdens}).
    The dot-dashed pink curve corresponds to the PI mass, assuming an MMSN-like initial disk. 
    The dashed horizontal lines represent the total mass within 1 au given the radial profile from Equation~\ref{eq:solidsurfdens} and the listed solid surface density normalization for the CI masses. 
    Black dashed lines encompass the range explored in \citetalias{MacDonald2020} for the CI mass. $\Sigma_{z,1} = 50 ~\mathrm{g~cm^{-2}}$ represents the average solid surface density among transiting systems in their mock observed sample.    
    A range of initial spacings within these total mass bounds allow us to vary the total embryonic mass in the system similar to how solid surface density is varied in \citetalias{MacDonald2020}.
    \label{fig:hillrange}}
\end{figure}

Similar to the MFI mass, the PI mass does not depend on modifiable parameters which directly alter the total mass within the inner disk, and so we instead vary the initial mutual Hill spacing of the embryos. 
We initialize 100 simulations by drawing $\Delta_0$ from the distribution $\mathcal{U} \left(2,12 \right)$ to capture as wide a range of total masses as possible (see Figure~\ref{fig:hillrange}). 
The upper Hill spacing limit is the largest value for $\Delta_0$ which still permits giant impacts between bodies, and so is chosen for our upper limit as well. 

To summarize, we explore a range of parameters for five separate isolation mass cases.
Two \change{suites }investigate the FI mass: one where we assume $\dot{M} = 10^{-9}~\mathrm{M_\odot~yr^{-1}}$ and draw $St$ from a log-uniform distribution between $10^{-5}-10^{-2}$ \change{for a total of} 86 simulations (FI1), and another where we assume $\dot{M} = 4\times10^{-8}~\mathrm{M_\odot~yr^{-1}}$ with 
$St$ between $10^{-5}-10^{-3}$ \change{for a total of} 78 simulations (FI2). 
Two other \change{suites} are governed by the MFI mass. 
\change{In MFI1,} embryos form within an undepleted, MMSN-like disk, and we explore a uniform range of \change{initial embryo} spacings within $1<\Delta_0<18$ \change{across 150 simulations}. 
\change{In MFI2,} embryos are formed within a gas disk depleted by a factor of 100 relative to the MMSN, and through 100 simulations we explore a uniformly distributed range of initial embryo spacings of $1-5$.
The final suite we investigate is governed by the PI mass, wherein we uniformly explore the initial embryo spacing range $2<\Delta_0<12$ across 100 simulations.

We \change{summarize} the initial simulation conditions for these five \change{suited} in Table~\ref{tab:init_table}, and in Figure~\ref{fig:mass_distributions}, we show the isolation mass as a function of radial distance for each model against the range of CI mass distributions from \citetalias{MacDonald2020}.
In the case of the PI and both MFI models, the mass distributions do not change as a function of $\Delta_0$, only the number of embryos.

\begin{table*}
    \centering \ra{1.3}
    \begin{tabular}{*9ll}    \toprule
    \textbf{Model} & $e_0$ & $i_0~\left( ^\circ \right)$ & $St$ & $\Delta_0$ & $N_{\rm{sims}}$ & $N_{\rm{emb},0}$ & $N_{\rm{final}}$ & $M_{\rm{tot}}~ (M_\oplus)$ & $\rm{NAMD_0}$ \\\midrule
        CI & 0 & $0.01h/ \sqrt{3}$ & N/A & 3 & 100 & 70 -- 370 & 2 -- 20 & 5.2 -- 68.9 & $1.4\times10^{-3}-1.5\times10^{-3}$\\
        FI1 & 0 & 0 -- 0.01 & $\mathcal{U} \left(\mathrm{\log{10^{-5}}}, \log{10^{-2}} \right)$ & 3 & 86 & 82 -- 770 & 6 -- 19 & 4.0 -- 103.2 & $1.3\times10^{-3}-1.3\times10^{-2}$\\
        FI2 & 0 & 0 -- 0.01 & $\mathcal{U} \left(\mathrm{\log{10^{-5}}}, \log{10^{-3}} \right)$ & 3 & 78 & 70 -- 728 & 4 -- 17 & 4.6 -- 71.3 & $1.1\times10^{-3}-1.2\times10^{-2}$\\
        MFI1 & 0 & 0 -- 0.01 & N/A & $\mathcal{U}\left(1, 18 \right)$  & 150 & 19 -- 304 & 3 -- 29 & 9.3 -- 85.6 & $3.0\times10^{-4}-5.2\times10^{-3}$\\
        MFI2 & 0 & 0 -- 0.01 & N/A & $\mathcal{U}\left(1, 5 \right)$ & 100 & 147 -- 778 & 7 -- 15 & 4.77 -- 26.5 & $2.4\times10^{-3}-1.3\times 10^{-2}$\\
        PI & 0 & 0 -- 0.01 & N/A & $\mathcal{U}\left(2, 12 \right)$ & 100 & 20 -- 54 & 3 -- 13 & 38.6 -- 59.1 & $4.0\times10^{-4}-4.4\times10^{-4}$ \\\bottomrule
     \hline
    \end{tabular}
    \caption{Initial conditions for each pebble accretion isolation mass model, as well as the CI model.
    Here $ h= \left ( \frac{M_{p,1} + M_{p,2}}{3M_*}\right)^{1/3}$. 
    Every model exhibits a wide range of initial embryo masses and distributions. NAMD is the Normalized Angular Momentum Deficit first presented in \citet{Chambers2001}. The limits of $M_{\rm{tot}}$ are determined by the maximum mass of all remaining systems for a given suite.}
    \label{tab:init_table}
\end{table*}

\subsection{Mock Detection of Transiting Planets}
\label{sec:detect_transit}

We compare the results of our simulations to the DR25 \emph{Kepler} catalog \citep{Thompson2018} as an additional baseline. 
We restrict the observable dataset to systems with host stars $4100~K~<~T_{\mathrm{eff}}~<~6100~K$, $\mathrm{log}~g > 4$, Kepler magnitude $< 15$, and those which contain at least one planet with $R < 4~R_\oplus$, remaining consistent with the limits chosen for the \emph{Kepler} dataset in \citetalias{MacDonald2020}. 
We refer to this subset of the DR25 catalog as ``the \emph{Kepler} catalog''.

\change{Within our simulations,} we exclude parts of parameter space that produce planet masses greater than $\mathrm{M_{p, ~max}} = 30 ~\mathrm{M_\oplus}$. 
For the FI model, we remove all simulations with $St$ greater than the minimum value that produces planets with $\mathrm{M_{p,~max}}$; for the case in which $\dot{M} = 10^{-9} ~\mathrm{M_\odot ~yr^{-1}}$, we remove no simulations.
For $\dot{M} = 4 \times 10^{-8} ~\mathrm{M_\odot ~yr^{-1}}$, we remove simulations with Stokes numbers greater than $5.722 \times 10^{-4}$, which excludes 24 of the original 78 simulations. 

For the MFI suites, \change{neither of the depleted (MFI2) or undepleted (MFI1) suites result in unrealistically} massive planets.
In the latter case however, beyond $\Delta_0 \sim 11.08$ the initial embryos remain totally isolated from each other, failing to dynamically interact and undergo giant impacts.
34 of 150 simulations occur at initial Hill spacings at or greater than this value, and are subsequently removed from consideration.

In the case of the PI mass, only those simulations with $\Delta_0 \geq 6.99$ produced planets with low enough mass to meet our criteria, removing 38 out of 100 simulations. 
The remaining systems' initial spacings are all below $\Delta_0 = 12$, meaning the embryos still undergo giant impacts.

The radius of each planet in the surviving simulations of each model is then determined by applying the mass-radius relationship presented in \citet{ChenKipping2017}.\change{\footnote{We verify that our results are robust to our choice of mass-radius relationship by testing two other relationships.}}

Next, we duplicate each final simulated system $10^4$ times, randomly orienting the mean orbital plane with respect to a fixed observer and labeling the planets within as transiting if their impact parameters $b$ meet the criteria $b < 1$.
We then compute detection probability over a grid of impact parameters, planet radii, orbital period, and photometric precision values following \citetalias{MacDonald2020}, who adapt the results from \citet{Burke2015,Christiansen2015,Christiansen2016}. 
For any realizations that result in at least one transiting planet, we draw a random number from a uniform distribution between 0 and 1 for that planet. 
Every transiting planet in that realization with detection probability above that drawn value is considered to be detected. 
For a more detailed explanation of this mock observation process, see Section 2.3 of \citetalias{MacDonald2020}. 


\subsection{Kepler Catalog Observables} \label{sec:keplerobservables}

To compare our mock observed planets to the known \emph{Kepler} catalog distribution, we follow \citet{Dawson2016} and \citetalias{MacDonald2020} to compare four primary observables, each exhibiting their own imprint of formation history and/or orbital architecture. 
These observables are the adjacent planet period ratio, adjacent planet mutual Hill spacing, transit multiplicity, and transit duration ratio normalized by orbital period. 
The distribution of period ratios offers some insight into the orbital structure of planetary systems, with smaller ratios being indicative of tightly spaced systems. 
Directly related to this parameter is the mutual Hill spacing, which indicates planet-planet separation as a function of the adjacent planets' masses.
The Hill spacing is generally more important for orbital evolution when planets are not near a mean-motion resonance or in a very closely spaced system. 
A system's transit multiplicity indicates the total number of planets observed to transit the host star and is set by the true underlying multiplicity distribution and mutual inclinations between adjacent planets. 
The normalized transit duration ratio is given by the equation \citep{FangMargot2012, Fabrycky2014}
\begin{equation}
    \label{eq:transdurratio}
    \xi = \frac{T_{dur,1}}{T_{dur,2}} \left( \frac{P_1}{P_2} \right)^{-\frac{1}{3}},
\end{equation}
where $T_{dur}$ is the transit duration, $P$ is the orbital period, and $1$ and $2$ represent the inner and outer planets in an adjacent planet pair, respectively. 
The distribution of $\mathrm{log_{10}} \xi$ peaks close to zero and is skewed positive by coplanarity. 
Non-zero mutual inclinations and eccentricities widen the distribution and reduce skewness.

\subsection{Re-weighting the Underlying Distribution of Formation Conditions} \label{sec:reweight}

In each of our simulations, we choose particular parameters to vary, such as $St$ for the flow isolation mass or $\Delta_0$ for the migration feedback isolation mass, to explore the distribution of initial conditions for a given model. 
These parameters are drawn from independent log uniform and uniform distributions, respectively, but in reality, we do not have any prior knowledge about the underlying distribution of these properties.
\citetalias{MacDonald2020} draw their solid surface density normalization ($\Sigma_{z,1}$) from a log uniform distribution, and to account for this aforementioned lack of knowledge, they re-weight the distribution of $\Sigma_{z,1}$ to improve the qualitative fit to the \emph{Kepler} catalog observables. 
The re-weighting function is given as 
\begin{equation}
    \label{eq:reweighting}
    W = \mathrm{exp} \left[- \frac{\left(\mu - \mu^-\right)^2}{2 {\sigma^-}^2}\right],
\end{equation}
where $\mu$ is the parameter distribution being re-weighted and $\mu^-$ and $\sigma^-$ are constants that dictate which parameter values are preferentially included in the re-weighted distribution. 

\change{\citetalias{MacDonald2020} compared the results of various depletion factors of the gas disk and different slopes $\alpha$ of the solid distribution.
To truly compare these parameters, and not at a fixed amount of mass, the reweighting parameters acted as additional degrees of freedom in their model.
}
For each of the $10^4$ system realizations described in Section~\ref{sec:detect_transit} with at least one detected planet, we draw a value from a uniform distribution $\mathcal{U} \left(0,1\right)$, and include the simulation in the weighted data set if the random number is less than $W$. 
For the PI, FI, and MFI models, the best fitting parameters are not able to be determined by hand. 
Instead, we determine the best fitting values by minimizing the summation

\begin{equation}
\label{eq:EMD_min}
    l_{T}= \sum_i\frac{l_{i}}{d_{i}},
\end{equation}
where $l_i$ is the Earth-mover's distance (EMD) calculated between any of the four observable property distributions and the \emph{Kepler} catalog counterpart, and $d_i$ is that property's inherent distance dispersion, determined by calculating the EMD for two randomly bootstrapped distributions of the property for 1000 samples, and then taking the standard deviation about the mean of this distribution of distances.
We discuss the Earth-mover's distance in more detail in Section~\ref{sec:EMD}. For consistency, we apply this minimization to the unweighted results of \citetalias{MacDonald2020}, finding $\mu^- = -8.6 \times10^{-4} ~\mathrm{g~cm^{-2}}$ and $\sigma^- = 88.1~\mathrm{g~cm^{-2}}$ to provide the lowest $l_T$. 
These weighting parameters are quite close to the by-eye reweighting of $\mu^- = 0$ \text{g~cm$^{-2}$} and $\sigma^- = 65$ \text{g~cm$^{-2}$} from \citetalias{MacDonald2020}, and so in subsequent figures the CI distribution is represented using the EMD-minimization reweighting method, unless otherwise noted. 

\change{In this work, we use the reweighting function in the case where the observable distributions of the suites themselves do not match the \emph{Kepler} catalog. 
The degree of reweighting is itself indicative of the difficulty for a model to match observations, with more dramatic reweighting indicating a model which likely cannot match observation without exploring additional factors.}

\subsection{Using the Earth Mover's Distance to Quantify Matches} \label{sec:EMD}



To properly quantify whether the observables of any particular pebble accretion isolation mass model match the distributions within the \emph{Kepler} catalog, we use the ``Earth Mover's distance'' \citep{Rubner1998}, also known as the Wasserstein metric, to determine a distance between the modeled and observed distributions. 
For one-dimensional distributions, this distance is measured by evaluating the expression
\begin{equation}
    W_1(\mu_1, \mu_2) = \int_\mathbb{R} \left | F_1(x) - F_2(x)\right | ~dx, 
\end{equation}
where $\mu_1$ and $\mu_2$ represent the probability distributions we seek to find a distance between and $F_1(x)$ and $F_2(x)$ are their cumulative distribution functions. 
The metric quantifies the minimal cost of modifying the distributions $\mu_1$ and $\mu_2$ such that they become identical; in other words, the distance one would need to move dirt from one pile to another in order for them to appear the same.
Thus, ``Earth Mover's'' distance (henceforth referred to as the ``EMD''). 
For our purposes, we calculate the EMD between the mock observed parameters of each of our isolation mass models and the true observed distributions from the \emph{Kepler} catalog to quantify how close each model gets to making a match. 
We also calculate the EMD between \emph{Kepler} catalog data and the CI results from \citetalias{MacDonald2020} as a basis for comparison to previous work. 
Furthermore, since we draw our observed planets from a random uniform distribution of viewing angles for a particular system, we recalculate EMD for 1,000 bootstrapped samples of a varying number of planets/planet pairs each to \change{construct an }uncertainty estimate on the EMD calculated for the total distribution.
\changethree{We then resample each observable of the \emph{Kepler} catalog 1,000 times with replacement to generate EMD distributions which represent the intrinsic variation in EMD for the catalog itself. These are used for direct comparison of our models to observation.}

To determine if our models more closely match the observed distributions with respect to the CI results, we compare their bootstrapped EMD distributions to those of the latter case, our updated EMD minimization reweighting of the results from \citetalias{MacDonald2020} described in Section~\ref{sec:reweight} (dotted, blue).
We include this distribution in Figures~\ref{fig:flowiso_emd},~\ref{fig:migration_emd} and~\ref{fig:lamb_emd}.

\subsection{Long-term Stability of Systems}

\change{\citet{Dawson2016} performed additional integration on three of their ensembles to test the long-term stability of their planetary systems. 
Integrating for 300 Myr, they find that the eccentricities remained approximately unchanged, whereas the median values of $\Delta$ increased by $\sim10\%$.
However, they used the CI mass for their initial embryo masses.}

\change{As a result, we select a random sample of 10 simulations from all five suites and integrate each out to 500 Myr.
None of the systems went unstable during this extended integration, and the distributions of the observables were not significantly different.
We are therefore comfortable with assuming that our integration time of 30 Myr is sufficient to produce a population to compare to observations.}



\section{Properties of Flow Isolation Planets} \label{sec:flowisoplanetprop}

\subsection{Properties Assuming $\dot{M} = 10^{-9} ~\mathrm{M_\odot ~yr^{-1}}$} \label{sec:flowiso_1e-9_props}
\begin{figure}
    \centering
    \includegraphics[width = 0.5\textwidth]{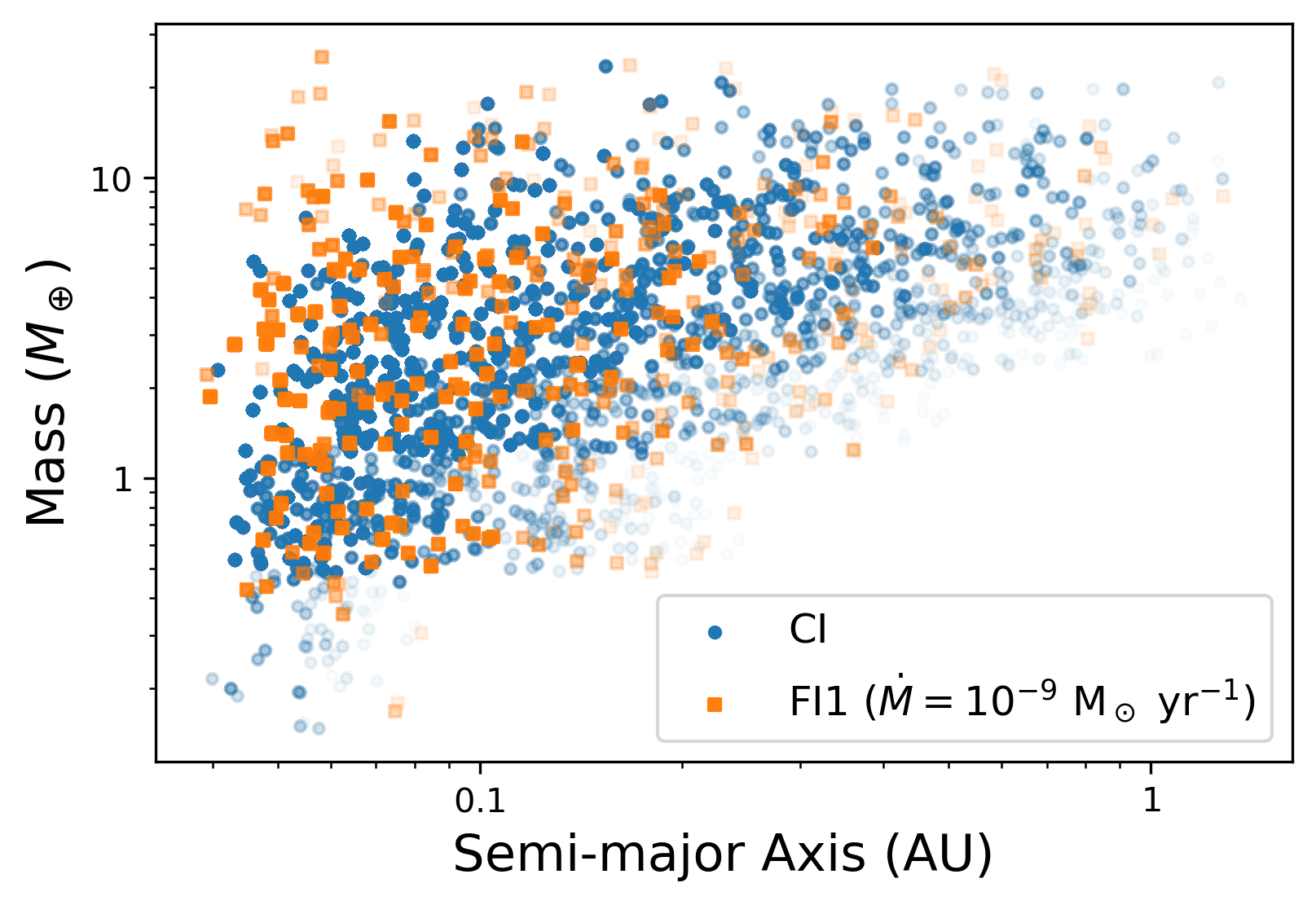}
    \caption{Mass as a function of semi-major axis for the forward modeled transiting planets from the FI1 (orange squares) model and \citetalias{MacDonald2020} CI with $\alpha=3/2$ (blue circles) model. 
    Mass tends to increase with semi-major axis for both cases, with the FI1 model producing fewer planets than CI at $a>0.4$. 
    More opaque points are detected more frequently as transiting planets.}
    \label{fig:flowiso_1e-9_mass_vs_a}
\end{figure}

We compare the final observed FI1 planetary mass as a function of semi-major axis in Figure~\ref{fig:flowiso_1e-9_mass_vs_a} to planetary mass assuming CI from \citetalias{MacDonald2020}. 

\begin{figure*}
    \centering
    \includegraphics[width = 3.2in, height = 2.1in]{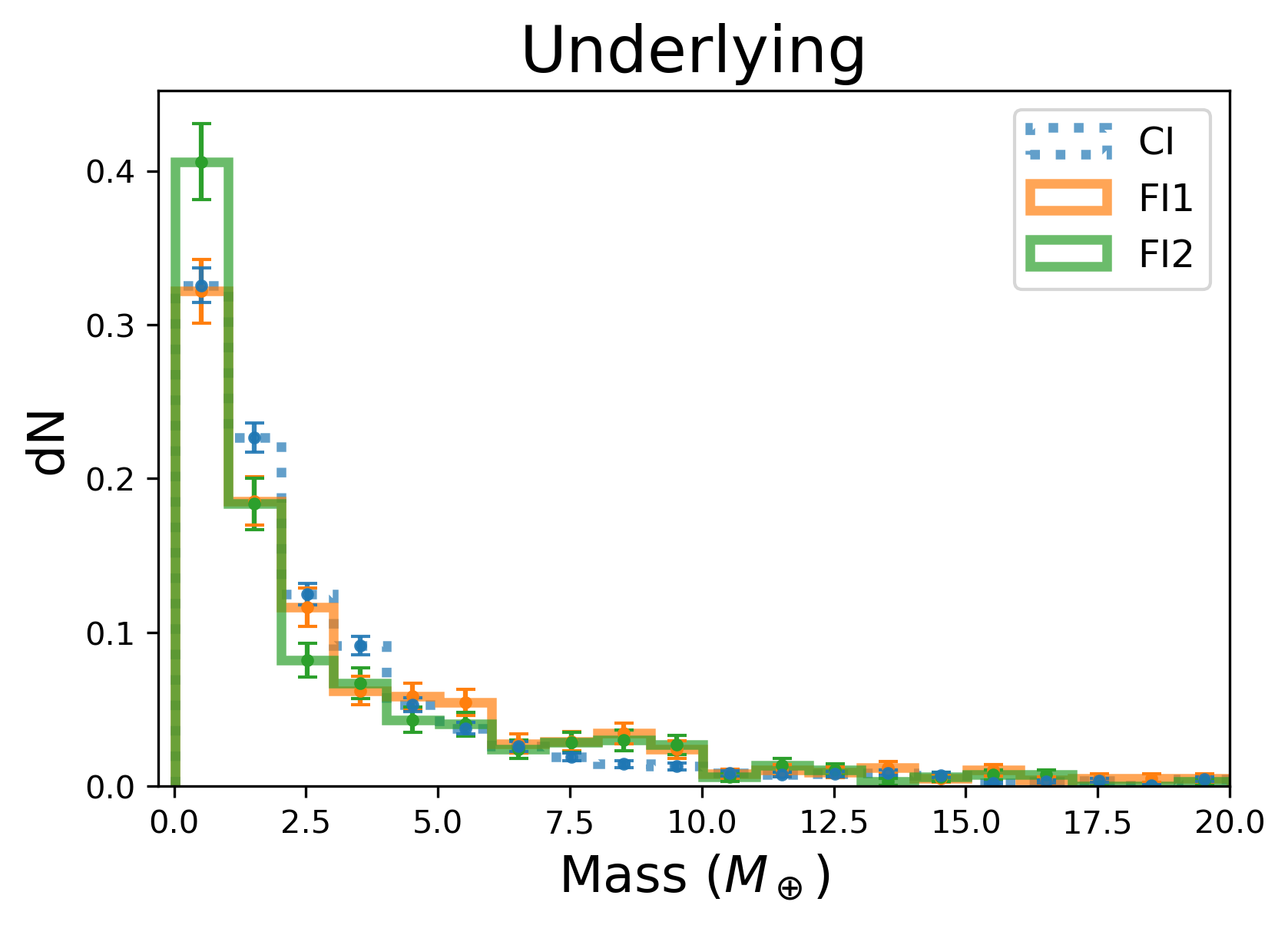}
    \includegraphics[width = 3.2in, height = 2.1in]{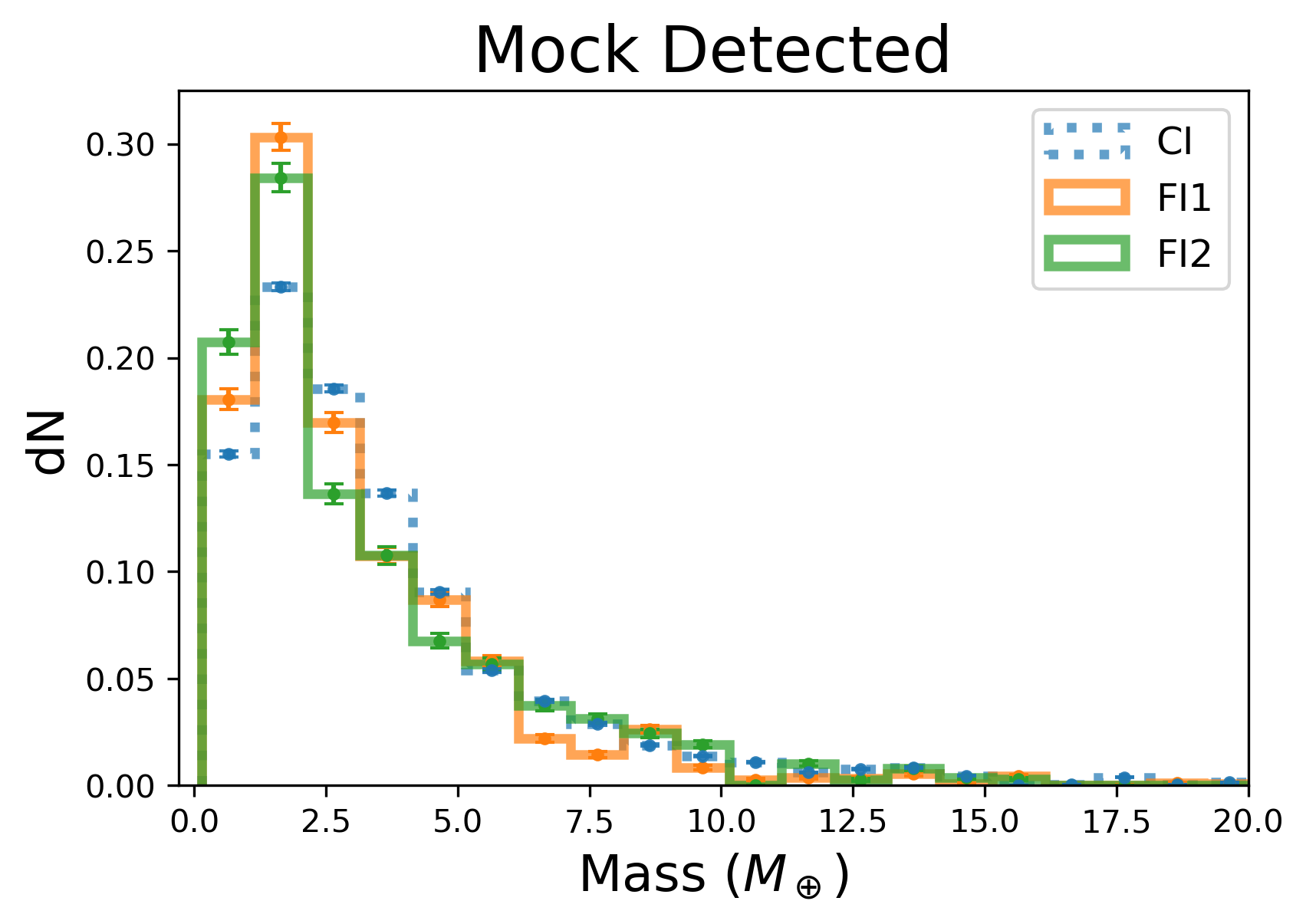}
    \includegraphics[width = 3.2in, height = 2.1in]{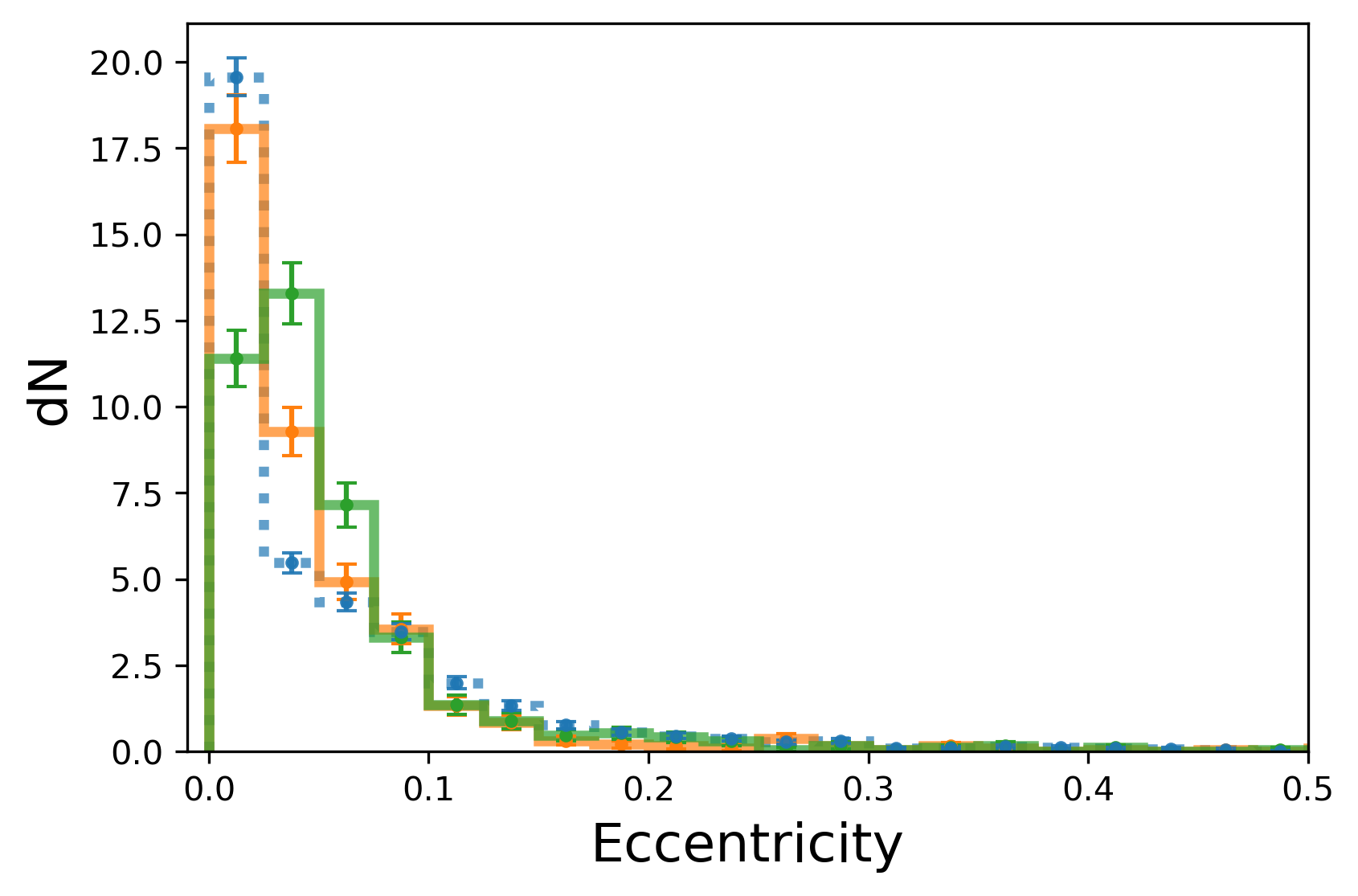}
    \includegraphics[width = 3.2in, height = 2.1in]{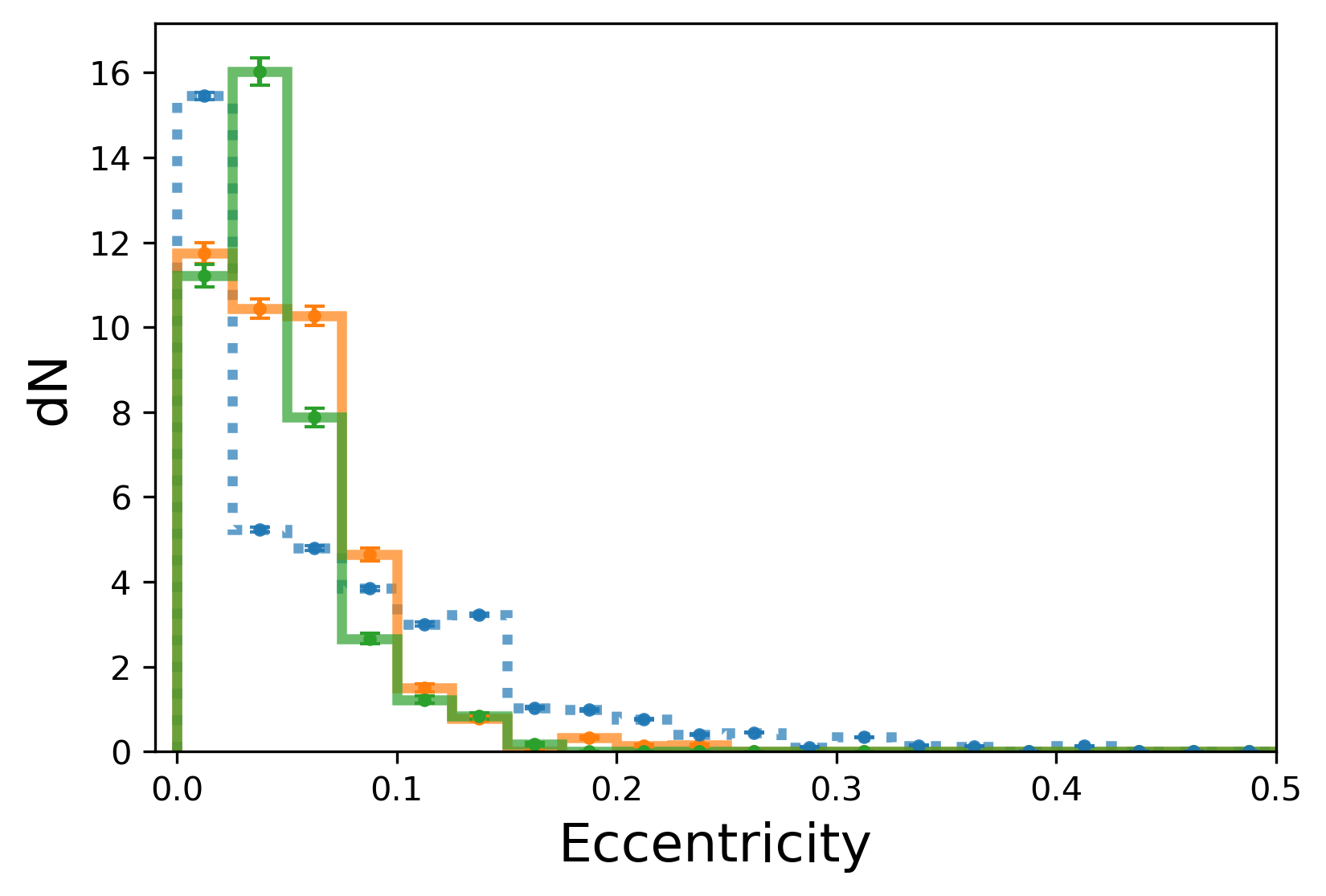}
    \includegraphics[width = 3.2in, height = 2.1in]{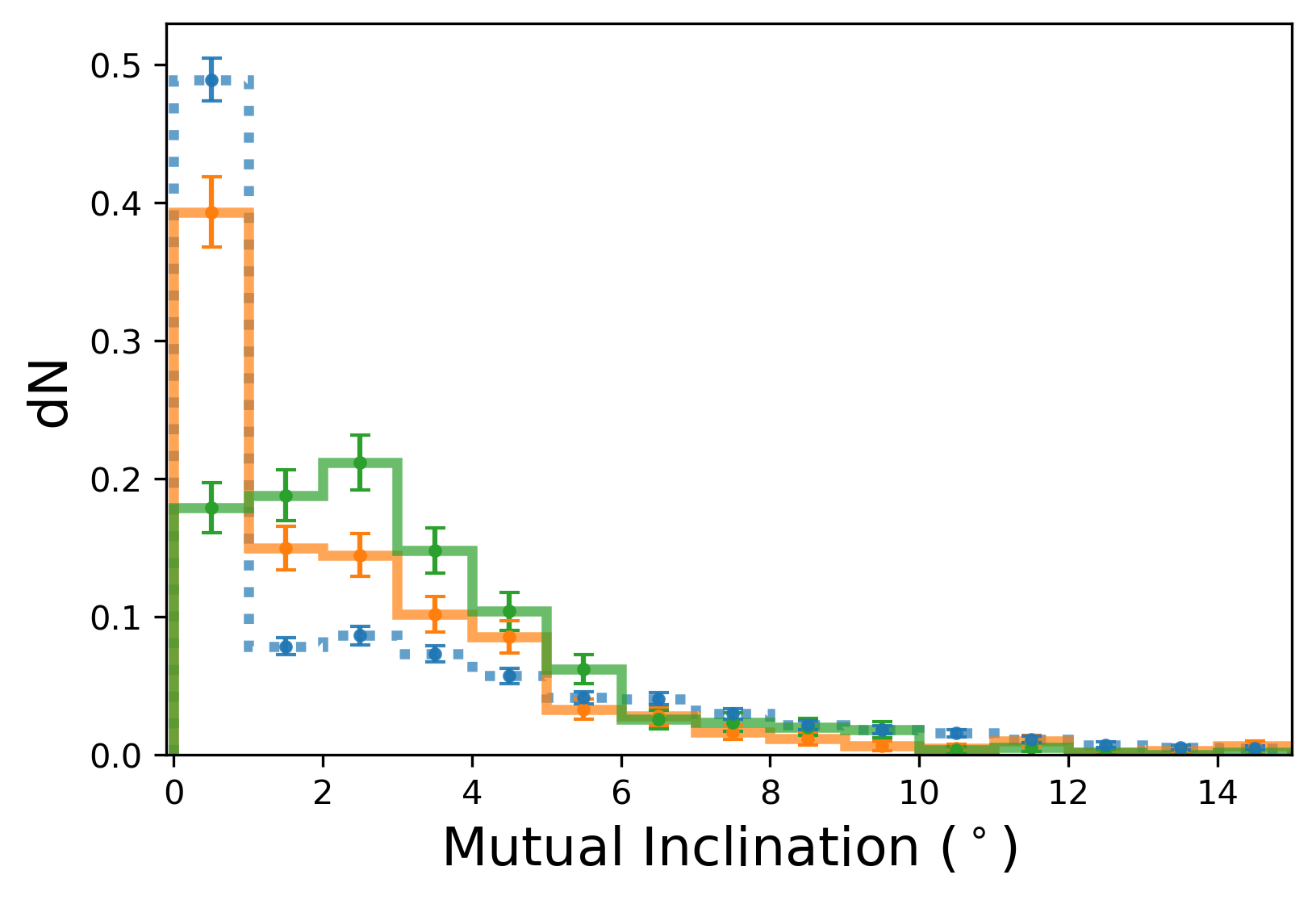}
    \includegraphics[width = 3.2in, height = 2.1in]{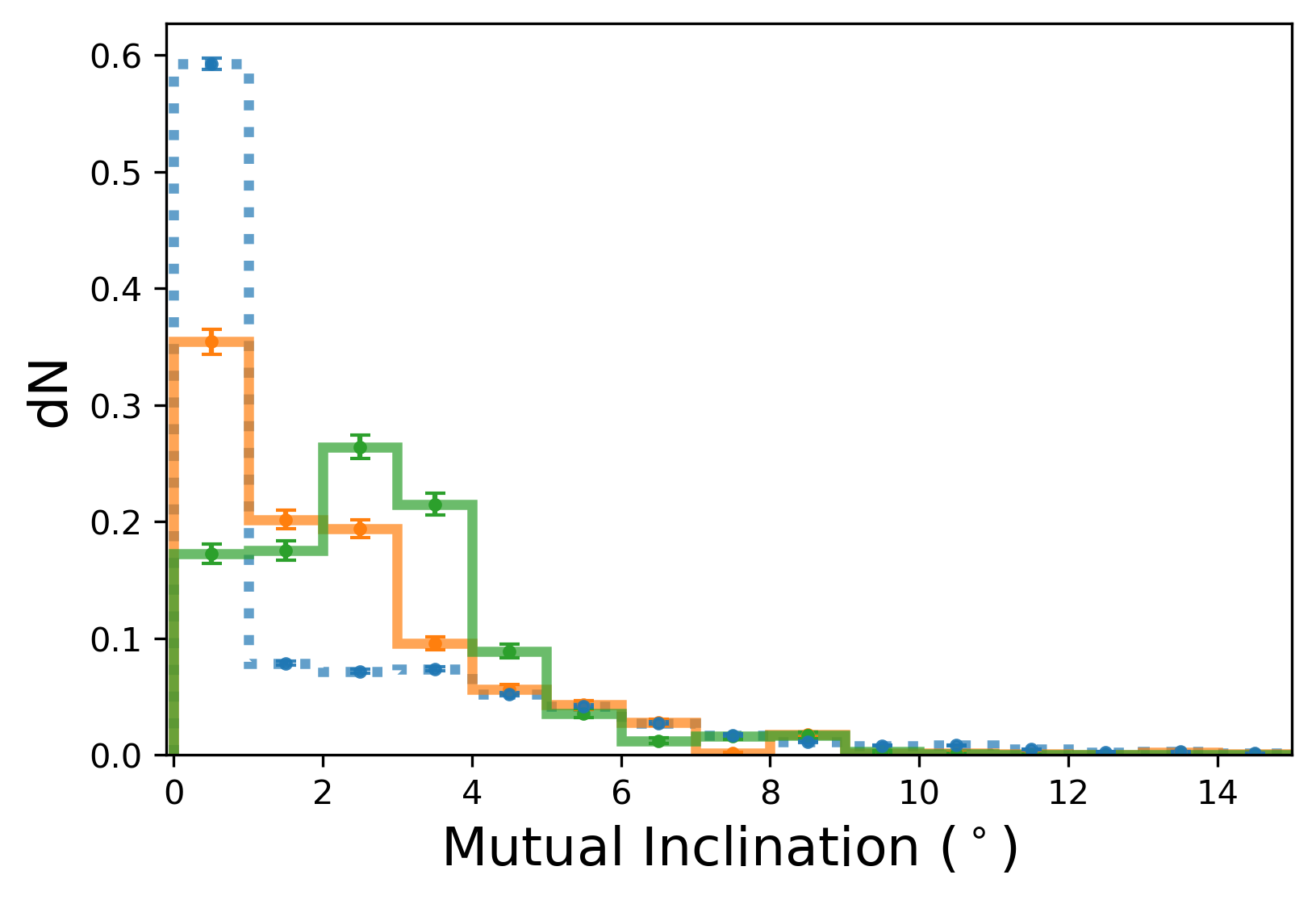}

    \caption{Distributions of mass (\emph{top}), eccentricity (\emph{middle}), and adjacent planet mutual inclination (\emph{bottom}) for both the underlying simulations (left) and the mock observed planets (right) for FI1 (orange), FI2 (green) and CI (dotted blue).
    Both the FI1 model and the FI2 model produce qualitatively similar mass distributions to CI, with the similarities retained after mock detection.
    The distributions of underlying eccentricity for FI1 and CI match, with mock detected FI1 planets following slightly more elliptical orbits than the majority of CI planets.
    The underlying FI2 eccentricity distribution shows fewer planets at $e\sim 0$ relative to CI. 
    After mock detection, the frequency of low $e$ FI2 planets increases, dropping off rapidly at $e >0.1$.
    The FI1 mutual inclination distribution remains relatively unchanged before and after mock detection, with mutual inclinations $i_{mut} >1^\circ$ becoming slightly more common.
    In contrast, the frequency of planet pairs $i_{mut} <1^\circ$ increases post-mock detection for CI.
    Thus, FI1 planet pairs are more mutually inclined relative to CI planet pairs.
    Underlying FI2 planet pairs are more mutually inclined relative to CI planet pairs, with this mutual inclination distribution maintained after mock detection.
    FI2 planets are thus similar to CI planets in mass and eccentricity distributions, but are more mutually inclined.}
    \label{fig:underlying_vs_observed_flowiso}
\end{figure*}

We plot the distributions of the masses, eccentricities, and adjacent planet mutual inclinations of both the underlying and the mock observed, re-weighted systems for the FI1 model in Figure~\ref{fig:underlying_vs_observed_flowiso}. 
These planet properties are not usually directly observable for a transiting system but are useful for understanding the properties that are observable. 

Compared to the CI mass simulations with $\alpha = 3/2$, the FI1 mass planets in Figure~\ref{fig:flowiso_1e-9_mass_vs_a} exhibit a qualitatively similar mass versus semi-major axis relationship. 
The majority of masses in both models span $\sim 0.5-20 ~\mathrm{M_\oplus}$, and less massive planets are less common at large $a$. 
FI1 produces fewer detectable planets $3~\mathrm{M_\oplus}<\mathrm{M_p}<10~\mathrm{M_\oplus}$ than CI at distances $a>0.4$.
When comparing to the top right panel of Figure~\ref{fig:underlying_vs_observed_flowiso}, we see that the total normalized distribution of mass nearly matches that of the CI model.
The underlying FI1 planets (Figure~\ref{fig:underlying_vs_observed_flowiso}, top left panel) skew toward Earth mass and below, with a tail trailing off $>5~\mathrm{M_\oplus}$.
This distribution is consistent with the underlying distribution of CI planet masses.

Underlying FI1 eccentricities (Figure~\ref{fig:underlying_vs_observed_flowiso}, middle left panel) also appear similar to those for CI, peaking at near-circular values and trailing off at values $e \gtrsim 0.1$. 
The mock observed CI distribution (middle right panel, blue, dotted) retains its underlying peak at $e \sim 0$.
The FI1 distribution favors detection of planets with small but nonzero eccentricities. 
While similar to the CI distribution prior to detection, the eccentricity distribution shifts rightward after detection, still peaking near zero but showing an enhancement of eccentricities $0.025<e < 0.1$.
At eccentricities $>0.1$, planets are much less common, leaving the majority of FI1 planets with nearly circular orbits.
In combination with Figure~\ref{fig:flowiso_1e-9_mass_vs_a}, FI1 tends to produce moderate mass, low eccentricity planets, much like the ``dynamically cold'' systems in the \emph{Kepler} catalog.

The underlying mutual inclination distribution of CI and FI1 both peak near zero (Figure~\ref{fig:underlying_vs_observed_flowiso}, bottom left panel), particularly for the CI model. 
However, FI1 does not show nearly as large a spike in mutual inclinations at $~0^\circ$ and instead shows a small bump around $0.5^\circ \lesssim i_{mut} \lesssim3^\circ$ relative to the CI distribution. 
We see an drop in the frequency of $i_{mut}\sim0$ FI1 systems once mock detected (bottom right panel), but the nonzero bump is maintained and slightly enhanced in frequency relative to the underlying population. 
We find that the weighting coefficients $\mu^- = 10^{-5.95}$ and $\sigma^- = 10^{-1.14}$ provide the minimal Earth-mover's distance, and so we reweight our initial log-uniform distribution of $St$ using Equation~\ref{eq:reweighting}. 
\change{As a result, the simulations with high $St$ are heavily downweighted, implying that either they are less important in regards to matching the \emph{Kepler} catalog, or other conditions must be met before the original distribution will work.}

\begin{figure*}
    \centering
    \includegraphics[width = 3.4in, height = 2.3in]{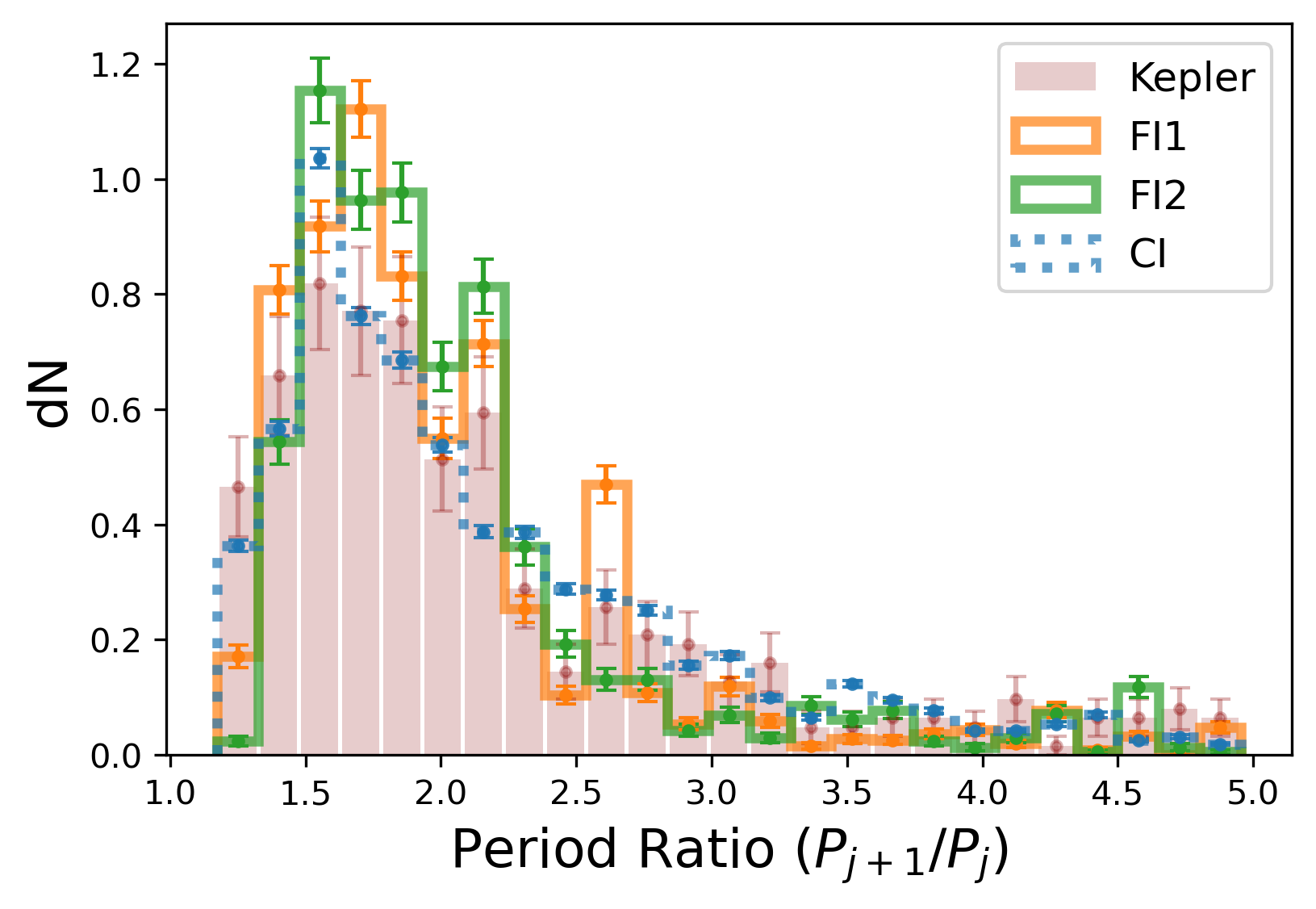}
    \includegraphics[width = 3.4in, height = 2.3in]{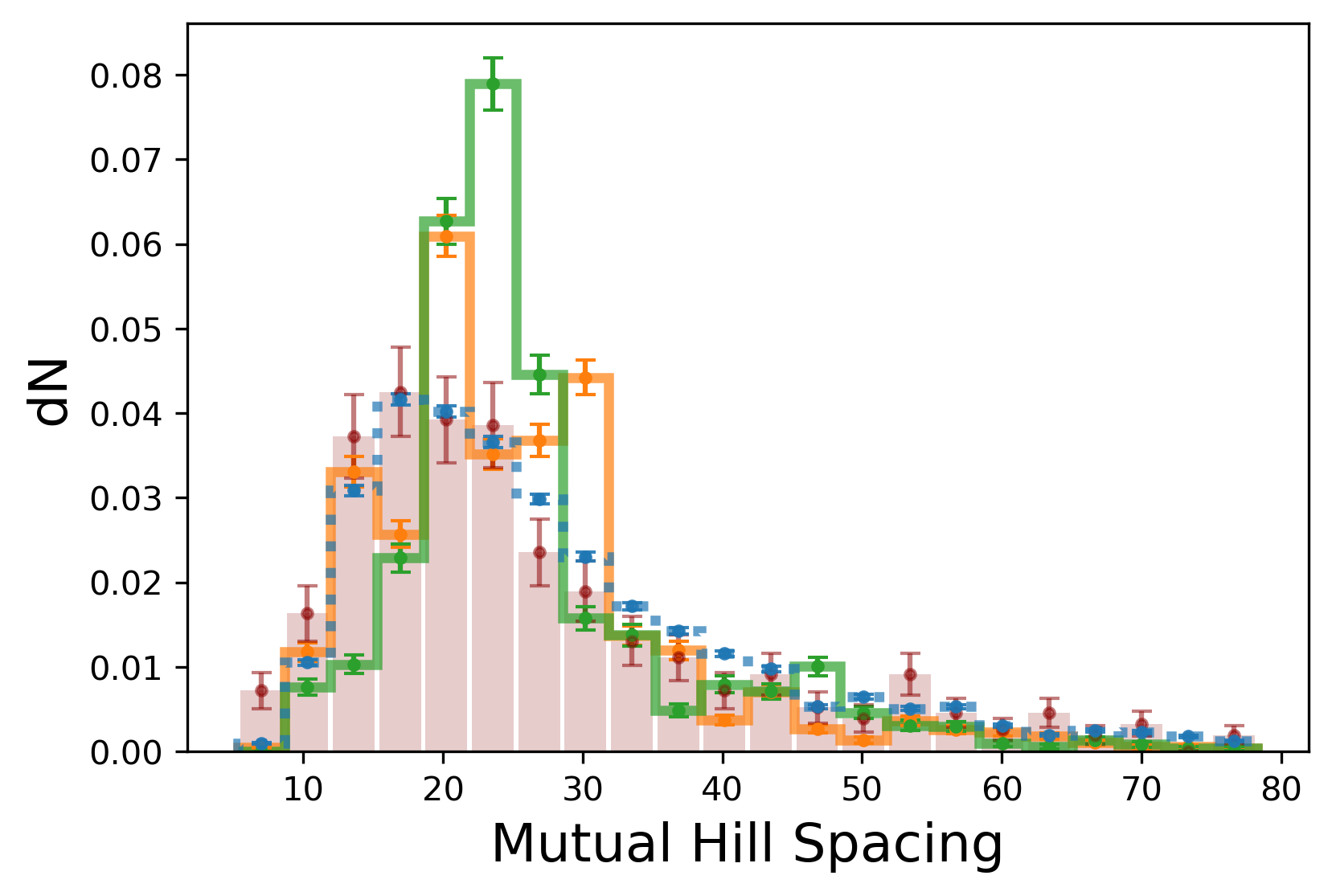}
    \includegraphics[width = 3.4in, height = 2.3in]{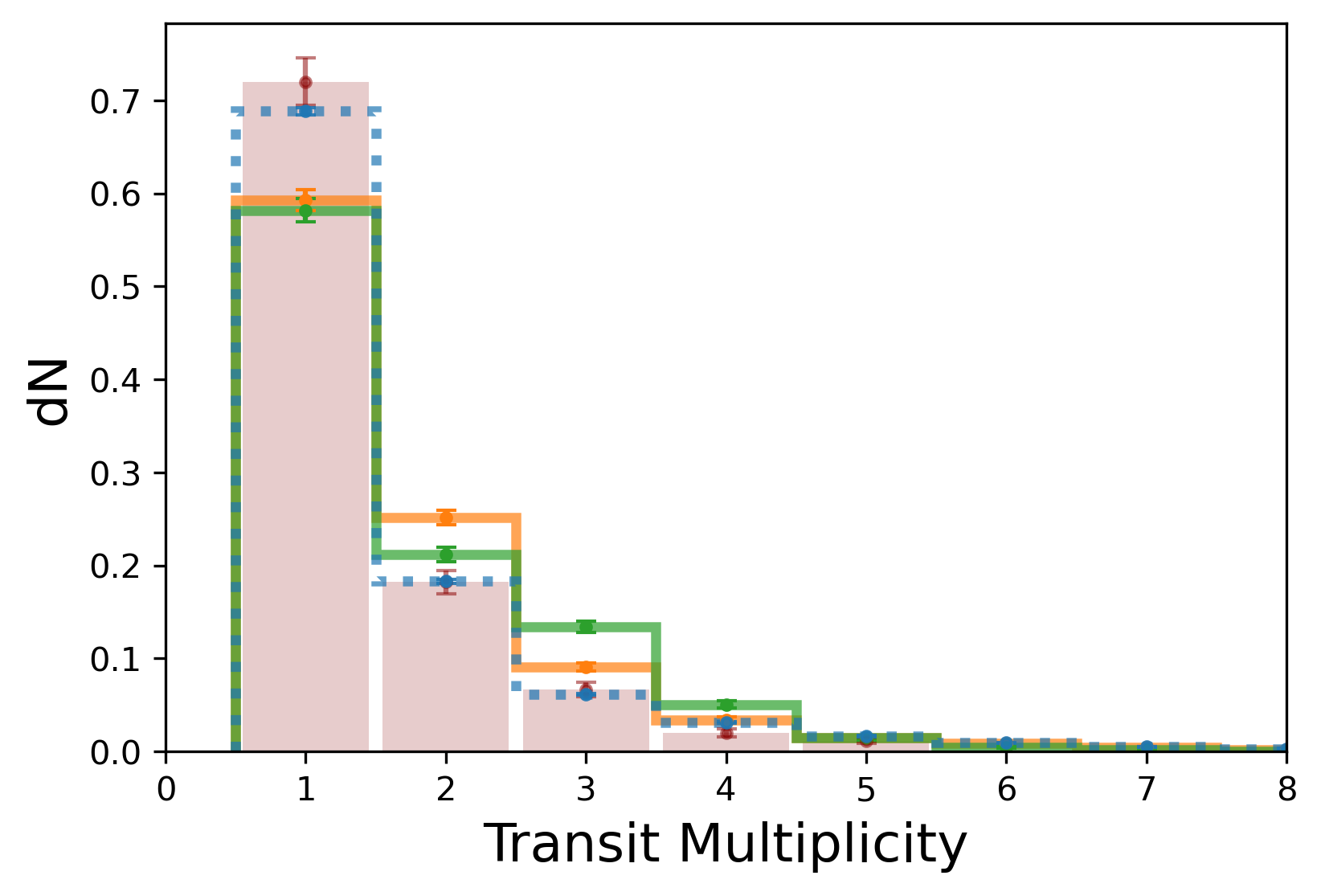}
    \includegraphics[width = 3.4in, height = 2.3in]{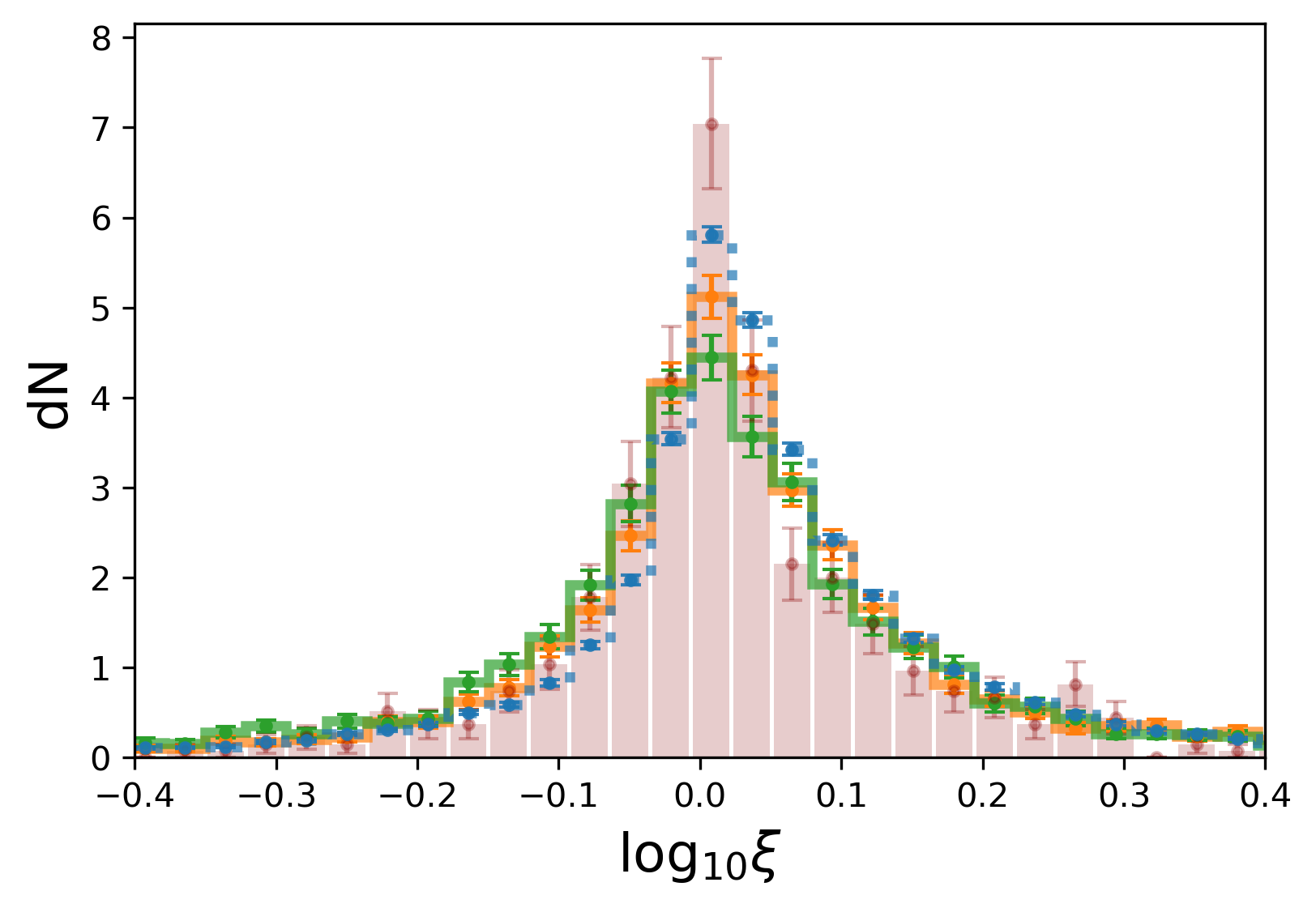}
    \caption{From upper left to lower right: distributions of period ratios of adjacent planets, mutual Hill spacings of adjacent planets, transit multiplicities, and transit duration ratios ($\xi$) of adjacent planets from the FI1 and FI2 mass models (solid orange and solid green respectively),  compared to the CI mass (dotted blue) and the \emph{Kepler} catalog distribution (solid pink). 
    As described in \citetalias{MacDonald2020}, the CI distributions are good qualitative fits to the true distributions.
    The FI1 distributions are also adequate matches, but a discrepancy emerges in the lack of single planets mock detected compared to the \emph{Kepler} catalog. The FI2 model lacks single planet system mock detections as well. 
    It also shows a mutual Hill spacing distribution which peaks at $\Delta\sim25$, much wider than the \emph{Kepler} catalog's peak at $\Delta\sim18$.}

    \label{fig:flowiso_observables}
\end{figure*}

We present the re-weighted results of our forward modeled detected systems governed by the FI1 mass, compared directly to the re-weighted CI results from \citetalias{MacDonald2020} and \emph{Kepler} catalog observables, in Figure~\ref{fig:flowiso_observables}. 
Here, we qualitatively compare the resulting distributions of period ratios (upper left), mutual Hill spacings (upper right), transit multiplicity (lower left), and transit duration ratio (lower right).
Simulations governed by the FI1 mass produce a period ratio distribution which peaks at $\sim 1.7$, nearly matching what is expected from the \emph{Kepler} catalog and \citetalias{MacDonald2020}'s CI simulations.  

When compared to the \emph{Kepler} catalog and the CI distribution, the FI1 model fails to capture planet pairs at the smallest period ratios (top left panel of Figure~\ref{fig:flowiso_observables}) $1.2 \lesssim \frac{P_{j+1}}{P_j} \lesssim 1.4$ and over-represents near the peak ($1.5 \lesssim \frac{P_{j+1}}{P_j} \lesssim 1.7$). 
Beyond $\frac{P_{j+1}}{P_j} \sim 2.6$, the frequency of pairs slightly falls off relative to the \emph{Kepler} catalog. 
However, the qualitative distribution of FI1 period ratios is a sufficient match to the \emph{Kepler} catalog distribution.

A similar relationship can be found between the FI1 mutual Hill spacings (top right panel of Figure~\ref{fig:flowiso_observables}) and the \emph{Kepler} catalog distribution.
Spacings less than $\Delta \approx 18$ and greater than $\Delta \approx 40$ are not represented as frequently for FI1 as for the \emph{Kepler} catalog. 
The FI1 distribution shows a few peaks at $\Delta \sim 22$ with a smaller peak at $\Delta \sim 30$.
The qualitative FI1 spacing distribution matches the \emph{Kepler} catalog, peaking at $\Delta\sim18$ and falling off to $\Delta\sim40$, beyond which we see few planet pairs.

Even after reweighting, the FI1 model appears to over-produce multi-planet transiting systems, particularly 2--3 planet systems (bottom left panel of Figure~\ref{fig:flowiso_observables}) and under-produce single planet transiting systems relative to the \emph{Kepler} catalog. 
The deficiency in single FI1 planets is greater than the corresponding bin error for either FI1 or the \emph{Kepler} catalog, and thus this is a true consequence of formation by the FI1 model \change{under our assumptions of gas damping and our initial range of $St$.}

In the bottom right of Figure~\ref{fig:flowiso_observables}, all three cases show a transit duration ratio peak at $\log_{10}\xi \sim 0.0$, which implies primarily coplanar systems.
There is no significant deviation between the FI1 and \emph{Kepler} catalog transit duration ratio distributions, implying a similar variety of orbital geometries.

\subsection{Properties Assuming $\dot{M} = 4 \times 10^{-8}~\mathrm{M_\odot ~yr^{-1}}$} \label{sec:flowiso_4e-8}

\begin{figure}
    \centering
    \includegraphics[width = 0.5\textwidth]{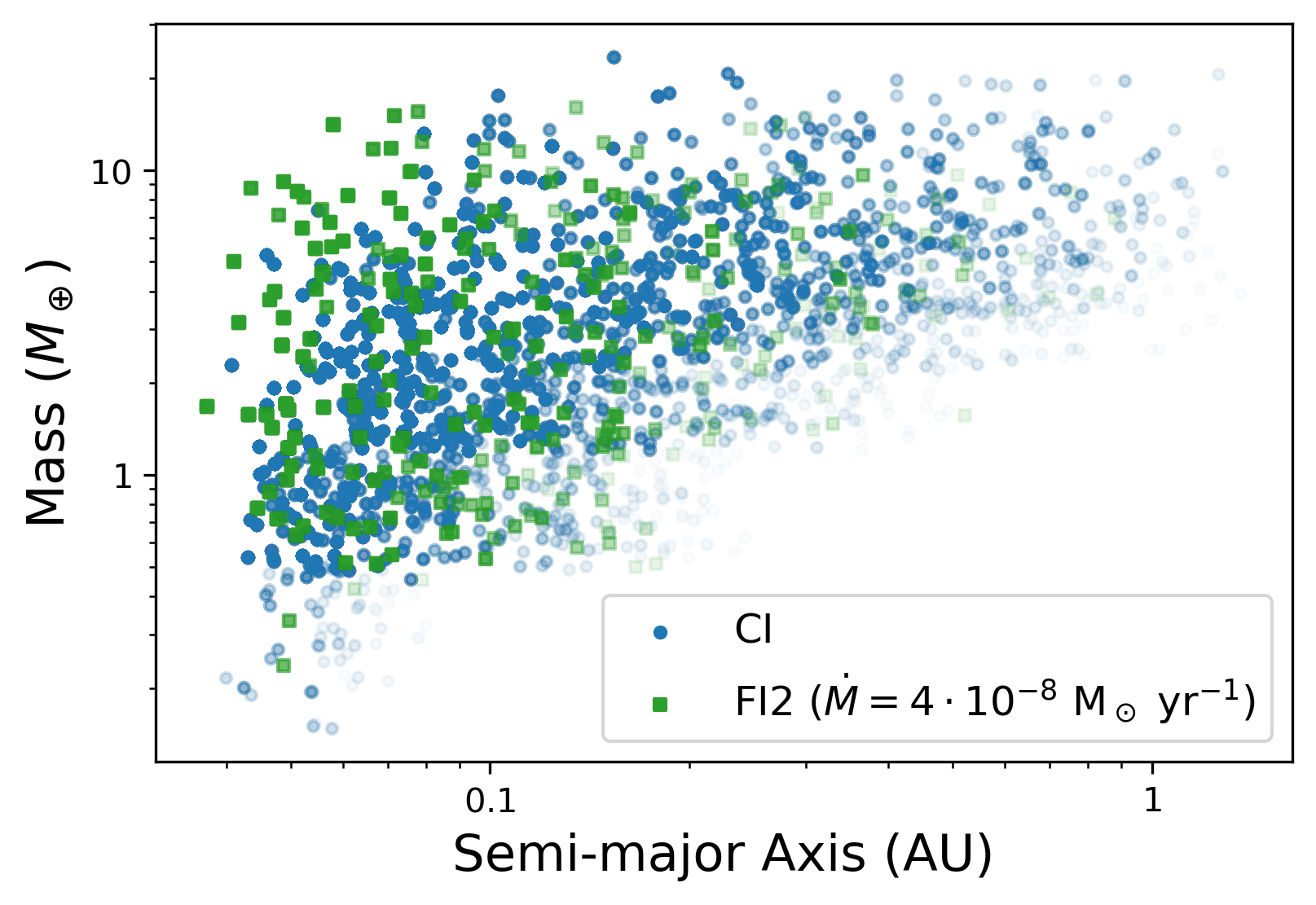}
    \caption{Mass as a function of semi-major axis for the mock detected transiting planets from the FI2 (green squares) model and CI with $\alpha=3/2$ (blue circles) model.
    Mass tends to increase with semi-major axis for both cases, with the FI2 model typically producing more massive planets at $a <0.1$.
    More opaque points represent more frequently detected planets.}
    \label{fig:flowiso_4e-8_mass_vs_a}
\end{figure}

We run an additional suite of flow isolation mass simulations with a higher accretion rate of $\dot{M} = 4 \times 10^{-8}~\mathrm{M_\odot~yr^{-1}}$ (FI2) to explore its effects on final orbital properties as well as the effects of a range of smaller Stokes numbers (see Figure~\ref{fig:stokesrange}).
We present the mass of these mock detected planets as a function of semi-major axis in Figure~\ref{fig:flowiso_4e-8_mass_vs_a}, where we compare them to the planets produced by \citetalias{MacDonald2020}'s CI model.
The planets produced by FI2 follow the same qualitative pattern as CI, with the majority of planets formed with mass $0.5-10 ~\mathrm{M_\oplus}$ and lower mass planets falling off in detectability as semi-major axis increases. 

The FI2 model's planets (Figure~\ref{fig:underlying_vs_observed_flowiso}, top left panel) exhibit much the same underlying mass distribution as the CI model. 
The majority of planets in the underlying distribution are $1.5 ~\mathrm{M_\oplus}$ or less, with a rapidly-tapering tail at values greater than $5 ~\mathrm{M_\oplus}$. 
The FI2 model produces more planets at $1~\mathrm{M_\oplus}$ than the CI model, but otherwise the distributions are effectively the same.
Once mock detection is performed, the discrepancy in the frequency of planets in the lowest mass bin is partially relaxed.
The FI2 model shows a slight enhancement in fraction of planets with $M_p < 2.5~\mathrm{M_\oplus}$ and fewer planets $2.5~\mathrm{M_\oplus}<\mathrm{M_p}<5~\mathrm{M_\oplus}$ relative to the CI distribution. 
This discrepancy is not enough to significantly differentiate the FI2 planet population from that of the CI mass, and thus we argue that they follow the same mass distribution.

The underlying eccentricity distribution (Figure~\ref{fig:underlying_vs_observed_flowiso}, middle left panel) of the FI2 case shows fewer planets with orbits $e\sim0$ than CI and more planets with $0.025<e<0.075$. 
Comparing the detected eccentricity distributions (middle right panel), we see the peak remain at slightly more elliptical values for FI2 (around $e \sim 0.05$), although the fraction of planets with $e\sim0$ increases.
However, we detect effectively no planets with eccentric orbits $e \gtrsim 0.2$, in contrast to the CI model which exhibits a small tail out to $e \sim 0.4$.
The FI2 model produces planets with orbits in a tight range of small eccentricities.

The distributions of mutual inclinations remain nearly unchanged between the underlying (Figure~\ref{fig:underlying_vs_observed_flowiso}, bottom left panel) and observed (bottom right panel) planets, with the CI model producing mostly coplanar systems and FI2 producing many more with small but significant misalignments, in the range $1^\circ ~<~i_{mut} ~<~ 4^\circ$. 
When applying our reweighting scheme, the lowest total EMD is given by the reweighting parameters $\mu^- = 10^{-6.30}$ and $\sigma^- = 10^{-1.07}$, \change{once again showing that these simulations must be weighted heavily toward low $St$ in order to match observations.}

Figure~\ref{fig:flowiso_observables} presents the distributions of FI2's observable properties, once again juxtaposed against the CI and \emph{Kepler} catalog distributions. 
The FI2 model's period ratio distribution (upper left) shows a peak at a value of $\sim 1.6$, roughly matching the \emph{Kepler} catalog peak.
The FI2 model produces fractionally more planet pairs in the range $1.5 \lesssim \frac{P_{j+1}}{P_j} \lesssim 2.3$ than both the CI and the \emph{Kepler} catalog data. 
Subsequently, this model fails to create detectable systems at much smaller ($<~1.3$) and much larger ($>~2.5$) period ratios. 
The shape of the FI2 period ratio distribution still roughly matches the \emph{Kepler} catalog, however.

The mutual Hill spacing distribution (upper right) shows a significant peak at $\Delta \sim 24$, shifted rightward from the \emph{Kepler} catalog and CI distribution peaks of $\Delta \sim 18$ with a fractional depletion of pairs $\Delta \leq 20$ when compared to the latter.
The FI2 spacing distribution matches the \emph{Kepler} catalog distribution for spacings $\Delta  > 30$.
While the spacing distribution matches for widely separated planets, FI2 does not produce planets as tightly spaced as the \emph{Kepler} catalog.

The FI2 model fails to produce enough mock-detected single planet systems to match the \emph{Kepler} catalog even after \change{the most extreme} reweighting, just like the FI1 model (Section~\ref{sec:flowiso_1e-9_props}.
It also fails to match the fraction of multi-planet transiting \emph{Kepler} catalog planets, instead exhibiting a higher fraction of two, three, and four planet systems. 

The transit duration ratio distribution for FI2 matches the \emph{Kepler} catalog and CI distributions, peaking at $\log_{10}{\xi}\sim0$. 
It does however exhibit wide wings, implying greater $e$ and $i_{mut}$.

\subsection{Flow Isolation EMD} \label{sec:flow_EMD}

\begin{figure*}
    \centering
    \includegraphics[width = 3.0in, height = 2.0in]{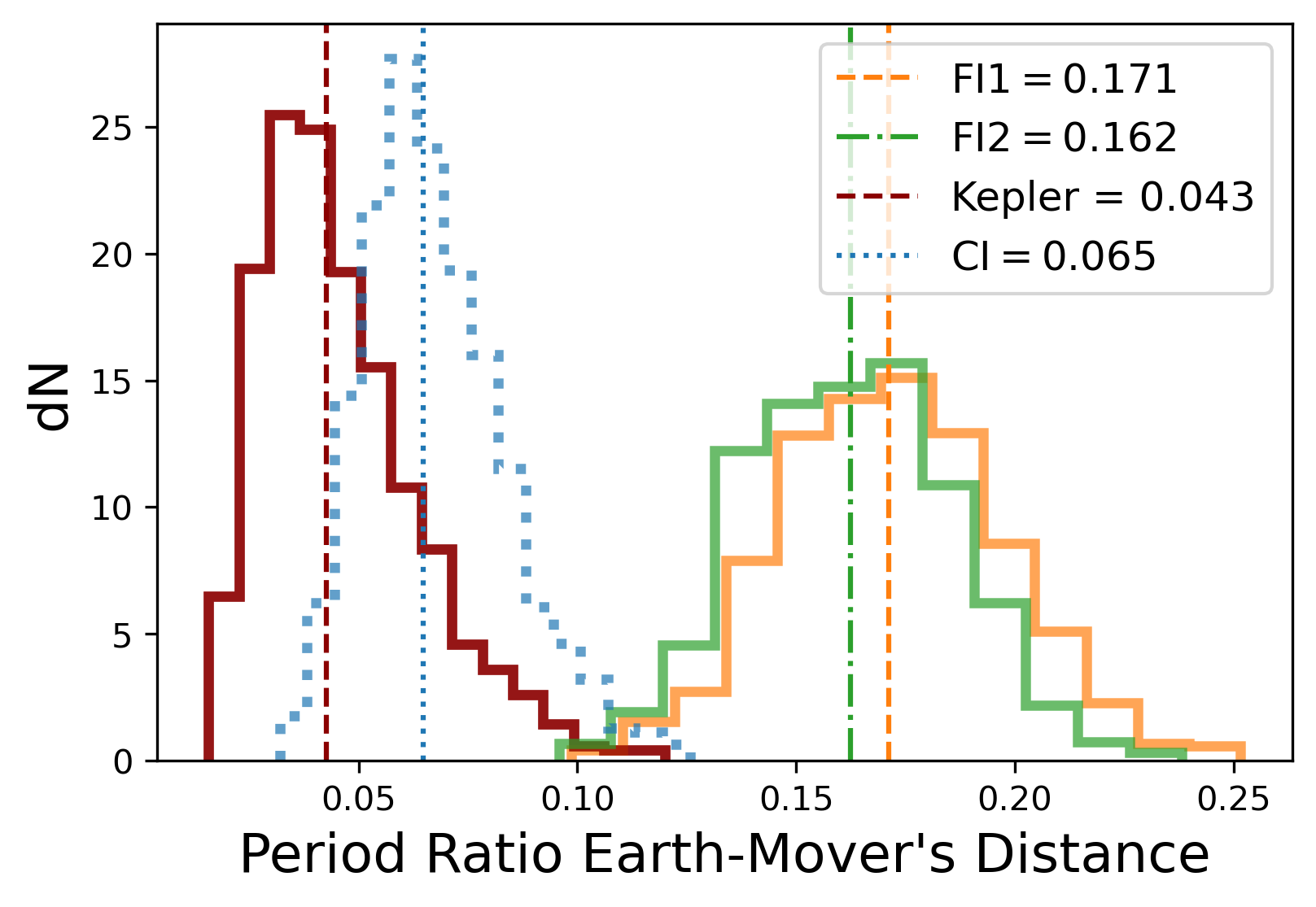}
    \includegraphics[width = 3.0in, height = 2.0in]{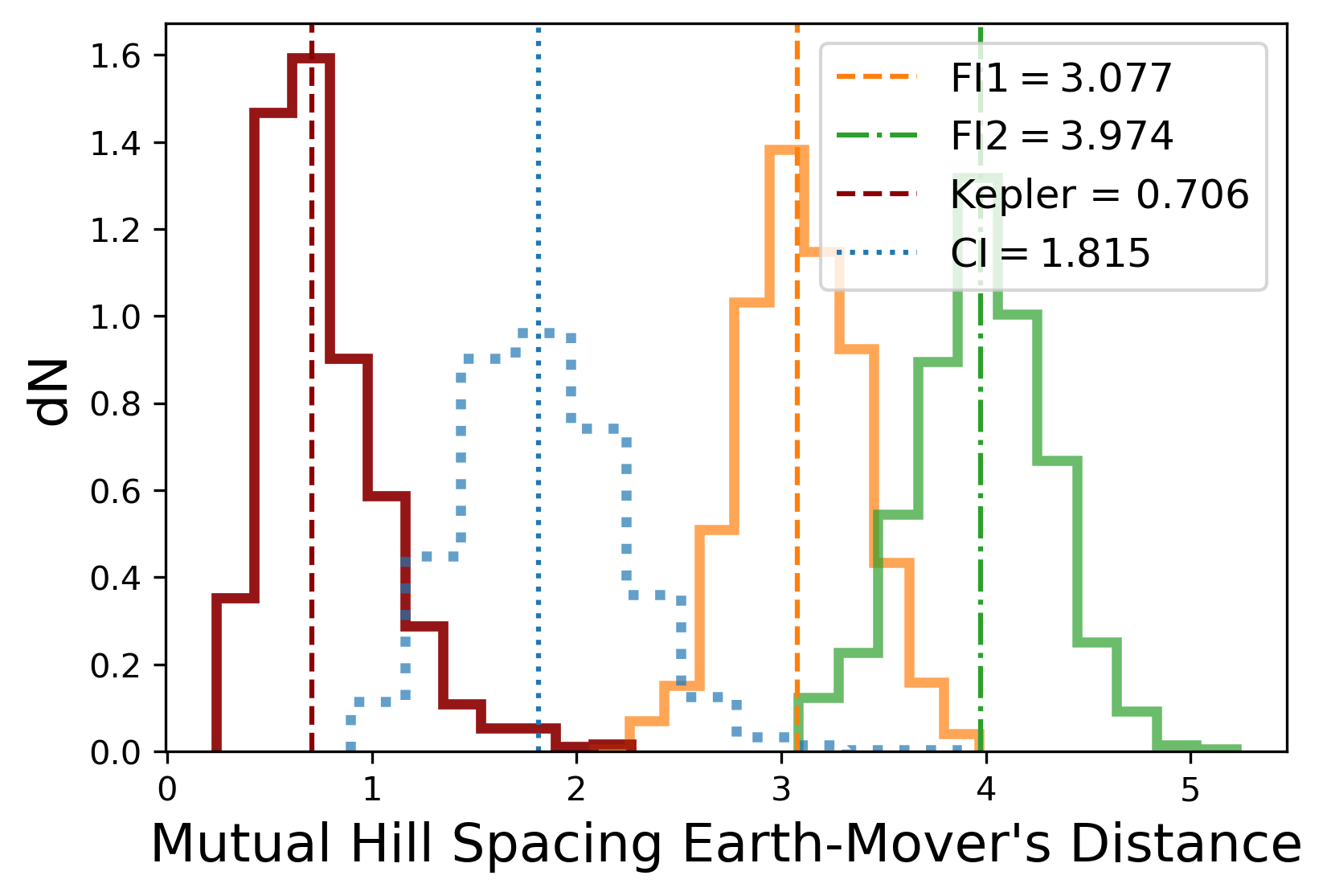}
    \includegraphics[width = 3.0in, height = 2.0in]{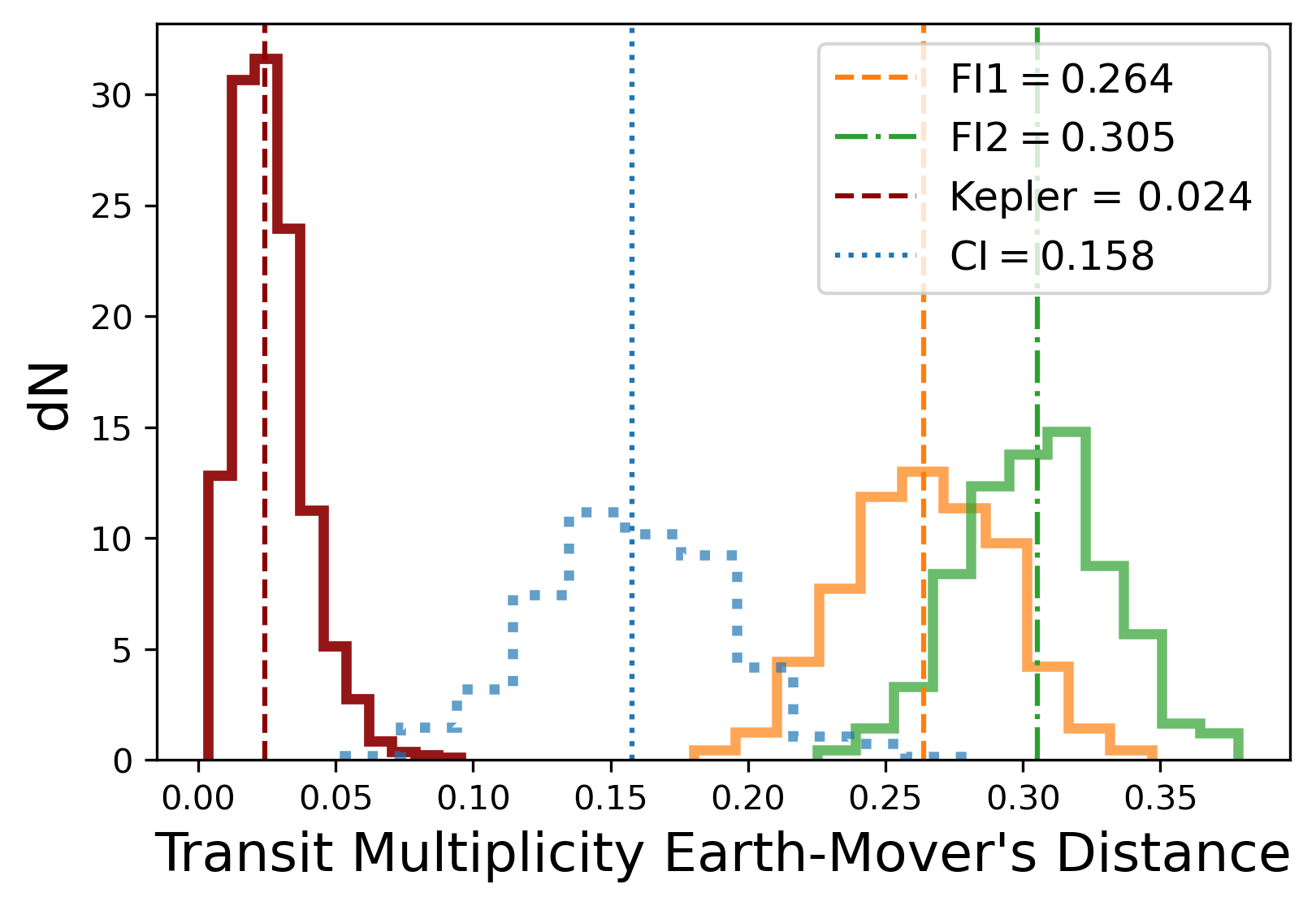}
    \includegraphics[width = 3.0in, height = 2.0in]{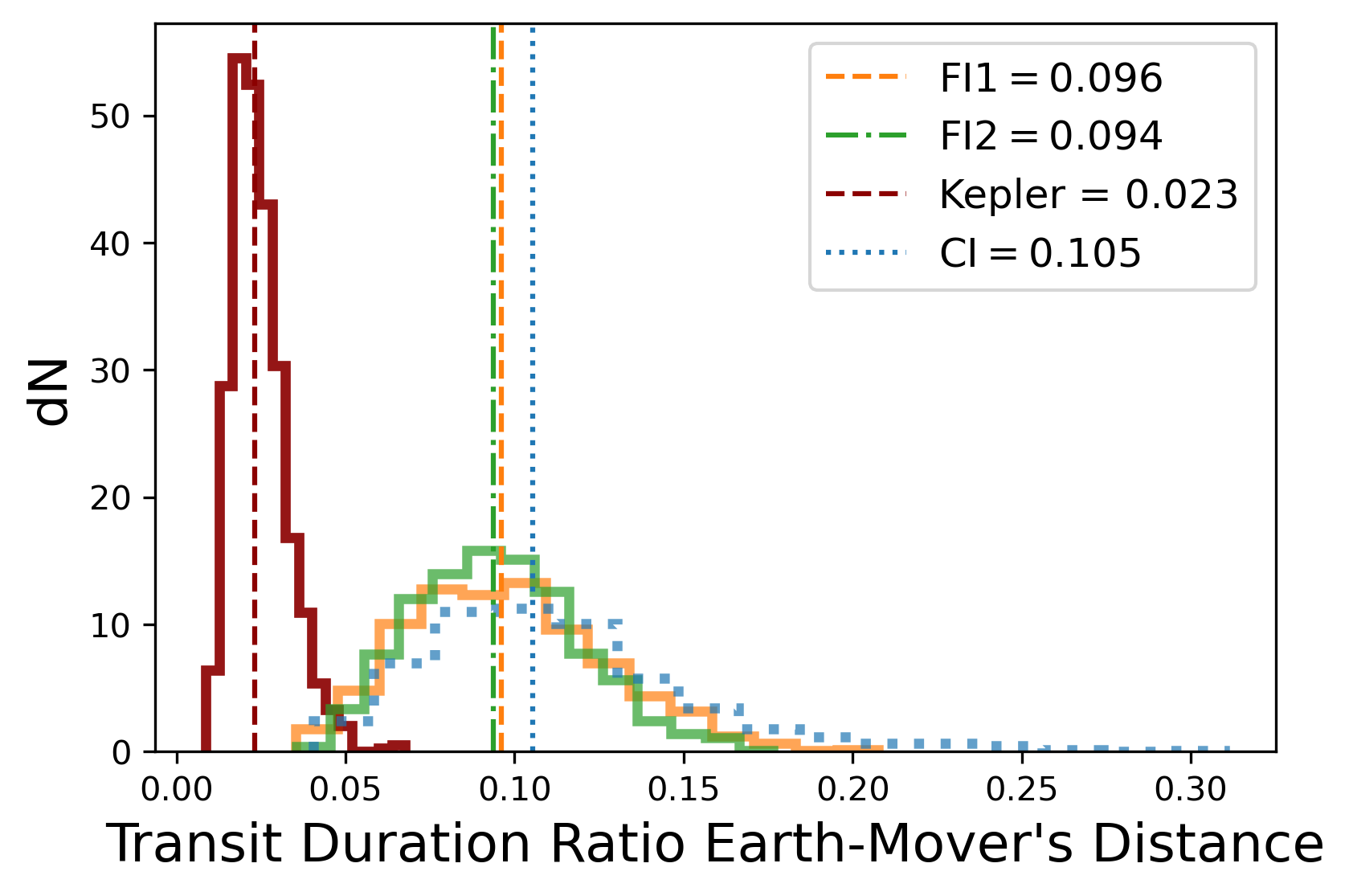}
    \caption{Earth Mover's distance distribution for the FI1 model (orange, solid) and FI2 model (green, solid) against the CI distribution (blue, dotted) and \emph{Kepler} catalog (red, solid). 
    In nearly all cases, the FI1 and FI2 EMDs are larger than the CI EMDs, except in transit duration ratio (bottom right), where the EMD distributions are effectively the same. 
    \changethree{There exists some overlap between FI1, FI2 and \emph{Kepler} in period ratio and transit duration ratio, but they are marginal at best.}
    The CI model is an overall better match to the observed data than the FI1 and FI2 models.}
    \label{fig:flowiso_emd}
\end{figure*}

We can make quantitative statements about how well each isolation mass model fits the distribution expected by the \emph{Kepler} catalog.
Beginning with Figure~\ref{fig:flowiso_emd}, we can see that the FI1 EMDs for any given viewing angle or observable tend to be larger than their CI counterparts.
In the case of period ratio (top left), the FI1 EMD distribution peaks at a larger value than the CI distribution.
With minimal overlap between the FI1 and \emph{Kepler} distributions relative to CI, we confidently state that the CI model matches the \emph{Kepler} catalog period ratios more closely than the FI1 model.
\changethree{The overlap between the FI1 and \emph{Kepler} EMD distributions for mutual Hill spacing (top right) is not significant compared to CI.}
The same can be said for transit multiplicity (bottom left): the CI model produces a closer overall match than FI1, \changethree{since it is within the uncertainty of the  \emph{Kepler} distribution.}
For transit duration ratio (bottom right), the FI1 and CI distributions overlap each other entirely. 
The FI1 model does however have lower uncertainties in duration ratio EMD and is within \emph{Kepler} uncertainties, and so by this metric is a marginally better match to the \emph{Kepler} catalog.
With this information in mind, we conclude that the FI1 model provides an overall poorer match to the \emph{Kepler} catalog distributions when compared to the CI model.

For the FI2 model, we see similar results as for FI1.
The FI2 period ratio EMD distribution (top left) is generally larger than that of the CI, \changethree{minimally overlapping with the \emph{Kepler} EMD distribution}. 
The FI2 mutual Hill spacing EMD distribution (top right) shows dramatically larger distances compared to CI.
While there is some overlap between FI2 and CI in transit multiplicity (bottom left), \changethree{FI2 does not overlap with \emph{Kepler}.} 
Finally, there is significant overlap between the FI2 and CI transit duration ratio distributions, \changethree{both of which extend into the tail of the \emph{Kepler} catalog distribution.}
Since the FI2 model has a smaller variance, it is more likely to produce smaller EMD values and is thus a better match.
When taking all four observables into consideration, the FI2 model is a poorer match to the \emph{Kepler} catalog than the CI model. \changethree{We provide a summary of the EMD statistics in Table \ref{tab:summary_stats_table}.}

\subsection{Implications of Formation by Flow Isolation} \label{sec:flowiso_implications}




The planetary systems formed by the flow isolation mass, regardless of the stellar accretion rate, tend to harbor mostly planets with $\mathrm{M_p} < 2.5~\mathrm{M_\oplus}$, with low eccentricities and moderate mutual inclinations (Figure~\ref{fig:underlying_vs_observed_flowiso}).
Comparing to the CI model, the FI1 and FI2 planets exhibit the same underlying mass distributions but with slightly higher mutual inclinations.
The mass and eccentricity distributions of the mock detected planets for FI1 and FI2 are consistent with CI, but the mutual inclinations remain slightly larger.
FI1 planets have period ratio, mutual Hill spacing, and transit duration ratio distributions that qualitatively match the \emph{Kepler} catalog, but produce fewer detected single planets. 
Although the FI2 period ratio and duration ratio distributions appear to match the \emph{Kepler} catalog, the mutual Hill spacing distribution is skewed higher, towards more widely-spaced planets, and too few single transiting planet systems are produced.
These large mutual Hill spacings are the result of FI2 embryos undergoing more mergers post disk dissipation compared to FI1 embryos.

To determine whether these discrepancies are a result of an incomplete range of Stokes numbers, we refer to Section 4.2 of \citet{RMC2020}, wherein they discuss the dependence of the flow isolation mass on fragmentation velocity. 
Within the inner disk, particles undergo frequent collisions at high velocities, suggesting that the primary factor that limits particle size is their tendency to fragment.
Using a simple fragmentation model from \citet{Birnstiel2009} in which a particle is thought to fragment if its collisional velocity surpasses a given fragmentation velocity $u_{frag}$, determine the maximum allowed Stokes number as a function of semi-major axis, stellar accretion rate, and fragmentation velocity. 

Lab experiments suggest that $u_{\rm{frag}}$ could range between $1-10~$\text{m/s} depending on material properties \citep{Blum2008, Stewart2009}.
The maximum Stokes number is $St_{max} \sim 10^{-2}$ for disks with $\dot{M} = 10^{-9}~\mathrm{M_\odot~yr^{-1}}$ and $St_{max} \sim 10^{-3}$ for disks with $\dot{M} = 4\times10^{-8}~\mathrm{M_\odot~yr^{-1}}$, if we assume a collisional velocity of $1$ \text{m/s} between particles.
These upper bounds are consistent with our choice of upper $St_{max}$ motivated by \citetalias{MacDonald2020} and Figure~\ref{fig:stokesrange}.
However, at higher collisional velocities, the maximum Stokes number increases to $St_{max} \sim 1$, which will create embryos that exceed a few Earth masses.
The planets that form from such large embryos would likely be much larger than the inner super-Earths we observe with \emph{Kepler}.
The lower limit of allowable Stokes numbers is less important than the upper limit, as the only mechanism which stymies their ability to accrete onto a body is the magnitude of their coupling to the gas disk. 
As a result, our lower limit of $St_{max} = 10^{-5}$ is sufficient to explore a significant range of maximum particle sizes and their corresponding isolation masses.

The noticeable deficiency of single-transiting FI systems, as well as the increase in frequency of high-multiplicity systems, is the clearest discrepancy between this model and observation.
Table~\ref{tab:init_table} illustrates a key factor in this discrepancy: the FI models in question produce few planetary systems with multiplicities of $< 4$, whereas the CI model produces many systems with multiplicities of $2-6$. 
The over-abundance of high multiplicity super-Earth systems produced by these models might be a result of the initial distribution of embryo mass.
Figures~\ref{fig:embryomass} and~\ref{fig:mass_distributions} show how the flow isolation mass scales with distance from the host star.
Interior to $r_{vis-irr}$, a characteristic radial distance at which the primary heating source in the disk switches from viscous heating to passive irradiation, mass is relatively constant with distance.

\citet{RMC2020} argue that a ``peas-in-a-pod'' system architecture \citep{Millholland2017, Weiss2018} should naturally arise from the planets formed by these close-in embryos: our results tentatively agree with this conclusion, given these systems are generally tightly-spaced and coplanar.
We leave a more thorough exploration of the peas-in-a-pod paradigm to future works. 

Exterior to $r_{vis-irr}$, mass begins to rapidly increase as the thermal mass \citep{LinPapaloizou1993}, on which the flow isolation mass is based, begins to itself increase.
In the outer disk, the FI mass will therefore produce embryo masses large enough to undergo runaway gas accretion which results in the formation of gas giants.

Should the assumption of embryos forming by isolation mass hold in the outer disk, these giant planets will greatly affect the dynamics of our tightly-spaced super-Earth systems during formation by exciting the eccentricities and inclinations of planets such that they no longer transit \citep[e.g.,][]{Huang2017, Livesey2025, Sandhaus2025}. 
As a result, we would see the fraction of single-transit flow isolation planets increase alongside a decrease in multi-planet transit detections, leading to a better match to the \emph{Kepler} catalog distribution.

The occurrence rate of exterior giant planets (``cold'' Jupiters) in systems which harbor tightly-spaced super-Earth planets has been studied extensively and suggests that the existence of gas giants in multi-super-Earth systems may be common.
For instance, \citet{Bryan2019} determined an occurrence rate of $39\% \pm7\%$ for planets with $0.5-20~M_J$ beyond $1$ au in systems with inner super-Earths by studying modulations in radial velocity data for these systems.
More recently, the high-precision radial velocity survey undertaken by \citet{Rosenthal2022} constrained an occurrence rate of $41^{+15}_{-13}~\%$ for outer giant planets ($30-6000~\mathrm{M_\oplus}$) in systems which also contain at least one inner small planet ($2-30~\mathrm{M_\oplus}$), with $17.6^{+2.4}_{-1.9}~\%$ of the sample of stars hosting an outer giant planet.
\citet{Rosenthal2022} also find that in the inverse case, $42^{+17}_{-13}~\%$ of stars that host an outer giant also host a small inner planet, with a general occurrence rate of $27.6^{+5.8}_{-4.8}~\%$ for small planets, regardless of the presence of a giant in the system. 
As \citet{RMC2020} predict and observations confirm \citep{ZhuWu2018, Bryan2019, Rosenthal2022}, it is not uncommon for inner super-Earth systems to host an outer giant planet, and so lacking these significant perturbers in our simulations may play a role in the discrepancies between our results and the \emph{Kepler} catalog observations.

Another possible explanation for the overabundance of observed multi-planet systems is the age discrepancy between our simulations and \emph{Kepler} systems as a whole.
We run our simulations for approximately 30 Myr, capturing the bulk of the giant impact era as well as dynamics due to the presence of a late-stage gas disk, whereas the average \emph{Kepler} system is on the order of a few Gyr old \citep[e.g.,][]{Aguirre2015}, nearly two orders of magnitude greater.
Many dynamical events can occur in this span of time that could significantly affect the likelihood of detecting multiple transiting planets, such as mergers, planet-planet scattering, photoevaporation, or secular chaos and subsequent inclination excitation.

Using observations from the Transiting Exoplanet Survey Satellite (TESS) \citep{Ricker2015} and \emph{Kepler}, \citep{Fernandes2025} perform a comparative analysis of the occurrence rates of young, short-period planets around both young and old stars. 
\citet{Fernandes2025} find that, among the younger TESS stars, the occurrence rate of sub-Neptunes around stars between $10-100$ Myr is $F_0=25.89^{+20.18}_{-11.67}\%$, compared to $F_0=7.98^{+0.37}_{-0.35}\%$ for sub-Neptunes around $\sim$Gyr-old \emph{Kepler} stars. 
Their occurrence rate is in agreement with prior work by \citet{Vach2024}, who found $F_0=22^{+8.6}_{-6.8}\%$ for planets with $2R_\oplus \leq R_p \leq 8R_\oplus$ around TESS stars younger than 200 Myr.
This drop in sub-Neptune occurrence may be due to factors such as photoevaporation or core-powered mass loss, but regardless suggests that the frequency of inner small planets in young exoplanet systems is higher than in the older \emph{Kepler} sample.

The FI model tends to produce planetary systems which are dynamically cold, with a few exceptions. 
Most of these systems contain multiple similarly-sized planets with tight spacings and small period ratios, coplanar orbits, and low eccentricities, properties which are exacerbated by increasing $St_{max}$. 
These planetary systems will become more coplanar as $log_{10}\xi$ shifts to positive values, all while the spacings get tighter and period ratios decrease.
The reweighting model heavily weights small Stokes numbers ($\mu^-=10^{-5.95}$ for $\dot{M} = 10^{-9}~\mathrm{M_\odot~yr^{-1}}$ and $\mu^-=10^{-6.30}$ for $\dot{M} = 4\times10^{-8}~\mathrm{M_\odot~yr^{-1}}$), as those systems which form from disks that are depleted in solid material produce more dynamically hot systems.
\citetalias{MacDonald2020} argue that the transition between preferentially dynamically hot and dynamically cold systems occurs at a solid surface density normalization of $\Sigma_{z,1}=50$ \textrm{g~cm$^{-2}$}. This value is equivalent to flow isolation systems formed with $St_{max} \lesssim 10^{-4}$, as they produce smaller planets with greater mutual inclinations and eccentricities, indicative of a dynamically hot system.
Thus, regardless of accretion rate, forming embryos with the FI mass in the inner disk, \change{without altering other model parameters such as total mass in the disk}, fails to fully capture the diversity of systems we observe in the \emph{Kepler} catalog.

\section{Properties of Migration Feedback Planets} \label{sec:migrationfeedback_planet_props}

\subsection{Undepleted Disk Model} \label{sec:undepleted_disk}

\begin{figure}
    \centering
    \includegraphics[width = 0.5\textwidth]{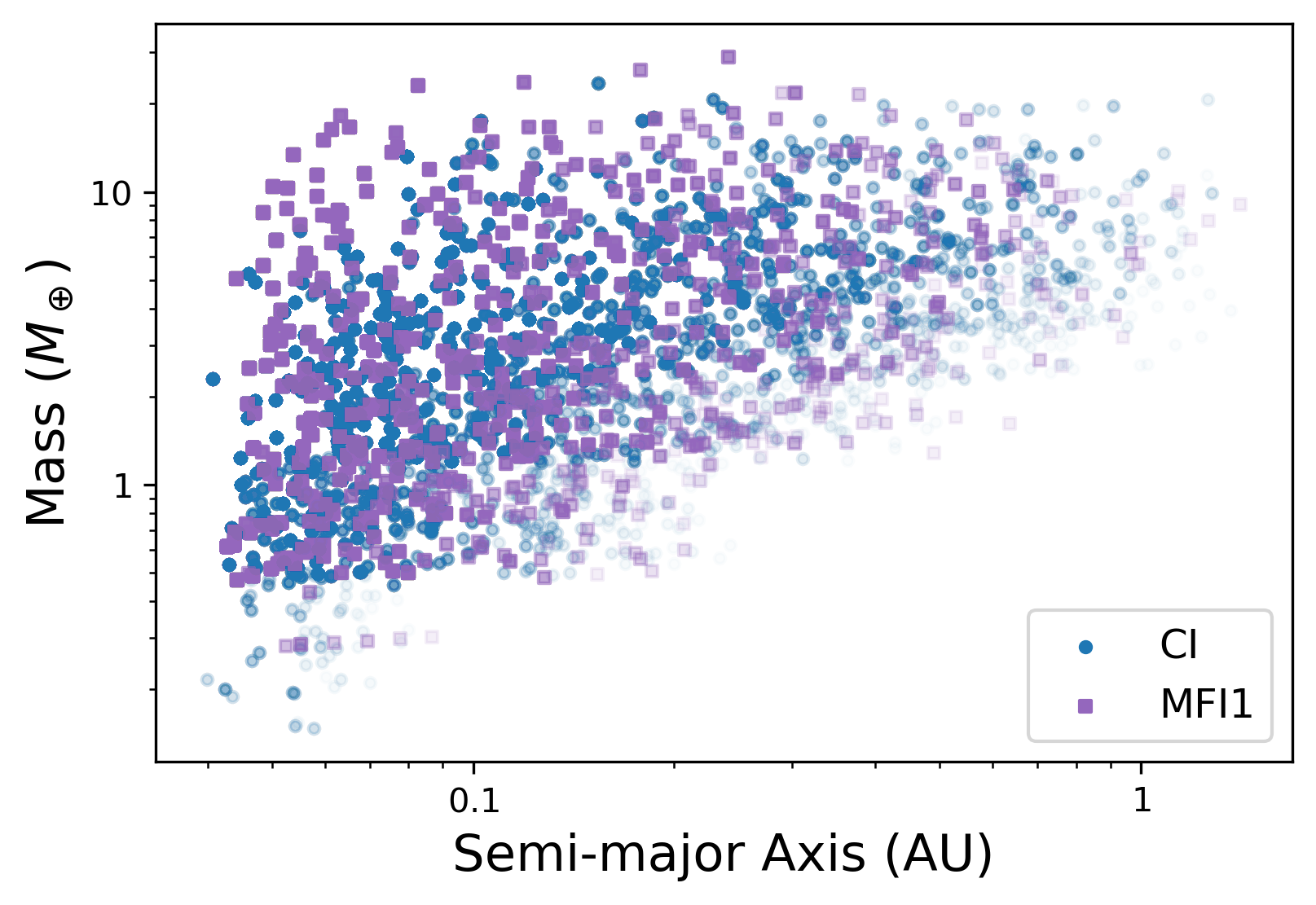}
    \caption{Mass as a function of semi-major axis for the MFI1 model (purple, squares), plotted alongside the CI planet masses. 
    The distribution qualitatively matches, implying similar initial conditions to place these planets at such locations.}
    \label{fig:fb_undepleted_mass_vs_a}
\end{figure}

\begin{figure*}
    \centering
    \includegraphics[width = 3.2in, height = 2.1in]{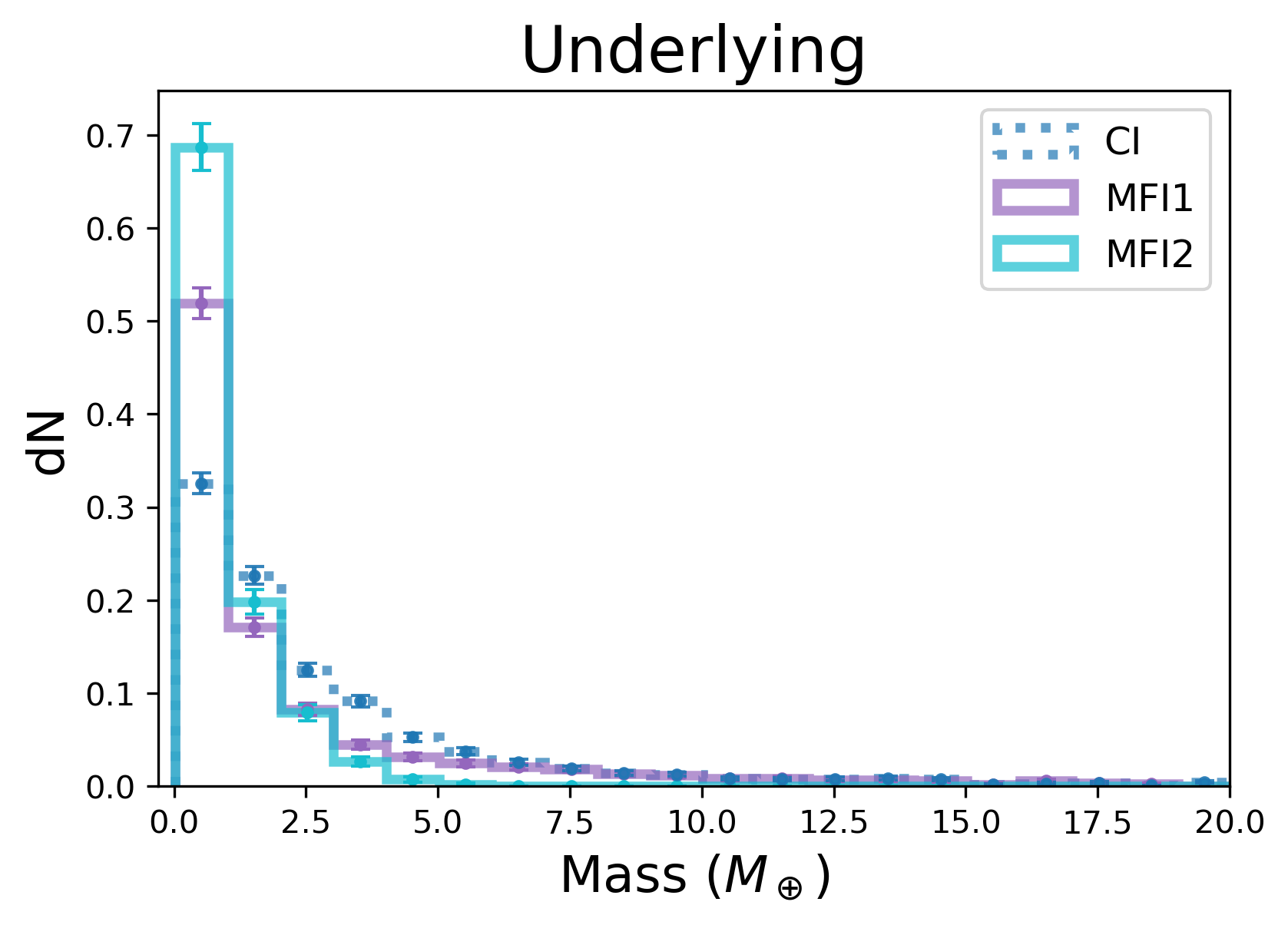}
    \includegraphics[width = 3.2in, height = 2.1in]{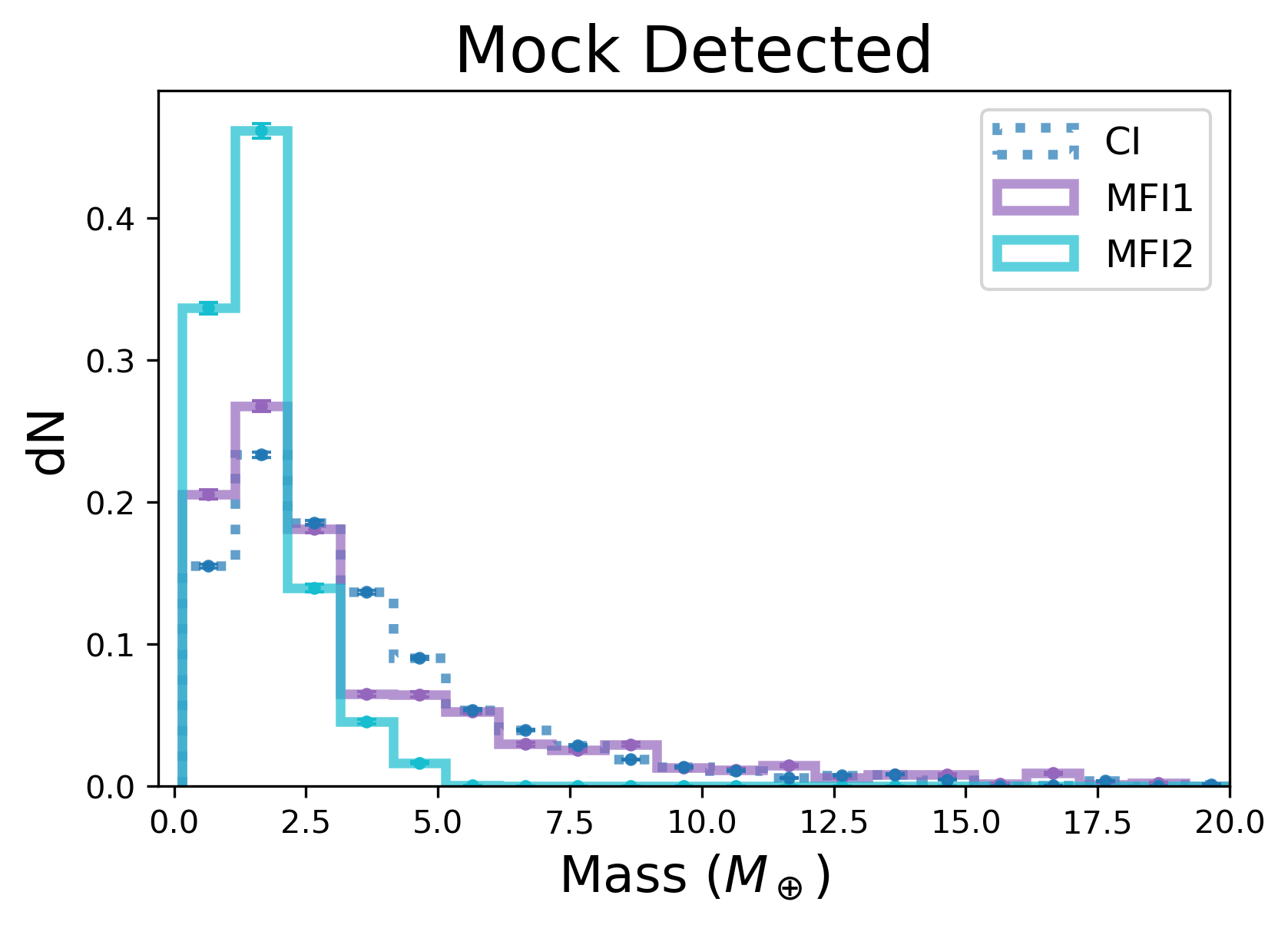}
    \includegraphics[width = 3.2in, height = 2.1in]{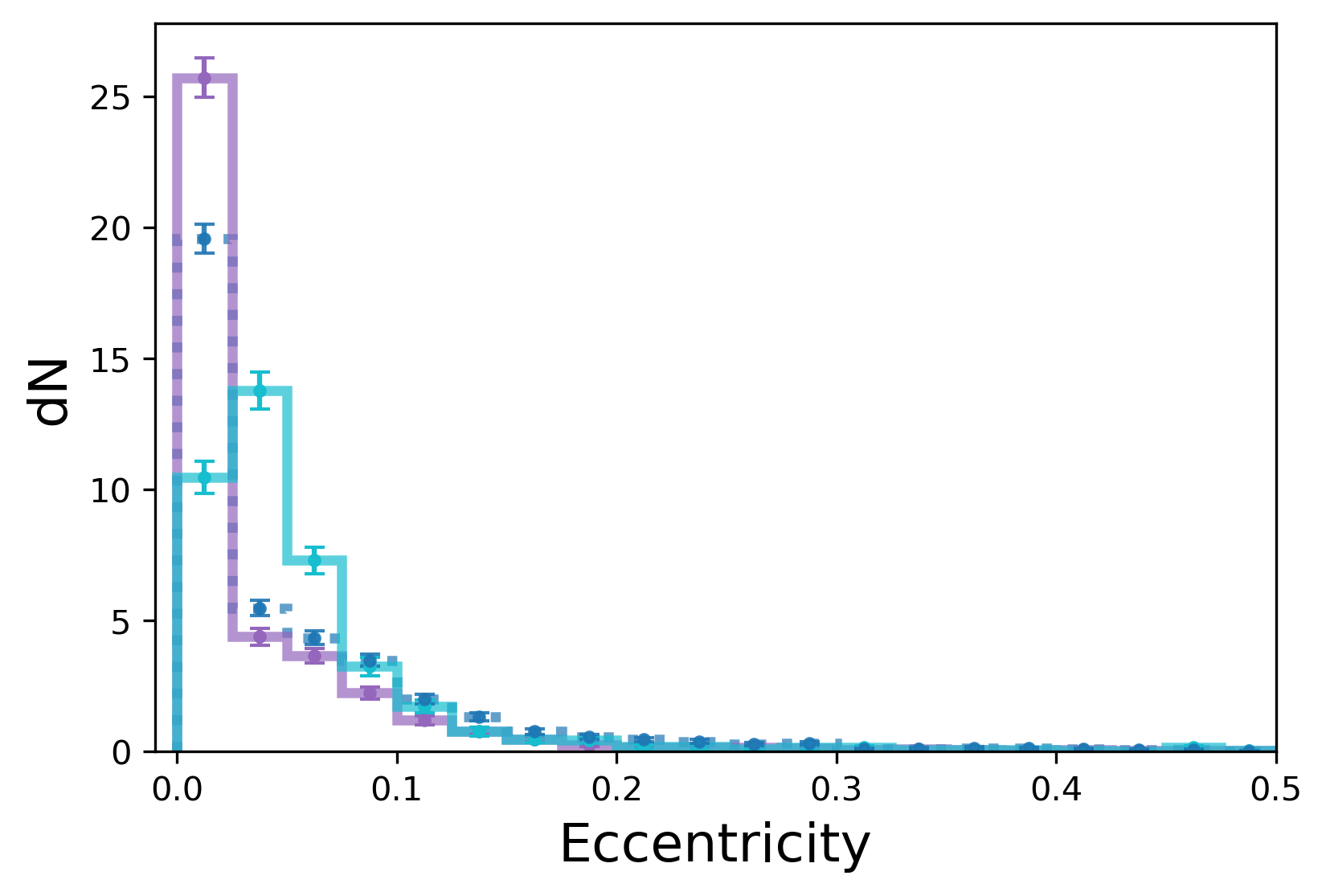}
    \includegraphics[width = 3.2in, height = 2.1in]{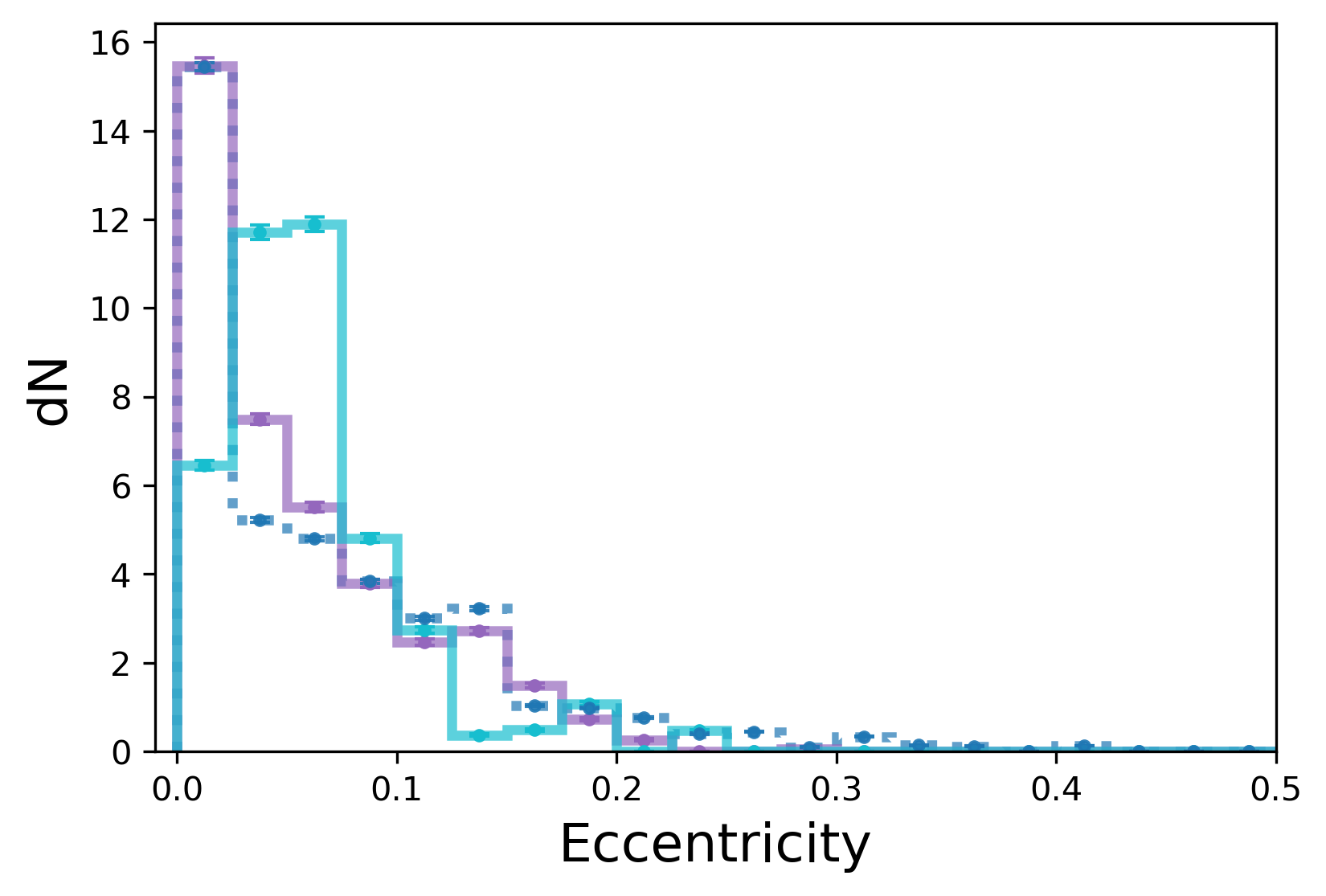}
    \includegraphics[width = 3.2in, height = 2.1in]{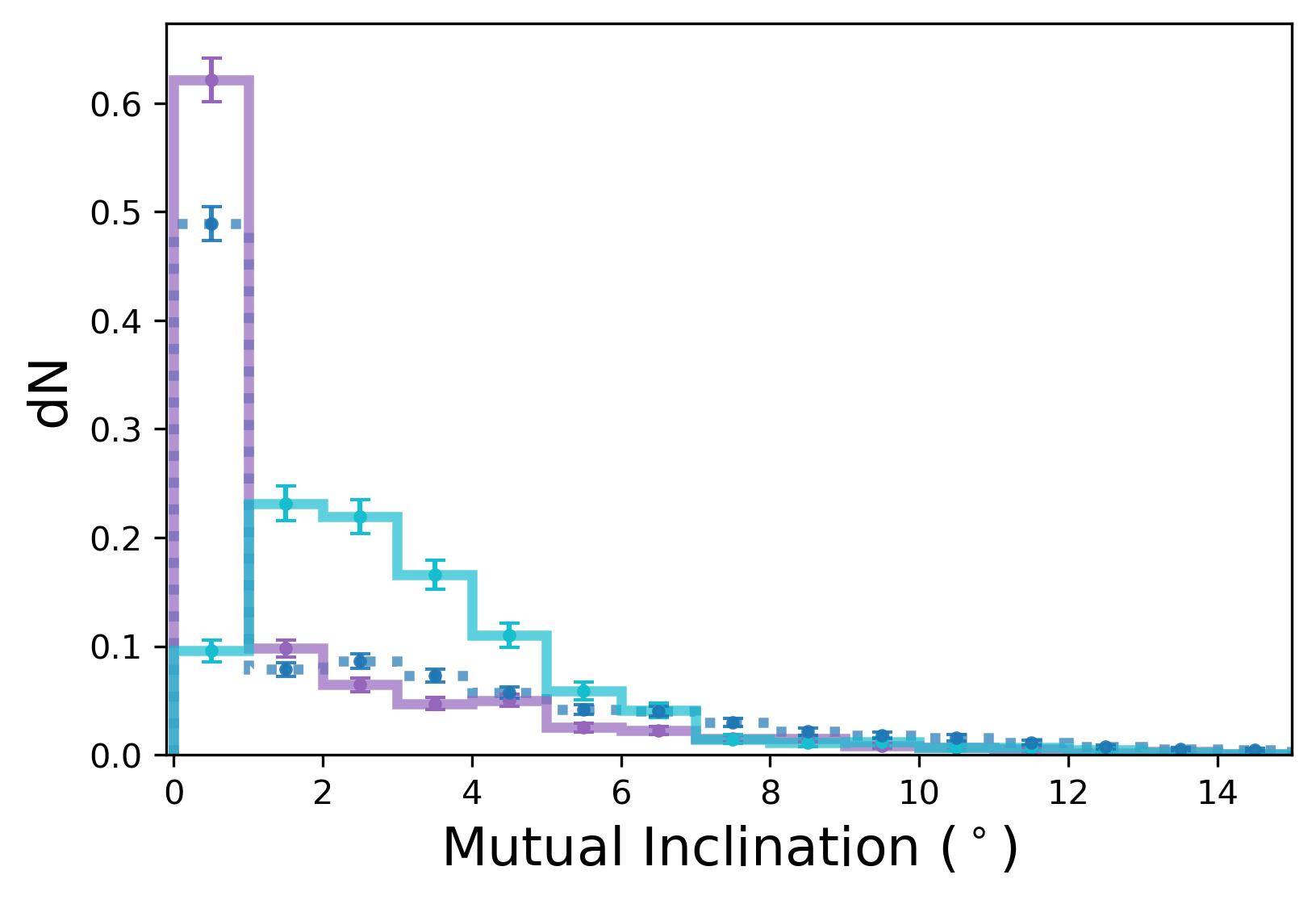}
    \includegraphics[width = 3.2in, height = 2.1in]{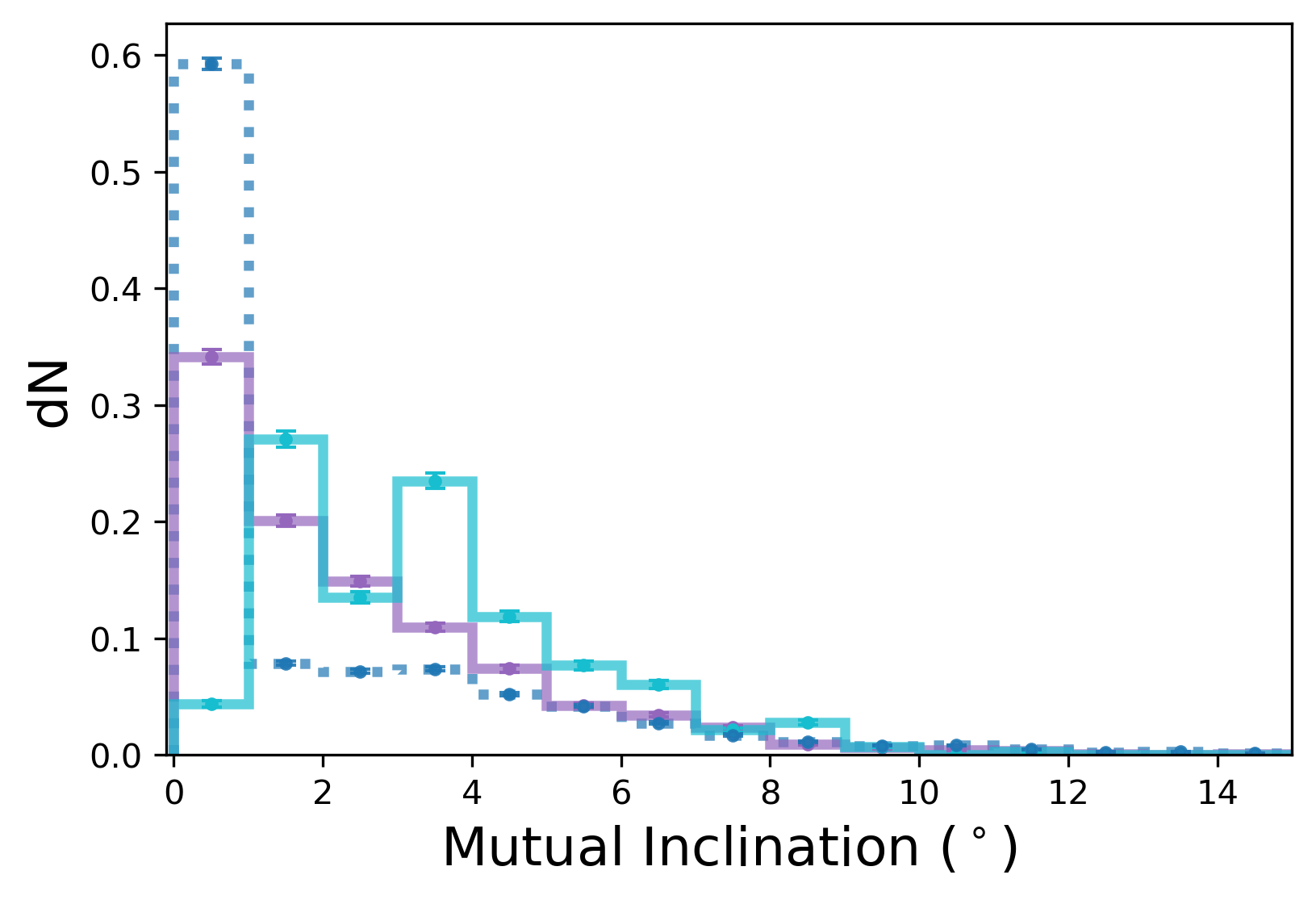}

    \caption{Distributions of mass (\emph{top}), eccentricity (\emph{middle}), and adjacent planet mutual inclination (\emph{bottom}) for both the underlying simulations (left) and the mock observed planets (right) of the MFI1 (purple, solid), MFI2 (cyan, solid) and CI (blue, dotted) models.
    The MFI1 model produces more terrestrial mass planets than the CI model, but after mock detection the mass distributions effectively match (except for a gap in the MFI1 distribution $2.5~\mathrm{M_\oplus}<\mathrm{M_p}<5~\mathrm{M_\oplus}$).
    The MFI2 model produces mostly planets with masses $\mathrm{M_p} < 2~\mathrm{M_\oplus}$, with the $1~\mathrm{M_\oplus}$ bin containing the majority of these planets.
    After mock detection, the MFI2 mass distribution shows a larger fraction of planets with $\mathrm{M_p} <2~\mathrm{M_\oplus}$ than the CI distribution, but the frequency of planets quickly falls off at $\mathrm{M_p}>5~\mathrm{M_\oplus}$.
    MFI1 and CI planets exhibit effectively the same eccentricity distributions before and after mock detection, with MFI1 showing slightly more planets on orbits with $e\sim 0$.
    MFI2 has fewer planets with $e\sim0$, but otherwise shows a similarly shaped eccentricity distribution before and after mock detection.
    The mutual inclination distributions of MFI1 and CI are consistent before mock detection, with the MFI1 model showing more planet pairs with $i_{mut}< 1^\circ$.
    After mock detection, MFI1 planets are more mutually inclined than CI planets.
    The MFI2 mutual inclination distribution shows many planet pairs with $i_{mut}>1^\circ$ in both the underlying and mock detected cases, which does not match the CI distribution.
    Thus, MFI2 planets are of terrestrial mass with moderately large mutual inclinations.
    }

    \label{fig:fb_underlying_vs_observed}
\end{figure*}

Figure~\ref{fig:fb_undepleted_mass_vs_a} plots planet mass as a function of distance from the host star for the population drawn from the MFI1 model ($d = 1$, $\Sigma_{g,1} =1700$ \textrm{g~cm}$^{-2}$). 
The MFI1 planet mass distribution qualitatively matches the CI distribution, with masses ranging between $0.4 \lesssim \mathrm{M_\oplus} \lesssim 25$ out to just over 1 au from the star.
As semi-major axis increases, mock detected masses also increase.
To explore these similarities further, we plot the distributions of underlying and observed masses, eccentricities, and mutual inclinations in Figure~\ref{fig:fb_underlying_vs_observed}.
Here, we see that the underlying MFI1 mass distribution (top left panel) peaks at roughly one Earth mass and falls off in frequency out to $15~ \mathrm{M_\oplus}$. 
While similar in distribution to the mass distribution of CI planets, the MFI1 model produces comparatively more Earth-mass planets and slightly fewer planets in the $1.0 < \mathrm{M_\oplus} < 6.0$ regime relative to CI.
By reweighting and applying the mock detection model, the detected mass distribution (top right panel) matches the CI distribution, with the largest deviation being fewer planets in the range $2.5 < \mathrm{M_\oplus} < 5.0$ relative to CI. 
We see a similar scenario with regards to the underlying eccentricities (middle left panel): the MFI1 model produces fractionally more circular orbits than the CI model, and a slight relative deficit in eccentricities $0.025 < e < 0.2$. 
The observed eccentricity distributions (middle right panel) of both models match relatively well. 
The underlying mutual inclinations (bottom left panel) exhibit the same trend-- MFI1 produces slightly more coplanar planet pairs, with a decrease in frequency at large mutual inclinations. 
Once mock detected (bottom right panel), the frequency of coplanar MFI1 pairs is halved, and we see a slight uptick in pairs with small mutual inclinations ($1^\circ < i_{mut} < 3^\circ$).

\begin{figure*}
    \centering
    \includegraphics[width = 3.2in, height = 2.1in]{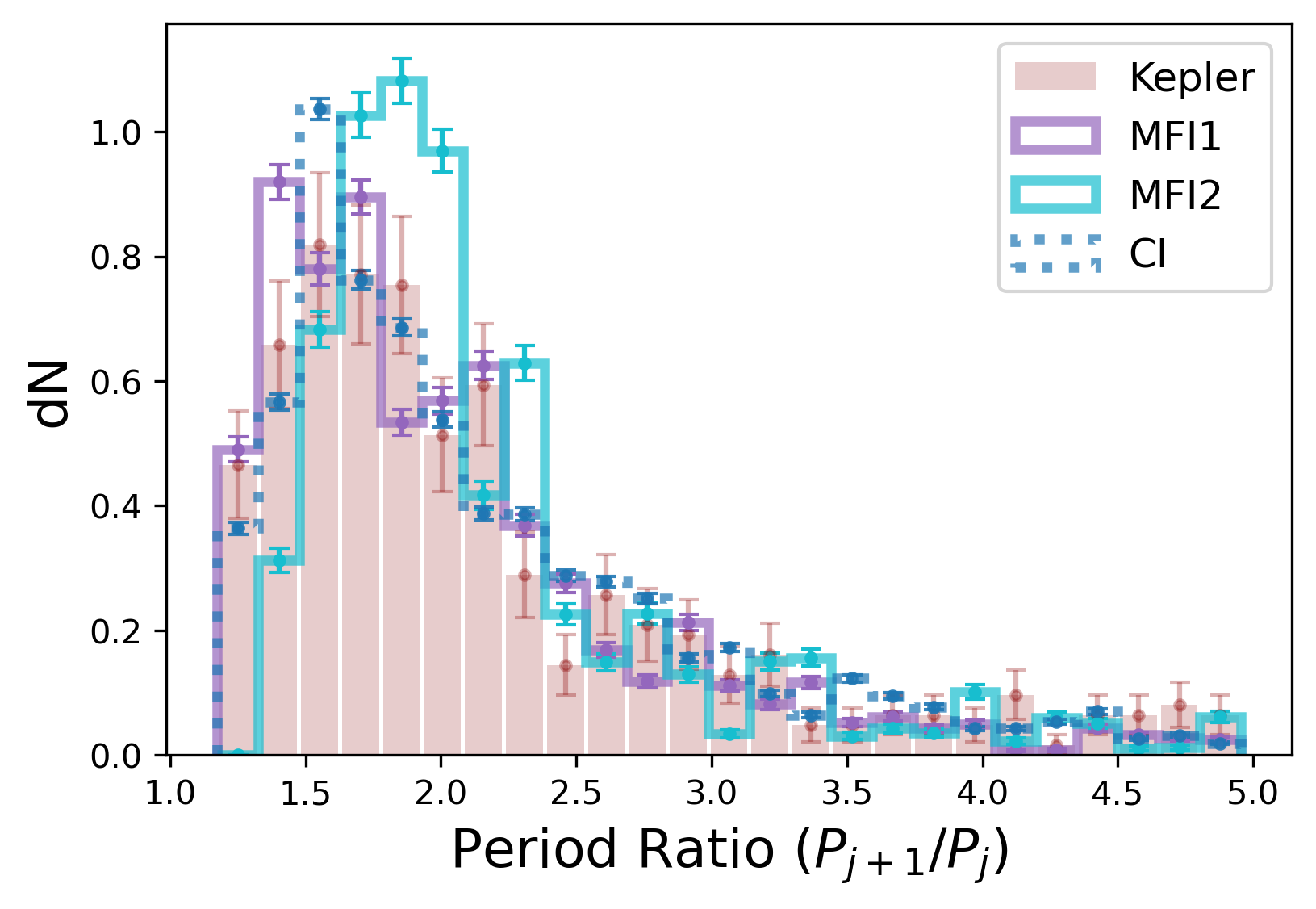}
    \includegraphics[width = 3.2in, height = 2.1in]{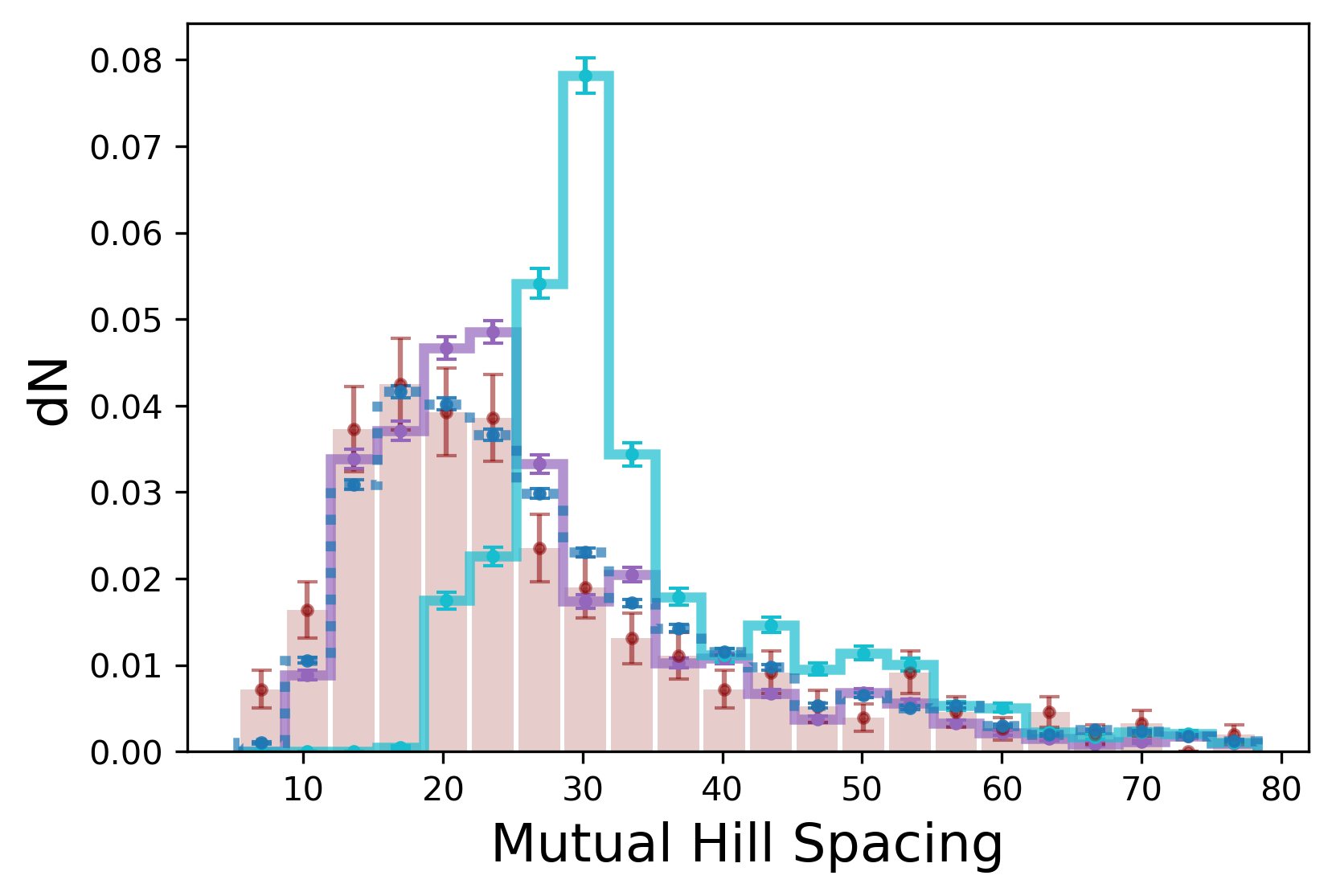}
    \includegraphics[width = 3.2in, height = 2.1in]{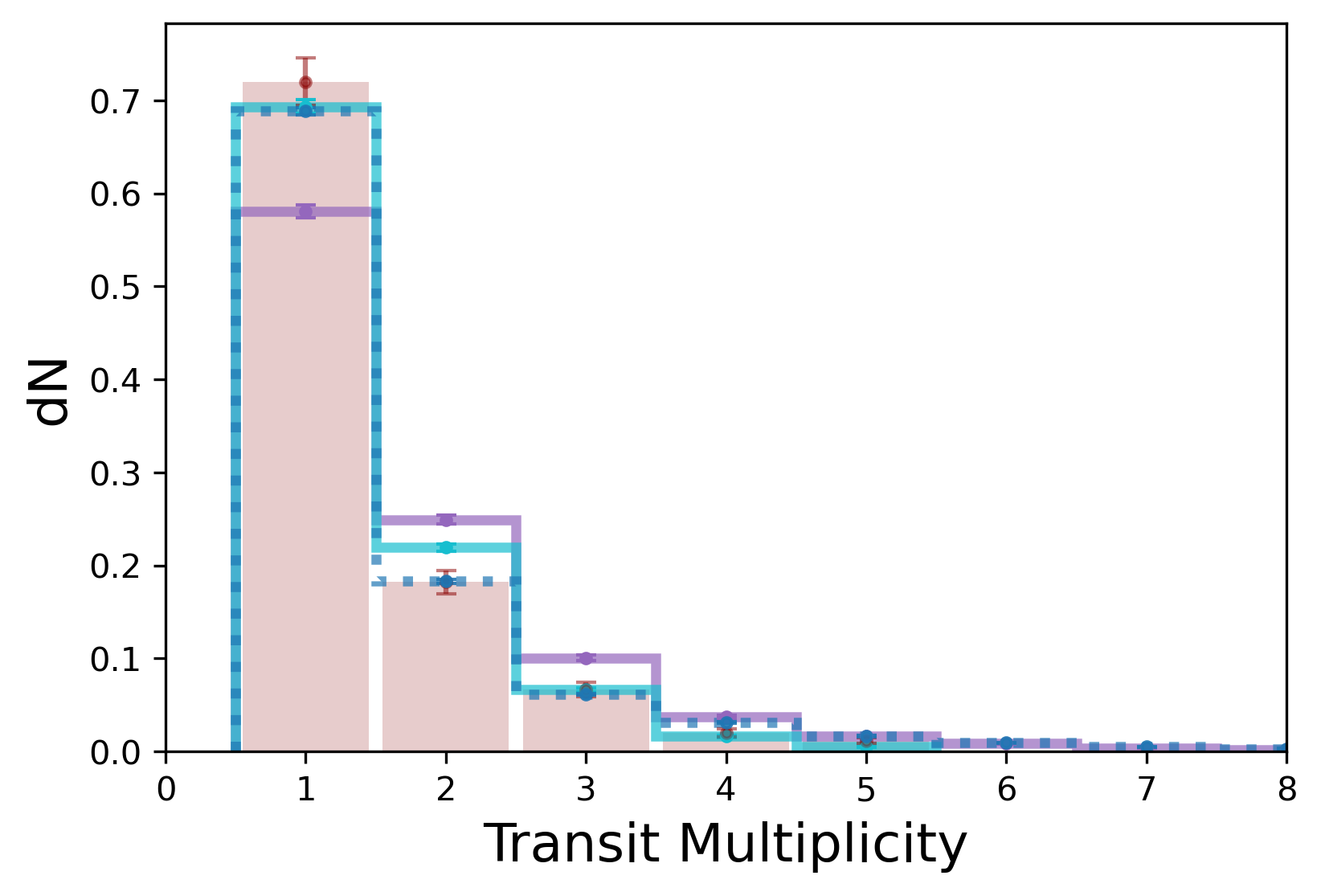}
    \includegraphics[width = 3.2in, height = 2.1in]{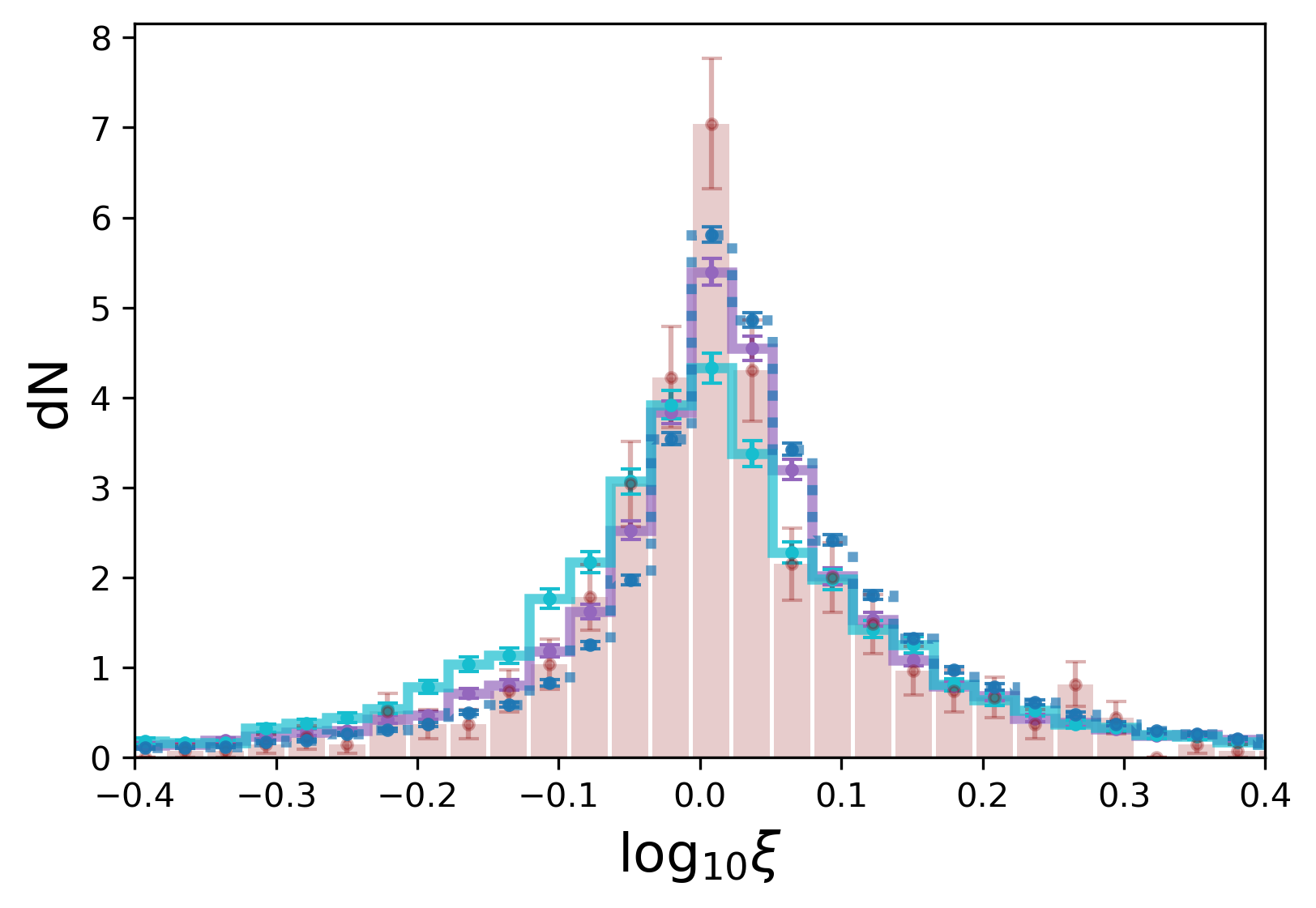}
    \caption{From upper left to lower right: Distributions of period ratios of adjacent planets, mutual Hill spacings of adjacent planets, transit multiplicities, and transit duration ratios ($\xi$) of adjacent planets from the MFI1 and MFI2 mass models (purple and cyan respectively), compared to the CI mass (blue) and the \emph{Kepler} catalog distribution (solid pink).
    The MFI1 distributions of period ratio, mutual Hill spacing, and log duration ratio all qualitatively match the CI and \emph{Kepler} catalog distributions.
    However, the MFI1 model produces fewer detected single planet systems, and too many multi-planet systems, relative to the \emph{Kepler} catalog.
    The MFI2 period ratio and mutual Hill spacing distributions peak at larger values than the \emph{Kepler} catalog, implying more widely spaced systems.
    The fraction of MFI2 single planet systems matches the \emph{Kepler} catalog within error, but we still see a slight enhancement of two planet systems.
    The MFI2 transit duration ratio distribution has wide wings relative to CI and the \emph{Kepler} catalog, implying larger mutual inclinations.}
    \label{fig:migration_observables}
\end{figure*}

In Figure~\ref{fig:migration_observables}, we compare the distributions of our four system observables for the reweighted MFI1 model, finding $\mu^- = 14.4$ and $\sigma^- = 5.50$ as our best fitting weight function parameters. 
\change{These weights significantly improve the degree to which the MFI1 suite matches observations, but dramatically skew the original distribution toward much wider initial separations, implying that the original choice of initial conditions fails to capture the whole population of dynamically hot systems.}

Comparing the MFI1 model's period ratio distribution to that of the CI model and the \emph{Kepler} catalog distribution (top left), we see a match between the distributions of the two models and the \emph{Kepler} catalog.
The mutual Hill spacing (top right) between MFI1 planet pairs follows much the same distribution as CI and the \emph{Kepler} catalog, peaking at $\Delta \sim 20$.
A significant fraction of MFI1 systems harbor only one detected planet (bottom left), yet we still see too few systems with this configuration, alongside a corresponding increase in 2--4 planet system frequency.

The transit duration ratio distributions (bottom right) are nearly identical for the MFI1 and CI models, both peaking at $\log_{10}\xi \sim 0$ and exhibiting wings which follow the \emph{Kepler} catalog distribution. 
The MFI1 model does however exhibit slightly wider wings and a slight leftward shift relative to the \emph{Kepler} catalog, as expected due to the broader distribution of mutual inclinations in the bottom right panel of Figure~\ref{fig:fb_underlying_vs_observed}.

\subsection{Depleted Disk Model} \label{sec:depleted_disk}

\begin{figure}
    \centering
    \includegraphics[width = 0.5\textwidth]{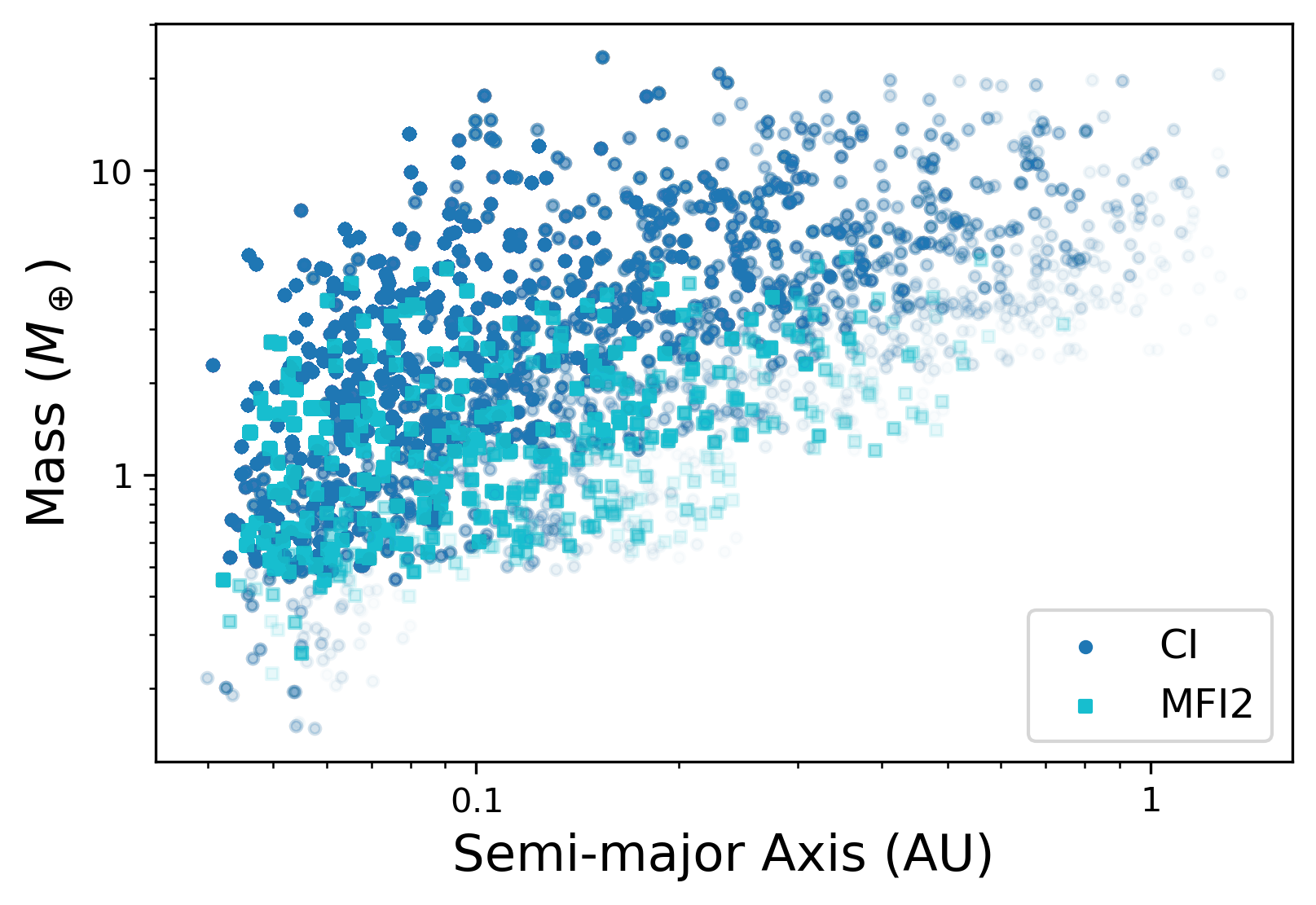}
    \caption{Mass as a function of semi-major axis for the MFI2 model (cyan, squares), plotted against the same for the CI model (blue, circles). 
    MFI2 tends to produce lower mass planets than CI at all semi-major axes, likely due to formation within a diffuse gas disk.}
    \label{fig:fb_depleted_mass_vs_a}
\end{figure}

Figure~\ref{fig:fb_depleted_mass_vs_a} plots mass as a function of semi-major axis for the MFI2 model ($d = 100$, $\Sigma_{g,1} =17$ \textrm{g~cm$^{-2}$}).
Mass shows a similar dependence on semi-major axis as the CI model, ranging between values of roughly $0.1-5~\mathrm{M_\oplus}$.
At masses $>5~\mathrm{M_\oplus}$, the frequency of MFI2 planets decreases drastically. 
The MFI2 model tends to produce a high fraction of sub-Earth mass, short-period planets. 
These planet masses are a result of the MFI2 mass's dependence on $\Sigma_g$ (see Equation~\ref{eq:feedbackmass}), as formation within a depleted disk will produce very low embryo masses.

We plot both the underlying and observed masses, eccentricities, and mutual inclination distributions of the planets produced by MFI2 along with those produced by CI in Figure~\ref{fig:fb_underlying_vs_observed}. 
The MFI2 model shows a preference for producing sub-Earth and Earth mass planets (Figure~\ref{fig:fb_underlying_vs_observed}, top left), with the majority $\mathrm{M_p}< 2~\mathrm{M_\oplus}$.
A small fraction of planets exhibit masses greater than this value, with few planets at masses greater than $5~\mathrm{M_\oplus}$.
This distribution is in contrast to the CI, which shows a non-negligible fraction of planets out to $10~\mathrm{M_\oplus}$.
After detection and reweighting (top right), the fraction of planets with $<1~\mathrm{M_\oplus}$ is halved and the frequency of planets $\sim2~\mathrm{M_\oplus}$ is more than doubled.
Beyond $3~\mathrm{M_\oplus}$, the number of detected planets drops dramatically, falling off entirely after $5~\mathrm{M_\oplus}$. 
Again, these low masses are likely a product of the assumption that the MFI2 embryos were produced within a diffuse gas disk, which creates much lower mass bodies than an undepleted MMSN disk (see Section~\ref{sec:undepleted_disk}). 

The underlying MFI2 planets also exhibit slightly more eccentric orbits than those given by the CI, peaking at $e \sim 0.05$, although with a significant fraction of planets maintaining circular orbits (Figure~\ref{fig:fb_underlying_vs_observed}, middle left panel). 
From the peak, the eccentricity distribution drops rapidly, becoming consistent with CI and trailing off at $e > 0.1$.
The eccentricity distribution remains qualitatively the same once we only consider mock detected planets (middle right panel), with a peak at $e \sim 0.05$ and rapid decrease in detection $e > 0.125$. 
Relative to CI, the MFI2 planets are more mutually inclined (Figure~\ref{fig:fb_underlying_vs_observed}, bottom left panel), with a unimodal distribution peaking at $i_{mut} \sim 2^\circ$ with a tail of higher mutual inclination values that taper off beyond $\sim 8^\circ$.
Mock detection changes the MFI2 mutual inclination distribution little, maintaining a high fraction of planet pairs with mutual inclinations $>1^\circ$.


In Figure~\ref{fig:migration_observables}, we compare the distributions of the primary four observables of the reweighted MFI2 model to the distributions of the mock CI planets, as well as to the \emph{Kepler} catalog distribution. 
\change{The best-fitting reweighting parameters for this model are $\mu^- = 2.62$ and $\sigma^- = 0.702$, suggesting only a slight preference for tighter spacings in order to match observations. 
However, even with reweighting we still do not capture dynamically cold systems, suggesting additional parameters must be varied.}


The mock detected planets produced by the MFI2 model have slightly larger period ratios (top left) than the planet pairs in the \emph{Kepler} catalog (peaking at $\frac{P_{j + 1}}{P_j} \sim 1.8$, shifted from the \emph{Kepler} catalog peak at $\frac{P_{j + 1}}{P_j} \sim 1.6$), exhibiting a deficit in the most tightly spaced systems at $\frac{P_{j + 1}}{P_j}~<~1.5$. 
At $\frac{P_{j + 1}}{P_j} > 2.4$, the distributions of \emph{Kepler} catalog and mock observed planets qualitatively match. 
The final mutual Hill spacings of the MFI2 planet pairs (top right) show a more dramatic deviation from the \emph{Kepler} catalog, with virtually no pairs at spacings of $\Delta < 20$, and a large peak at $\Delta \sim 30$. 
This peak at large spacings, following Eqs.~\ref{eq:Hillrad} and~\ref{eq:mutualhill}, likely manifests as a result of massive embryos colliding frequently during the dissipated gas disk stage, and the low mass embryos colliding only after dissipation.

The MFI2 model does however produce many single transiting planet systems, matching the \emph{Kepler} catalog peak to within $1\sigma$. 
While the fraction of two-planet MFI2 systems is still higher than that of the \emph{Kepler} catalog, the difference is small.
Thus, the MFI2 planets show a transit multiplicity distribution which is in good agreement with that of the \emph{Kepler} catalog.
The wings of the MFI2 transit duration ratio distribution are wider than those of the \emph{Kepler} catalog distribution, a signature of larger mutual inclinations and nonzero eccentricities; the overall shape of the distribution is maintained, but the frequency of negative $\log_{10}{\xi}$ planet pairs indicates high mutual inclinations and thus a failure to match the \emph{Kepler} catalog distribution \change{without potentially altering other model parameters, such as total system mass}.

\subsection{Migration Feedback EMD} \label{sec:migration_EMD}

\begin{figure*}
    \centering
    \includegraphics[width = 3.0in, height = 2.0in]{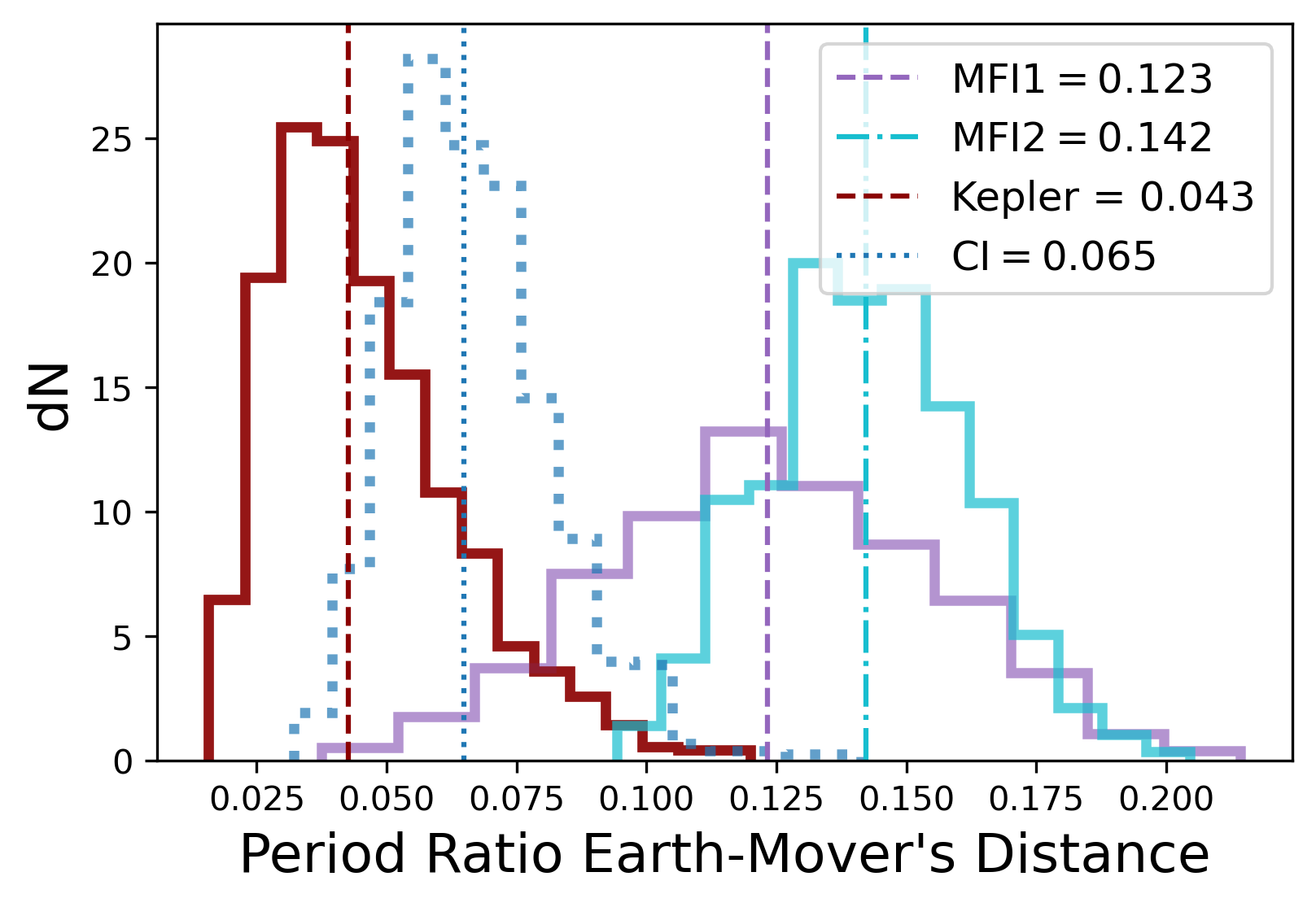}
    \includegraphics[width = 3.0in, height = 2.0in]{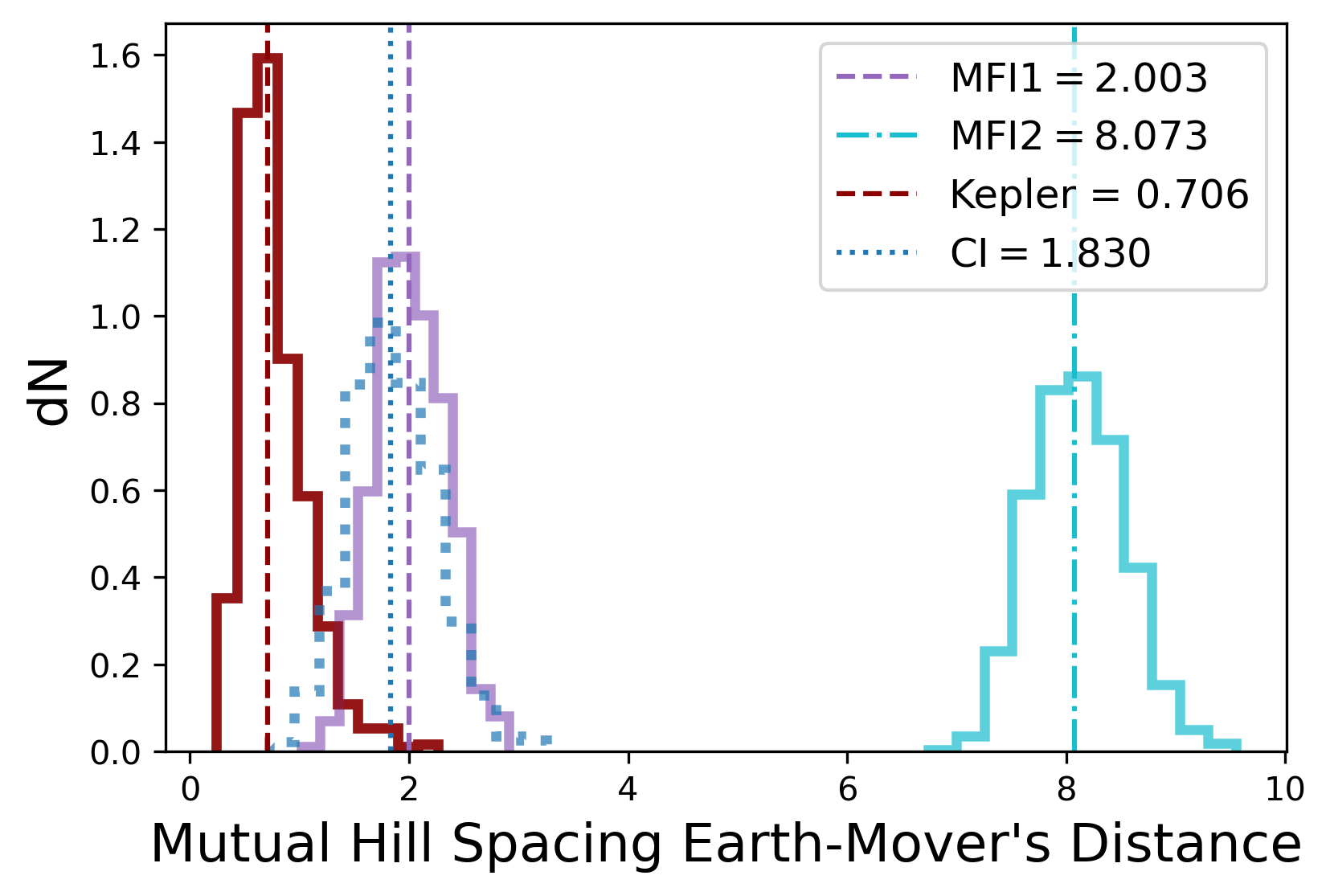}
    \includegraphics[width = 3.0in, height = 2.0in]{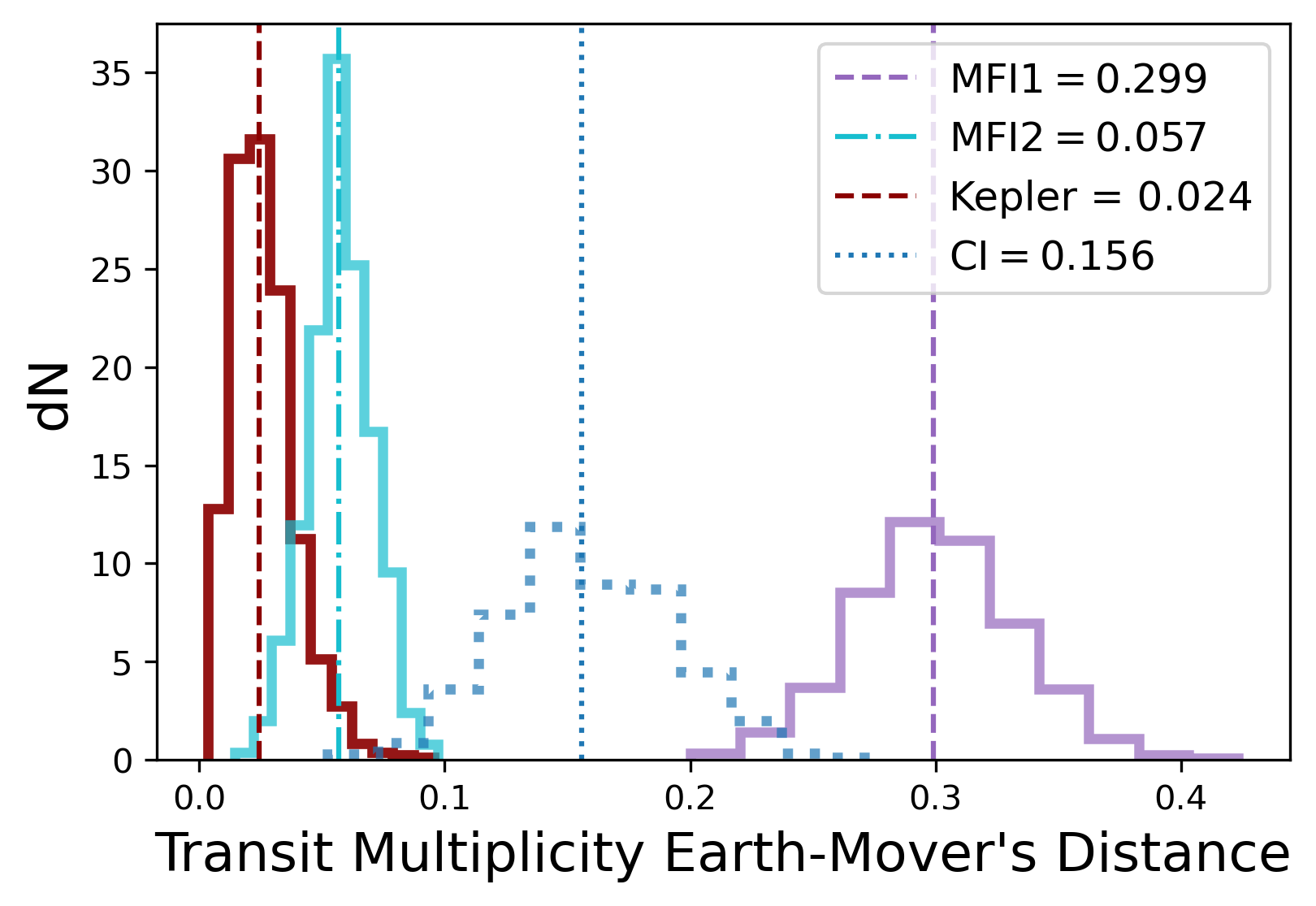}
    \includegraphics[width = 3.0in, height = 2.0in]{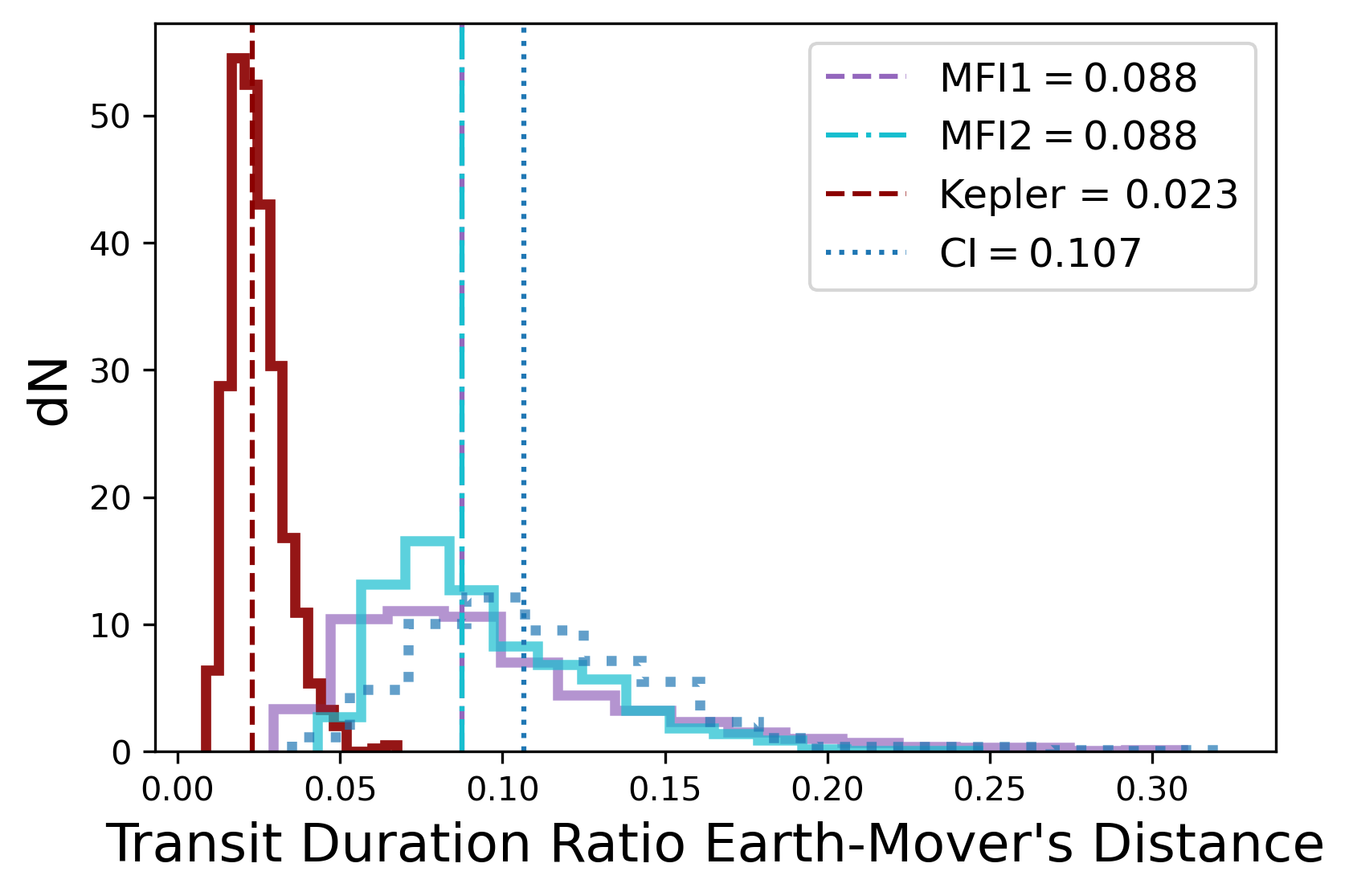}
    \caption{Earth Mover's distance distribution for the MFI1 and MFI2 models (solid purple and solid cyan, respectively) against the CI (blue, dotted) and \emph{Kepler} (dark red, solid) distributions.
    We see significant overlap between the EMD distributions for MFI1, CI and \emph{Kepler} in period ratio, mutual Hill spacing, and transit duration ratio, implying that the MFI1 model is an equally good fit to the \emph{Kepler} catalog or slightly better. 
    The MFI1 transit multiplicity match (bottom left) is poorer than that of the CI, having no overlap with \emph{Kepler}.
    This result implies the MFI1 model is a good match to the \emph{Kepler} catalog in three of four observables, but it fails to properly capture the multiplicity distribution of the \emph{Kepler} catalog.
    The MFI2 model shows greater overall distances in period ratio and mutual Hill spacing, implying that it provides a worse overall match to the \emph{Kepler} catalog distributions for these parameters.
    \changethree{However, it is nearly a match to the \emph{Kepler} distribution of transit multiplicity EMD, better still than the CI distribution.}
    Thus, while the MFI2 planets are more widely spaced and less massive than those in the \emph{Kepler} catalog, the transit multiplicity and duration ratio distributions match.}

    \label{fig:migration_emd}
\end{figure*}

In general, the MFI1 model produces systems which more closely resemble the \emph{Kepler} catalog systems with respect to the other four models. 
In the top left panel of Figure~\ref{fig:migration_emd} we see a wide distribution of period ratio EMDs for the MFI1 model, with the variance from the mean nearly fully encompassing the CI distribution. 
The CI distribution does still provide a closer match to the \emph{Kepler} catalog, but \changethree{MFI1 is by no means a poor fit.}
The MFI1 and CI mutual Hill spacing distributions (top right) encompass one another and nearly entirely overlap, \changethree{with means within the uncertainty of the \emph{Kepler} distribution}.
Both models provide an adequate match to the \emph{Kepler} catalog's distribution of mutual Hill spacing, with CI fitting marginally better.
In the case of transit multiplicity (bottom left), the MFI1 distribution peaks at a larger distance than CI \changethree{and does not fall within \emph{Kepler} uncertainties}.
As with other models, the transit duration ratio EMD distributions are nearly identical.
Overall, the MFI1 model produces distributions which are adequate matches to the \emph{Kepler} catalog, with the largest discrepancy being the lack of single transiting planet systems.

Through juxtaposition of the EMD observable property distributions for the MFI2 model (in which $d = 100$ in Equation~\ref{eq:gassurfdens}) in Figure~\ref{fig:migration_emd} against CI, a few properties become clear. 
For the period ratio EMD distribution (top left), the MFI2 model produces larger distances than CI nearly irrespective of viewing angle.
\changethree{There is still some overlap between the MFI2 and \emph{Kepler} distributions, however small.}
In the case of the mutual Hill spacing (top right), the MFI2 EMD distribution \changethree{peaks at values several times larger than \emph{Kepler}}.
However, the MFI2 model clearly matches the \emph{Kepler} catalog's transit multiplicity more closely than CI, \changethree{given the mean of the multiplicity EMD distribution is within \emph{Kepler} uncertainties}.
As above, the transit duration ratio EMD distributions are effectively identical for MFI2 and CI.
This distribution would lead one to believe that the MFI2 and CI duration ratio distributions appear equally similar in the bottom right panel of Figure~\ref{fig:migration_observables}, and yet this is not the case.
The MFI2 model produced more detected planets than CI at $log_{10}\xi < 0$ and under-produced at ratios $0 < \log_{10}\xi < 0.15$, which implies the detected MFI2 planet pairs have more significant mutual inclinations than CI pairs.
We can confirm this through the bottom right panel of Figure~\ref{fig:fb_underlying_vs_observed}, where we see a majority of planet pairs produced by this model with mutual inclinations $1^\circ < i_{mut} < 5^\circ$.
The duration ratio EMD distribution exemplifies a shortcoming of this method of comparison.
The MFI2 model is a poor match to the \emph{Kepler} catalog in regards to its majority of widely spaced, low mass planets. 
However, the model's tendency to produce systems such as this leads to a higher fraction of single planet systems compared to MFI1, FI1, FI2 and PI. \changethree{We provide a summary of the EMD statistics in Table \ref{tab:summary_stats_table}.}

\subsection{Implications of Formation by Migration Feedback Isolation} \label{sec:feedback_implications}

We compare the results of the MFI2 model to the \emph{Kepler} catalog in Figure~\ref{fig:migration_observables}. 
When taken as a whole population, the MFI2 systems do not qualitatively match the distributions expected from the \emph{Kepler} catalog. 
MFI2 planet pairs have generally larger period ratios and wider Hill spacings, implying that these planets are widely separated.
While the MFI2 model produces a similar fraction of single transiting planet systems as the \emph{Kepler} catalog, the MFI2 planets are more eccentric and mutually inclined when compared to the CI planets' distributions.
Combined with the planets' low masses (and as a result, small radii) and wide spacings, the enhancement in the fraction of single planets could be due to low detection probabilities among the underlying planet distribution.

Referencing Figure~\ref{fig:hillrange}, for nearly all spacings in the range $\Delta_0 = 1-5$, the total solid mass in the inner depleted disk is below the total mass threshold given by $\Sigma_{z,1} = 50 ~\mathrm{g~cm^{-2}}$. 
\citetalias{MacDonald2020} argue that this solid surface density normalization (when assuming formation in a disk structured like the MMSN) is the transition value between whether dynamically hot or cold systems are preferentially formed by the given model.
The more solid mass is present in the inner disk, the more likely the resulting planetary system will be dynamically cold: tightly spaced, coplanar, multi-planet systems. 
The MFI2 distribution is thus representative of primarily dynamically hot systems.
Figure~\ref{fig:fb_underlying_vs_observed} shows that planets throughout the suite exhibit a deficit in circular, coplanar orbits, instead preferentially orbiting in slightly eccentric and inclined configurations reflective of dynamically hot systems. 
As a result of the restrictive initial spacing window and low overall planet mass, the MFI2 model does not succeed in matching the \emph{Kepler} catalog, instead forming only a small subset of the population we expect.

The results of the MFI2 model suggest that classification of a planetary system by only one parameter is fraught with risk. 
The ``\emph{Kepler} dichotomy'' refers to how \emph{Kepler} systems are organized into either single-planet transit or multi-planet transit categories \citep{Lissauer2011}, and is a common metric used to categorize \emph{Kepler} systems.
MFI2 systems recreate the \emph{Kepler} dichotomy well, and so if multiplicity is the only observable considered, one may continue to analyze the data under the assumption that the other observable distributions match. 
However, the MFI2 planets are generally low mass, lower than expected for planets in the \emph{Kepler} catalog.
Thus we must also compare other observable properties to the \emph{Kepler} catalog when able, such as distributions of transit duration ratio and mutual Hill spacing.

If the MFI embryos form in an undepleted disk structured like the MMSN, a variety of system architectures emerge.
As initial mutual Hill spacing increases, systems trend toward higher multiplicities, lower masses, tighter spacings, and smaller mutual inclinations, like dynamically cold \emph{Kepler} systems.
Those systems with tighter initial spacings tend to produce low multiplicity systems with larger mutual inclinations and higher masses.


Despite this diversity of systems, the MFI1 model \change{is still unable to} fully capture the breadth of parameters in the \emph{Kepler} catalog. 
From Figure~\ref{fig:migration_observables}, we see that MFI1 underproduces final systems with mutual Hill spacings $\Delta < 16$, implying that these planets are not as tightly spaced as they should be.
We also see, like with the FI planets in Section~\ref{sec:flowiso_implications}, a deficit of single-planet systems and an overabundance of multis.

\citet{FungLee2018} posit that their isolation mass prescription should also form giant planets in the outer disk.
The primary purpose of the MFI model is to explain why gas giants are preferentially found at wide orbits, whereas smaller planets tend to orbit much closer to their host star. 
By neglecting to include giant planets in our simulations, we have left out important physical processes during the dissipated disk and giant impact phases that might play a significant role in the evolution of small planets in the inner disk (see Section~\ref{sec:flowiso_implications}).
The presence of outer giants might also affect our assumption of perfectly efficient pebble accretion and constant solid replenishment to the inner disk. 
We should then expect to see lower mass embryos at the beginning of our simulation. 
The presence of outer giant planets may thus lead to lower multiplicity in our final systems, and in turn fewer dynamically cold systems, aligning our results more closely with the \emph{Kepler} catalog.

The deficit of low multiplicity systems may also be due to a failure by our simulations to incite giant impacts at the widest initial spacings. 
We found in Section~\ref{sec:detect_transit} that the largest initial spacing at which MFI1 embryos still underwent giant impacts was $\Delta_0 = 11.08$.
Beyond this value, the embryos cannot significantly interact before the dissipation of the gas disk, and after dissipation are too far apart to be excited into collisions. 
Therefore, we removed these simulations from consideration, and in doing so removed all those which explore the inner disk total mass regime from $\sim5-10~\mathrm{M_\oplus}$ (Figure~\ref{fig:hillrange}). 
Were they to undergo giant impacts, these low mass systems could produce the low multiplicity, high eccentricity, and high inclination planets the full MFI1 simulation suite lacks.
However, with the incomplete exploration of the \citetalias{MacDonald2020} total mass regime, we cannot fully recreate the \emph{Kepler} catalog system diversity with this model through a continuum of initial embryo spacings.

This suite of simulations implies that the true observed diversity could be achieved through a combination of varying initial conditions, rather than a range of one particular parameter. 
In future work we may explore the dependence of final migration feedback planet properties on parameters such as disk scale height and gas surface density (see Equation~\ref{eq:feedbackmass}), the two of which were held constant in this work.


\section{Properties of Pebble Isolation Planets} \label{sec:pebbleiso_planetprops}

\begin{figure}
    \centering
    \includegraphics[width = 0.5\textwidth]{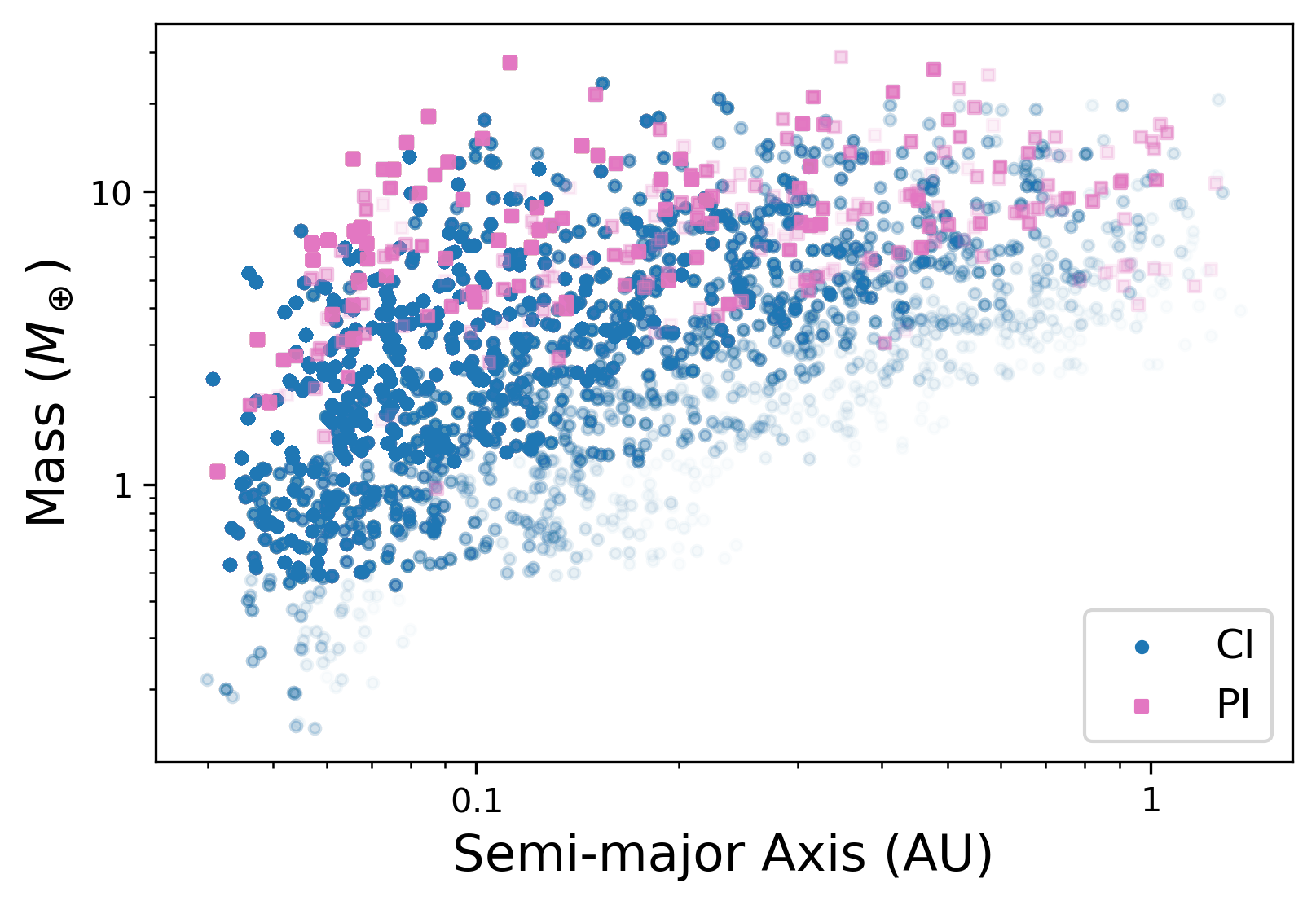}
    \caption{Mass as a function of semi-major axis for the PI model (pink, squares), plotted against the same for the CI model (blue, circles).
    The PI planets are all high mass, showing a weak increase in mass with $a$.
    The PI model fails to capture the lowest mass planets $a\leq 0.5$ au.}
    \label{fig:lamb_mass_vs_a}
\end{figure}

\begin{figure*}
    \centering
    \includegraphics[width = 3.2in, height = 2.1in]{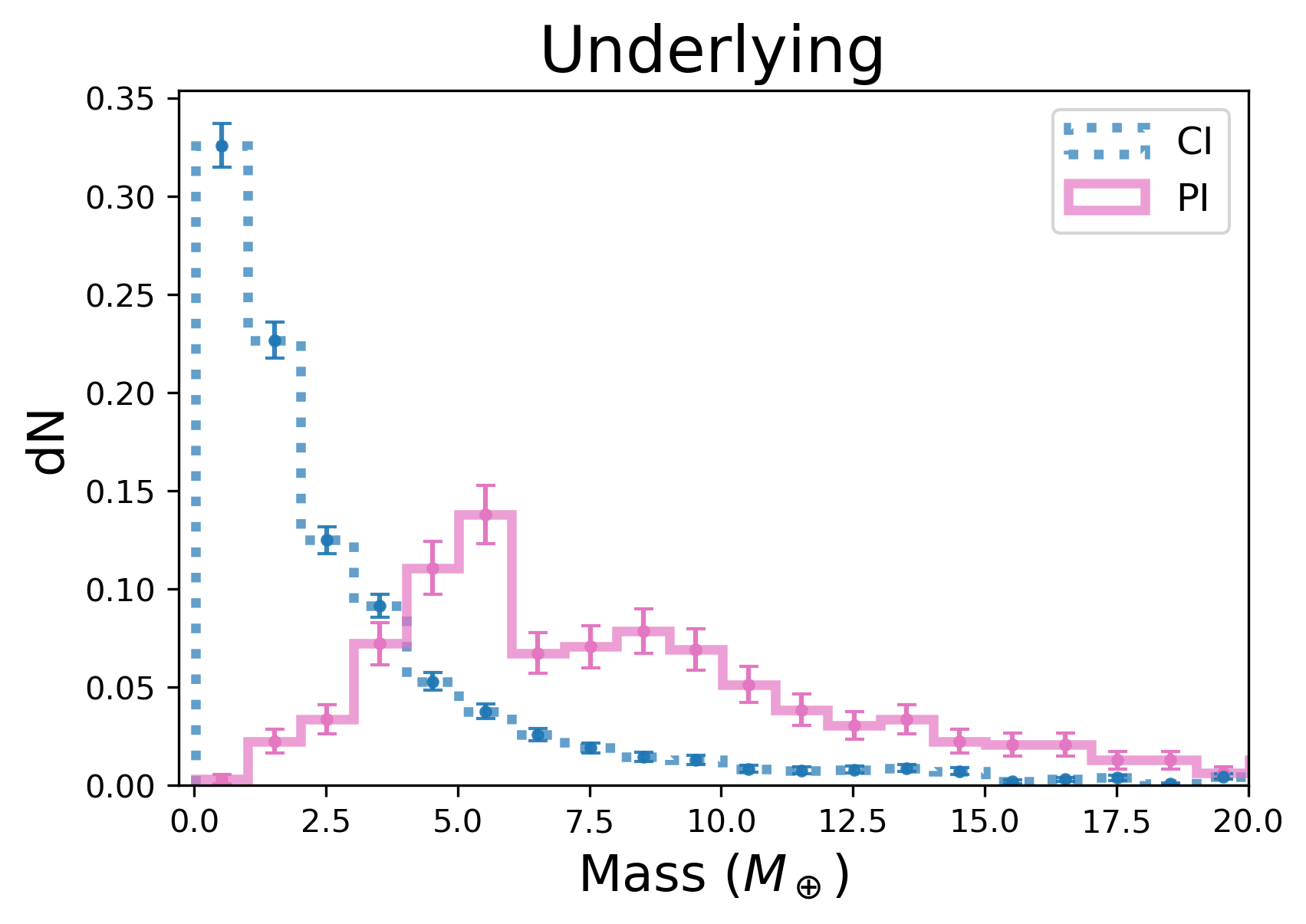}
    \includegraphics[width = 3.2in, height = 2.1in]{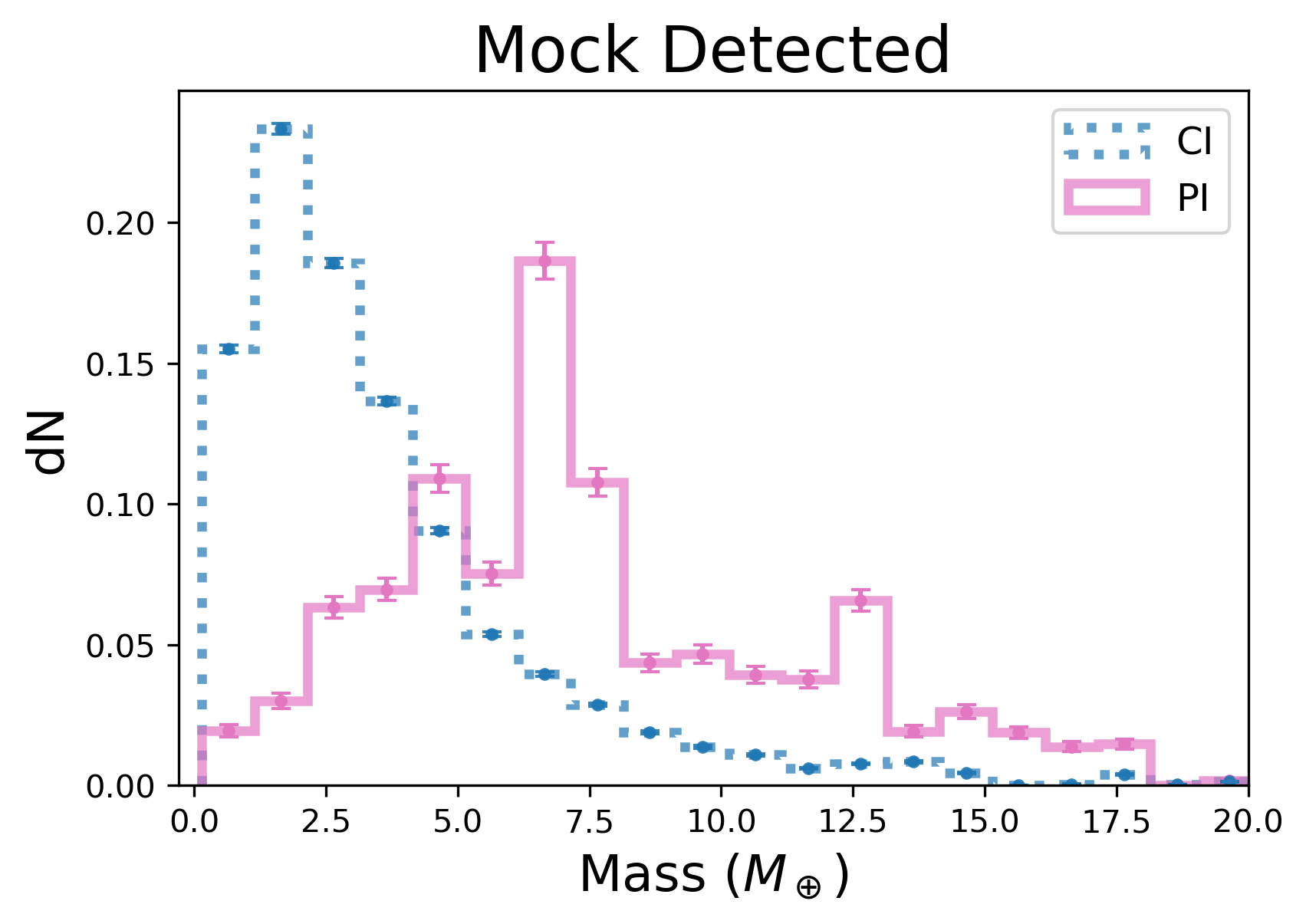}
    \includegraphics[width = 3.2in, height = 2.1in]{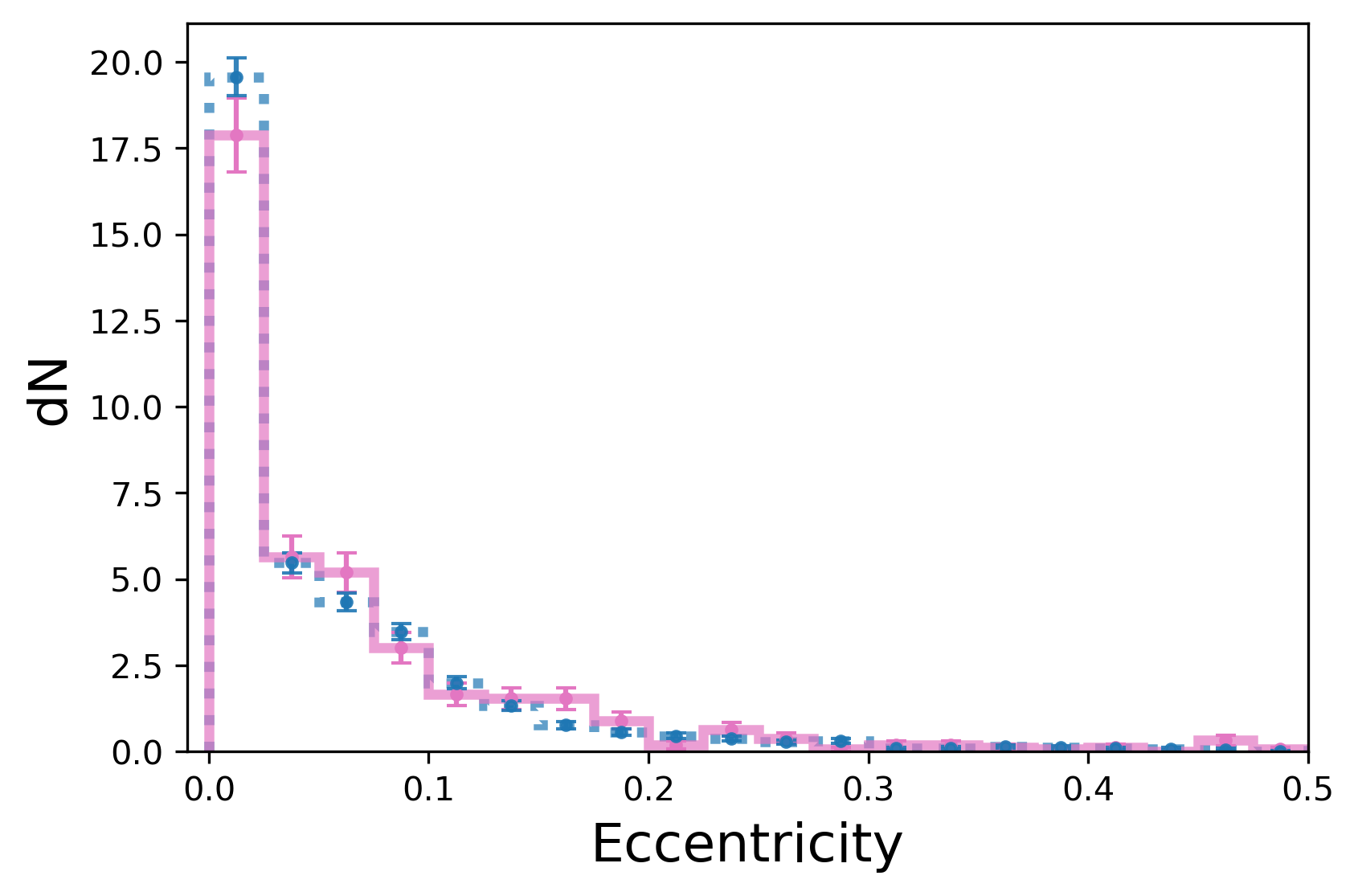}
    \includegraphics[width = 3.2in, height = 2.1in]{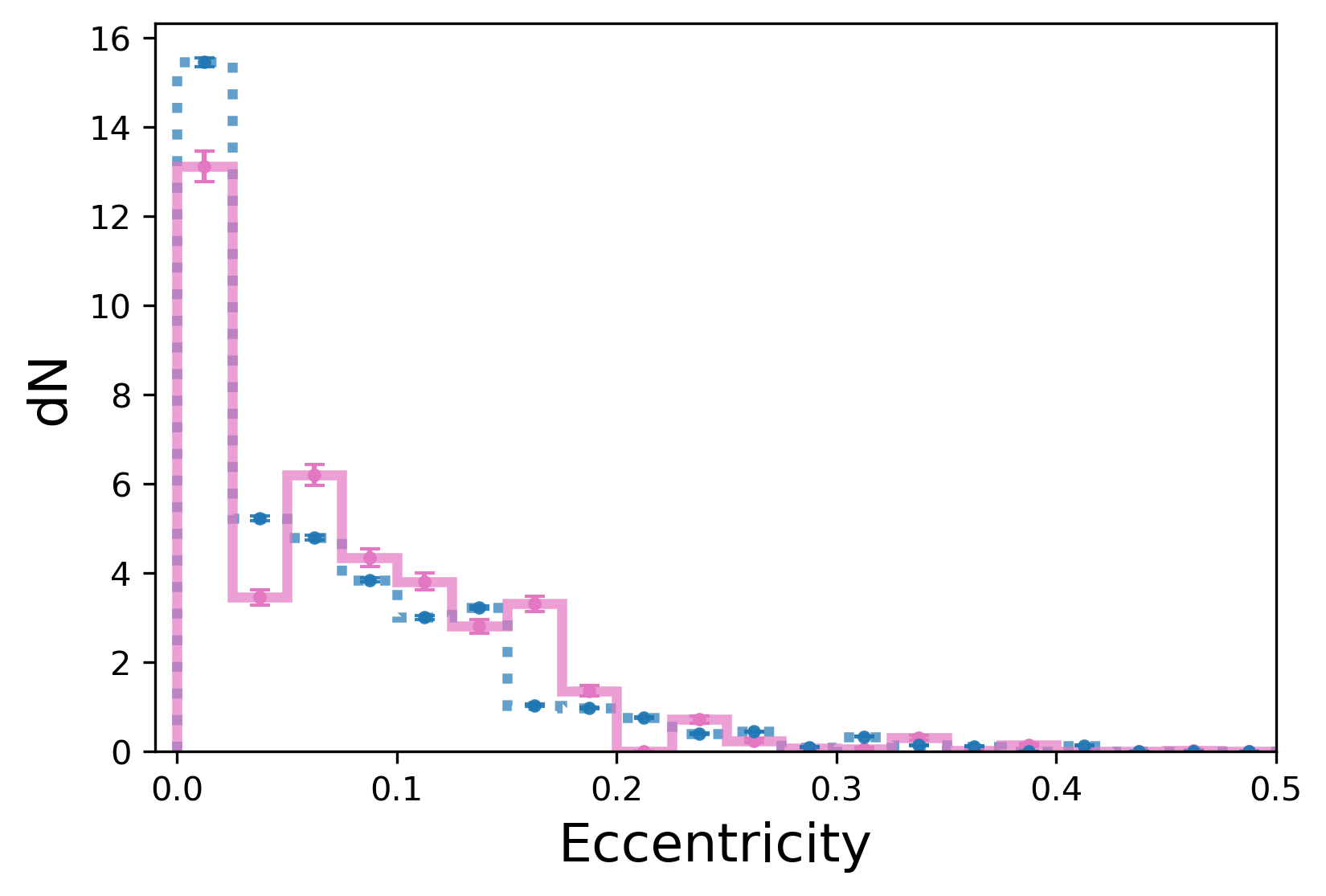}
    \includegraphics[width = 3.2in, height = 2.1in]{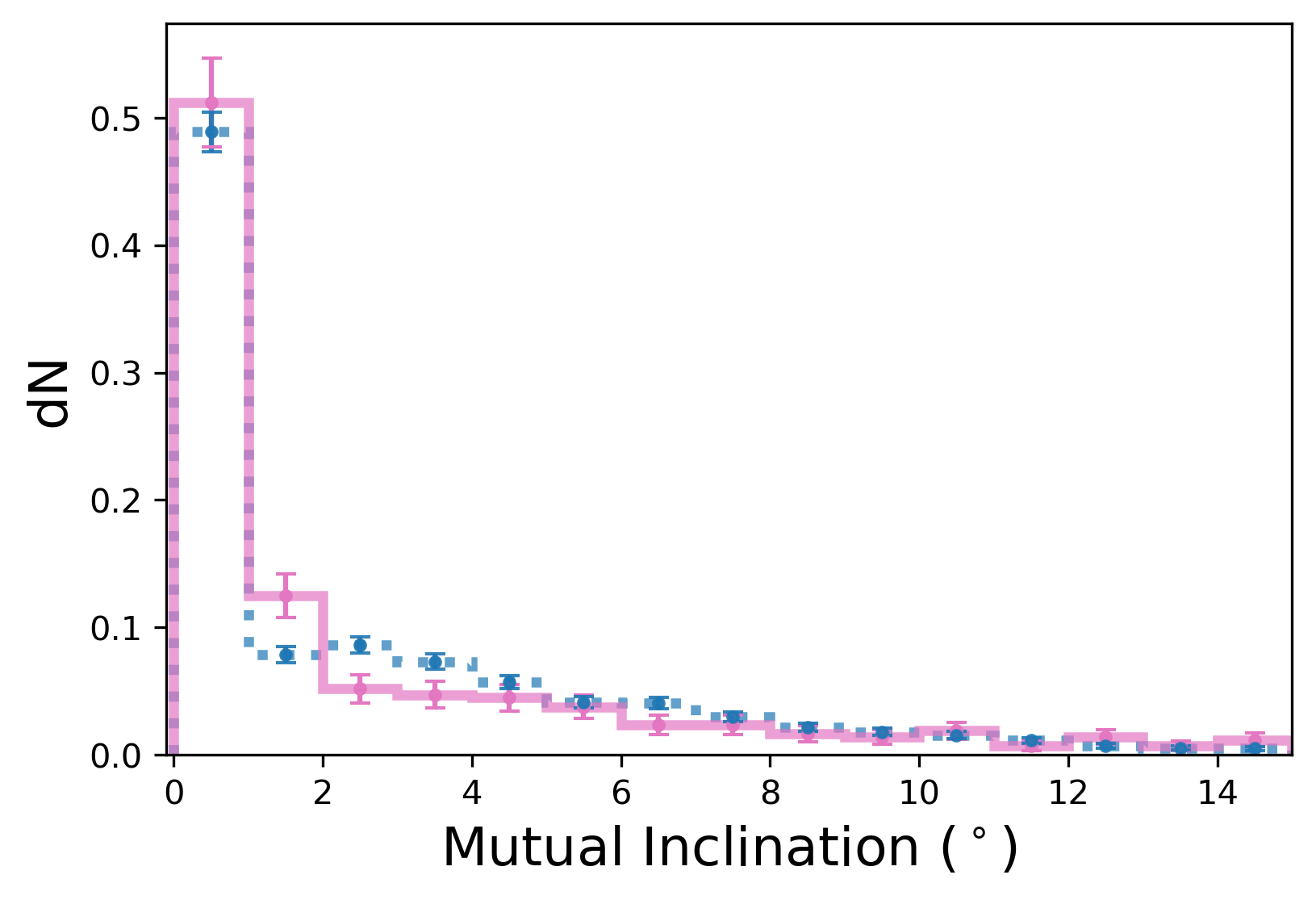}
    \includegraphics[width = 3.2in, height = 2.1in]{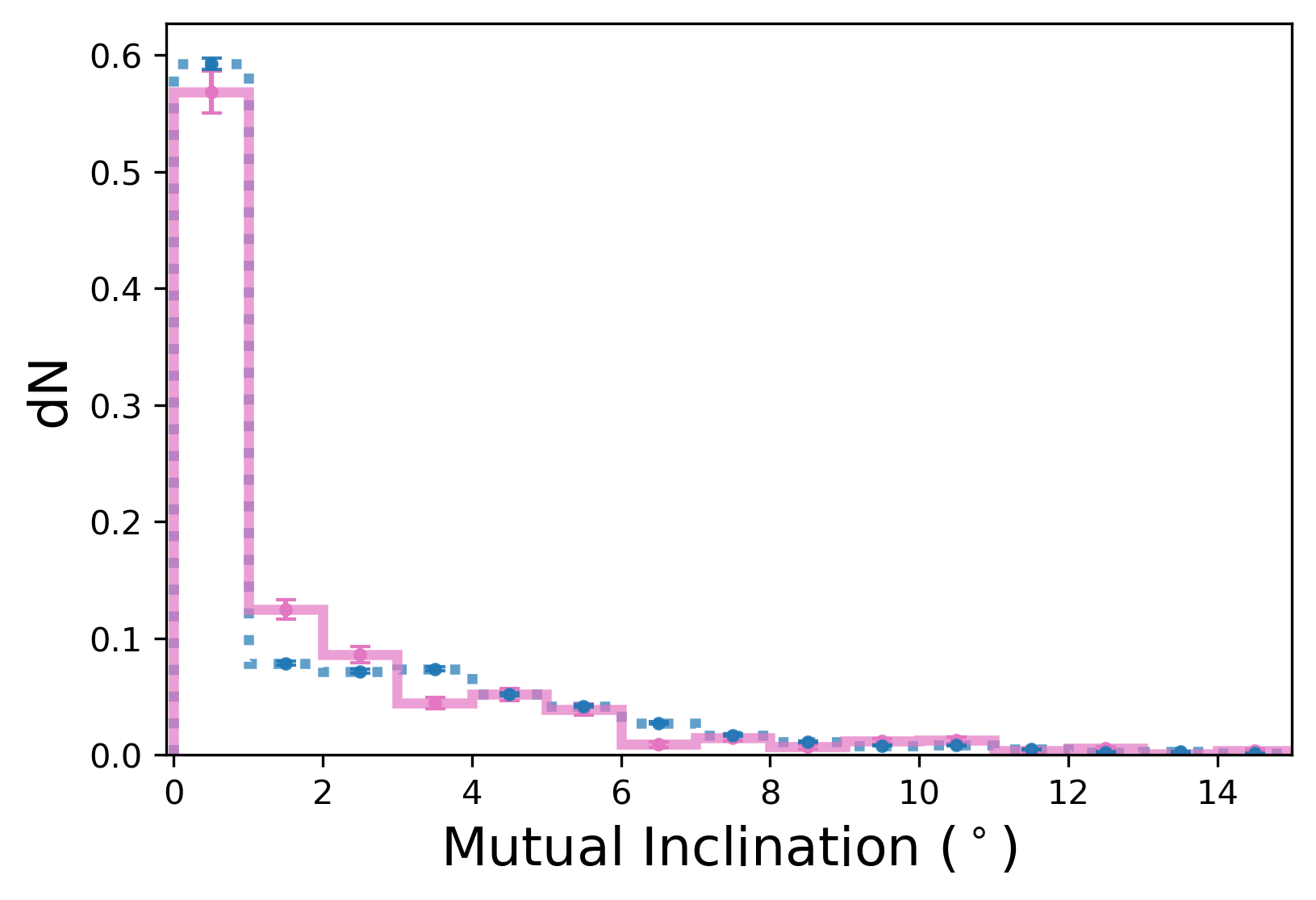}

    \caption{Distributions of mass (\emph{top}), eccentricity (\emph{middle}), and adjacent planet mutual inclination (\emph{bottom}) for both the underlying (left) and the mock observed planets (right) of the PI (pink, solid) and CI (blue, dashed) models.
    The PI model produces many high mass planets, and does not at all match the CI distribution.
    However, PI planets follow a similar $e$ distribution to the CI, in both the underlying and mock observed cases.
    PI planets are also coplanar, with underlying and observed mutual inclination distributions that match CI.}
    \label{fig:lamb_underlying_vs_observed}
\end{figure*}

We show the results of the PI simulations in Figure~\ref{fig:lamb_mass_vs_a}. 
The PI model of embryo formation differs greatly from the CI method, as its planets are detected at very high masses, clustering in the range $5~\mathrm{M_\oplus}< \mathrm{M_p} < 30~~\mathrm{M_\oplus}$. 
However, as discussed in Section~\ref{sec:detect_transit}, $30~\mathrm{M_\oplus}$ is a hard upper limit for planet masses in our simulations. 
In truth, when considering the full range of spacings explored, the PI model produces many planets above this mass limit in the range $6.99 < \Delta < 12$.
Runaway gas accretion is required to create planets this massive, and thus these bodies are not physically possible under the PI model.
The distribution of mass is mostly independent of semi-major axis, except at the smallest semi-major axes $0.04-0.08$ au, unlike the other models we explore. 
The top left panel of Figure~\ref{fig:lamb_underlying_vs_observed} shows that the underlying mass distribution of the PI model peaks at $5.5 ~\mathrm{M_\oplus}$, with very few planets with lower masses and virtually none at an Earth mass or below.
The PI model exhibits a significantly higher fraction of planets than CI with $> 4.0~ \mathrm{M_\oplus}$, only matching at $\sim 20 \mathrm{M_\oplus}$. 
Once detected, the mass discrepancies are only exacerbated (top right panel), with the peak of PI masses shifting to $7.0 ~\mathrm{M_\oplus}$ and masses greater than this being detected more frequently than in the CI systems.
This overabundance of massive planets is likely due to the range of initial mutual Hill spacings chosen for this suite of simulations, as Figure~\ref{fig:hillrange} shows that the total mass of these PI systems (for a range of $4 < \Delta_0 < 12$) is between $40-100~\mathrm{M_\oplus}$, all contained within 1 au. 
Since initial spacings greater than $\Delta_0 = 12$ produce systems which do not undergo giant impacts, increasing spacing further is ineffective in lowering overall embryo mass.

In contrast to the mass distribution, the underlying eccentricity distribution shown in the middle left panel of Figure~\ref{fig:lamb_underlying_vs_observed} is nearly an identical match to the distribution exhibited by the CI.
Each model produces many planets on nearly circular orbits, with the CI model producing slightly more, and both diminish in frequency dramatically beyond $e> 0.025$, exhibiting a small but nonzero number of planets with non-zero eccentricities. 
Beyond $e = 0.2$, both the PI and CI models produce virtually no planets.
\change{\citet{Sandhaus2025} found that allowing up to 10\% of CI systems to be dynamically sculpted by exterior giants led to matching this longer tail of the eccentricity distribution without altering the other observable properties of the population as a whole. 
The PI suite may thus benefit from a similar study.}
When applying the mock transit detection (middle right panel) we see much of the same effect as in the underlying distribution: the PI results in mostly circular orbits, but not as many as CI, and beyond $e = 0.2$ very few planets are detected. 
When comparing underlying mutual inclinations (bottom left panel), the distributions are qualitatively identical, preferentially producing coplanar planet pairs and infrequently producing those with mutual inclinations $>0.5^\circ$.
The mock detected mutual inclination distributions (bottom right panel) remain effectively unchanged from the underlying distribution.

\change{The best fit reweighting parameters we find for the PI model are $\mu^- = 2.13 $ and $\sigma^- = 2.77$. 
Like with MFI2, this suggests that while reweighting is minimal, additional parameters must be varied to fully match the \emph{Kepler} catalog.}

\subsection{Observables} \label{subsec:pebbleiso_obs}

\begin{figure*}
    \centering
    \includegraphics[width = 3.2in, height = 2.1in]{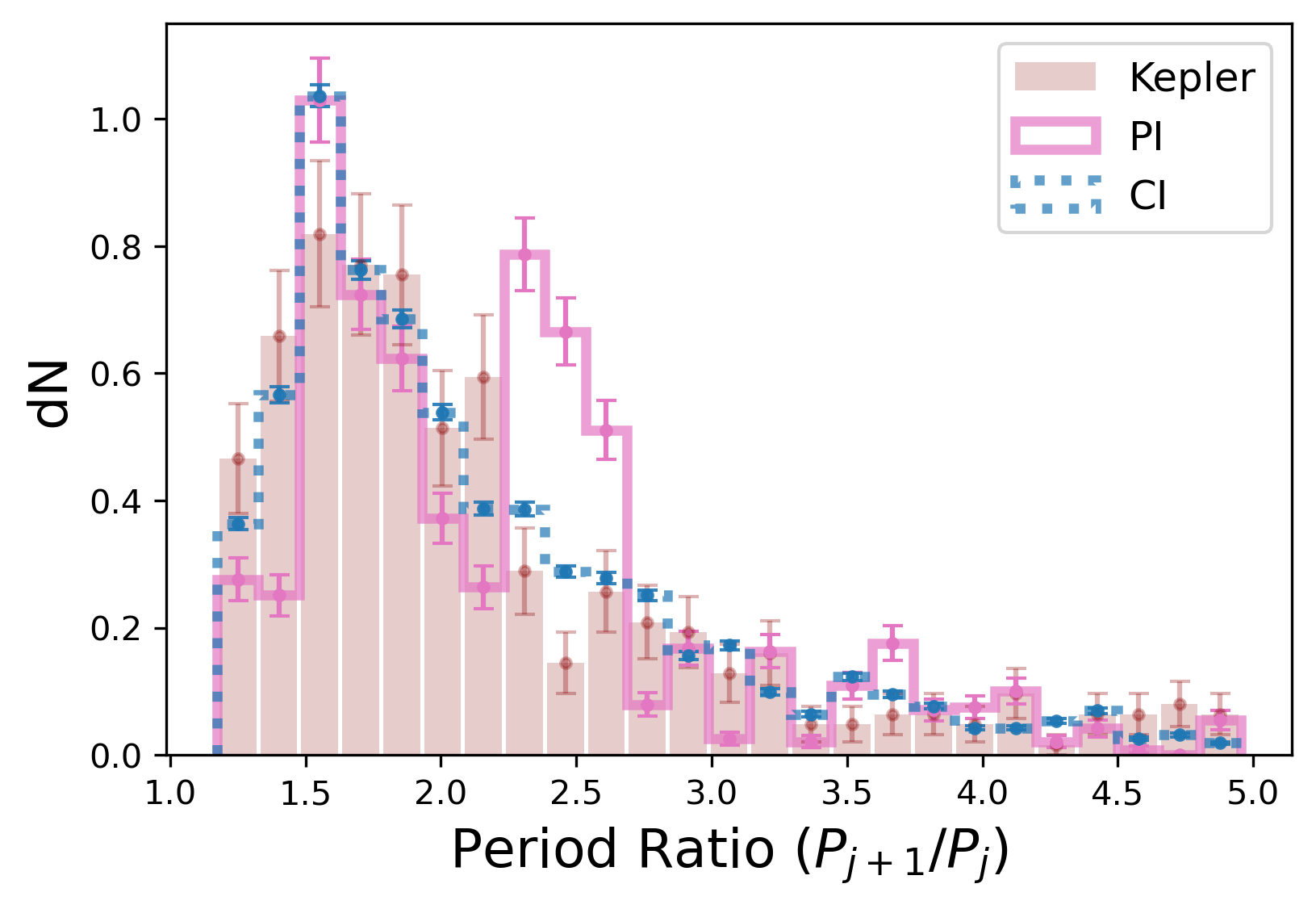}
    \includegraphics[width = 3.2in, height = 2.1in]{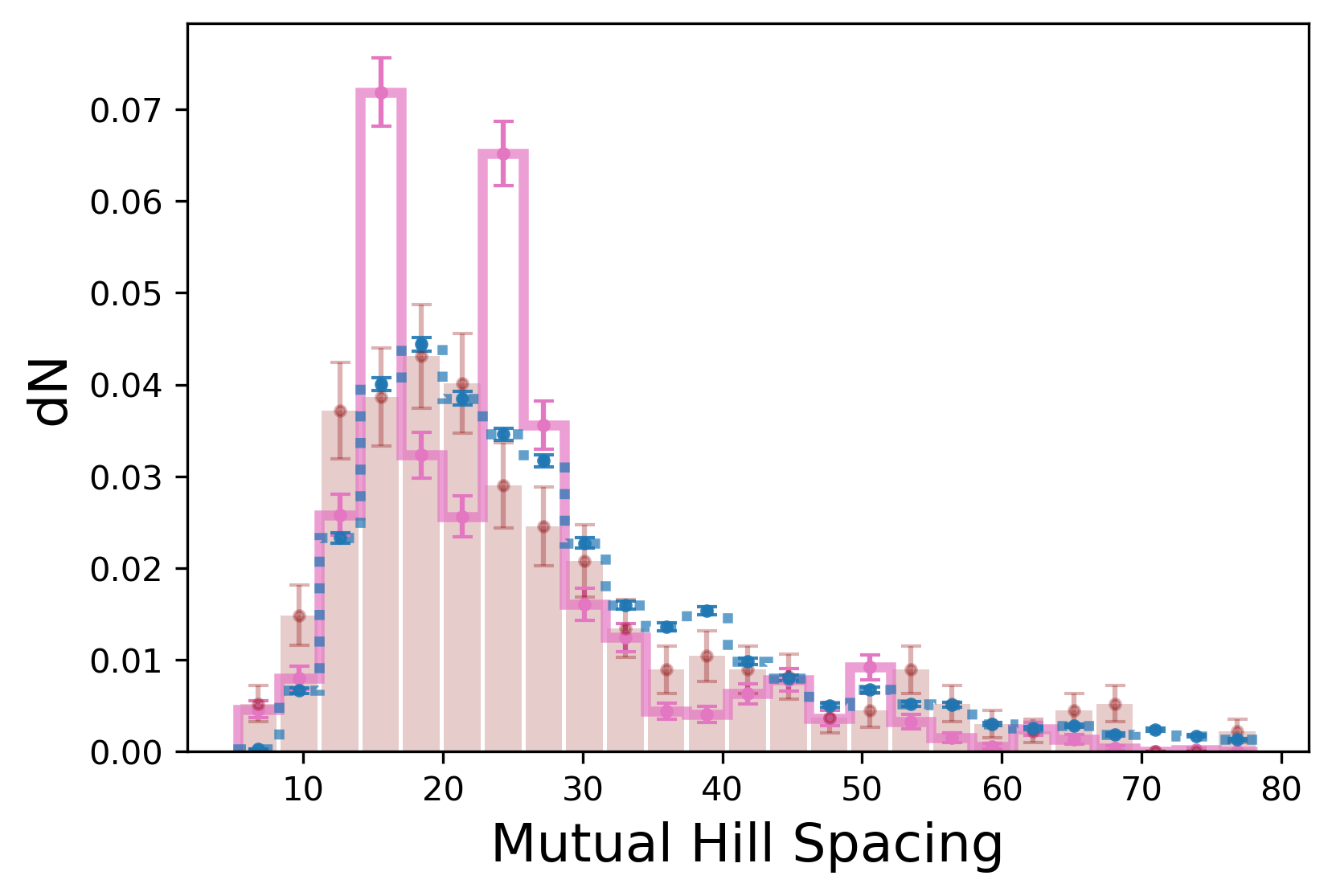}
    \includegraphics[width = 3.2in, height = 2.1in]{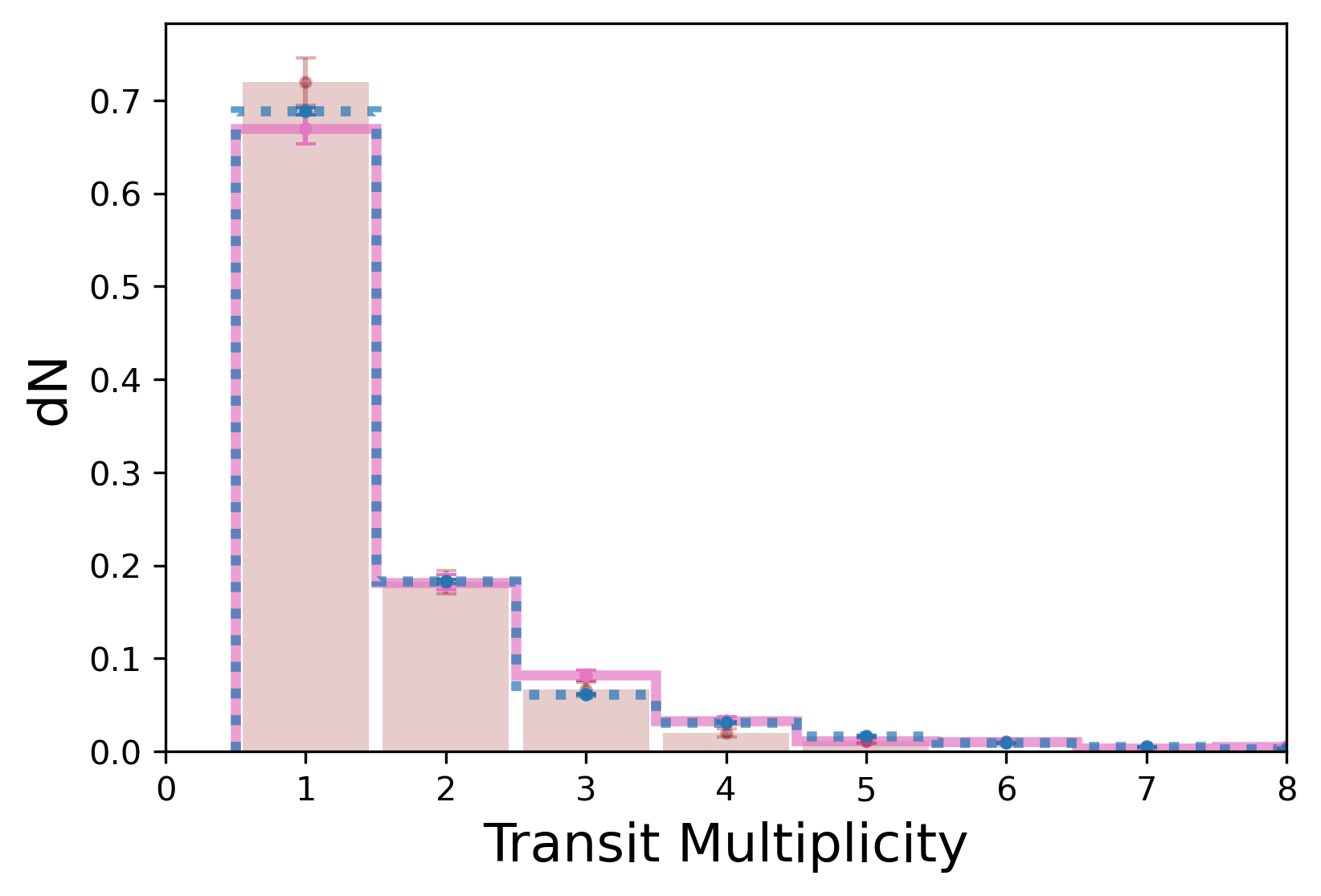}
    \includegraphics[width = 3.2in, height = 2.1in]{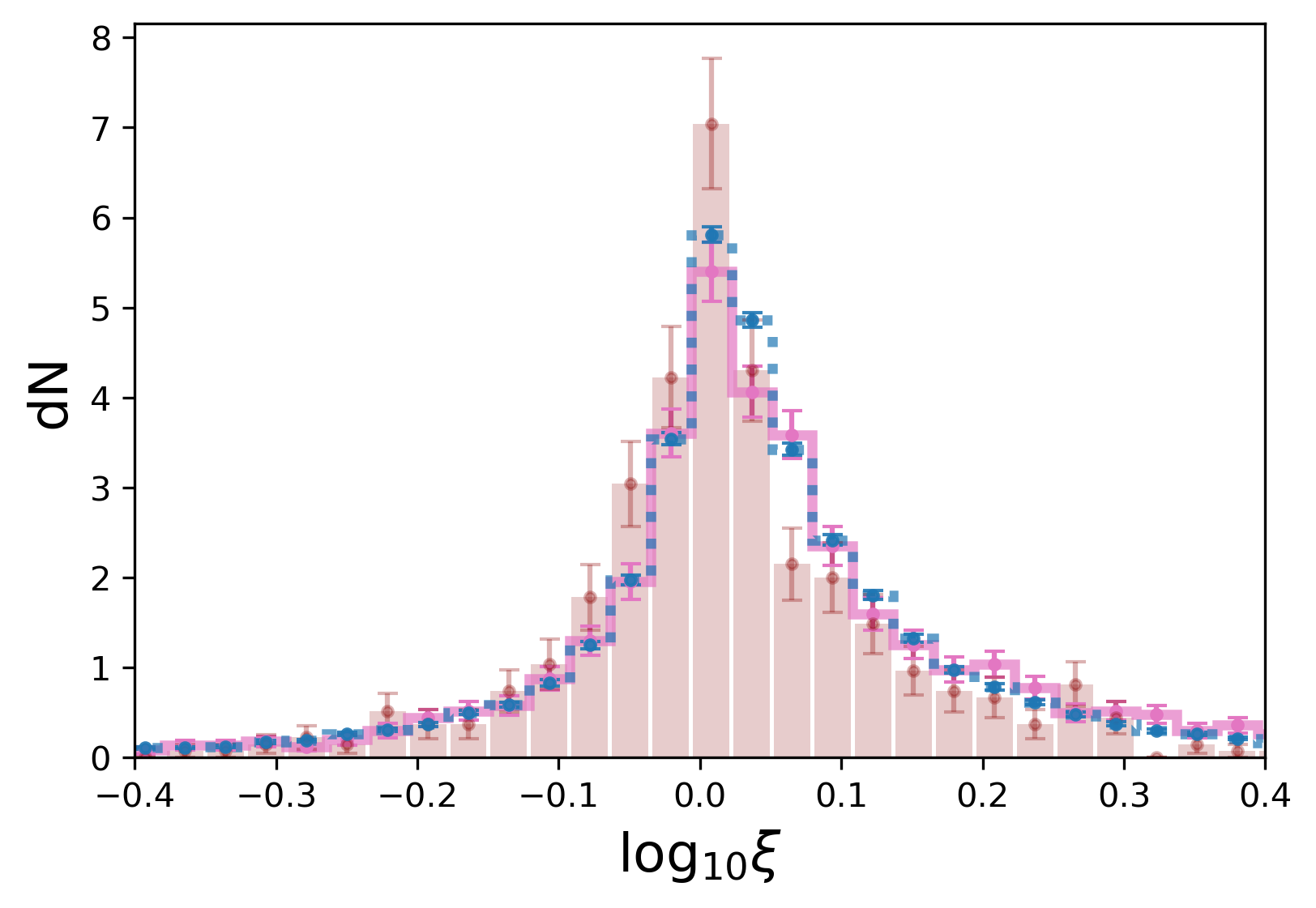}
    \caption{From upper left to lower right: Distributions of period ratios of adjacent planets, mutual Hill spacings of adjacent planets, transit multiplicities, and transit duration ratios ($\xi$) of adjacent planets from the PI mass model (pink, solid), compared to the CI mass (blue, dotted) and the \emph{Kepler} catalog distribution (solid pink).
    The PI model shows distinct bimodal period ratio and mutual Hill spacing distributions that match the \emph{Kepler} catalog peaks with the first of its two peaks.
    The transit multiplicity of the PI distribution, within error, matches the \emph{Kepler} catalog.
    The PI transit duration ratio distribution remains qualitatively similar to the CI distribution.
    Thus, PI planets have incorrectly spaced orbits, but an adequate distribution of multiplicity and coplanarity.}
    \label{fig:lamb_observables}
\end{figure*}

Figure~\ref{fig:lamb_observables} shows distributions of the period ratio (top left), mutual Hill spacing (top right), transit multiplicity (bottom left), and transit duration ratio (bottom right) for the PI simulation suite, after mock detection. 
Of those planet pairs that are detected, their period ratio distribution (top right) is visibly different from that of the \emph{Kepler} catalog.
We see bimodal peaks in period ratio at $\frac{P_{j+1}}{P_j} \sim 1.6$, which aligns with the peaks of CI as well as the \emph{Kepler} catalog, but also at $\frac{P_{j+1}}{P_j} \sim 2.4$, where both the CI and \emph{Kepler} catalog exhibit a deficit of planet pairs.

The PI mutual Hill spacing distribution (top right) is roughly bimodal, with two peaks at $\Delta\sim 16$ and $\Delta\sim25$ and a deficit of systems between.
The rest of the distribution is similar to that of the \emph{Kepler} catalog, but \change{it overall provides a poor match.}

Regarding transit multiplicity, however, we see a much closer match (bottom left).
The PI model produces nearly as many single planet systems as the \emph{Kepler} catalog, consistent within $1\sigma$ for these two distributions. 
Despite the slight enhancement in 3--4 planet systems when compared to the \emph{Kepler} catalog, the distribution of PI transit multiplicity qualitatively matches.

The transit duration ratio distribution (bottom right) matches the peak at $\log_{10}\xi = 0$ from observations, but detects fewer negative ratios and more positive ratios.
This shift toward positive values of $\log_{10}\xi$ and small wings around the peak imply coplanarity and near-zero eccentricities are common in these systems, respectively, which is also seen in the middle and bottom right panels of Figure~\ref{fig:lamb_underlying_vs_observed}.

\subsection{Pebble Isolation EMD} \label{sec:lamb_EMD}

\begin{figure*}
    \centering
    \includegraphics[width = 3.0in, height = 2.0in]{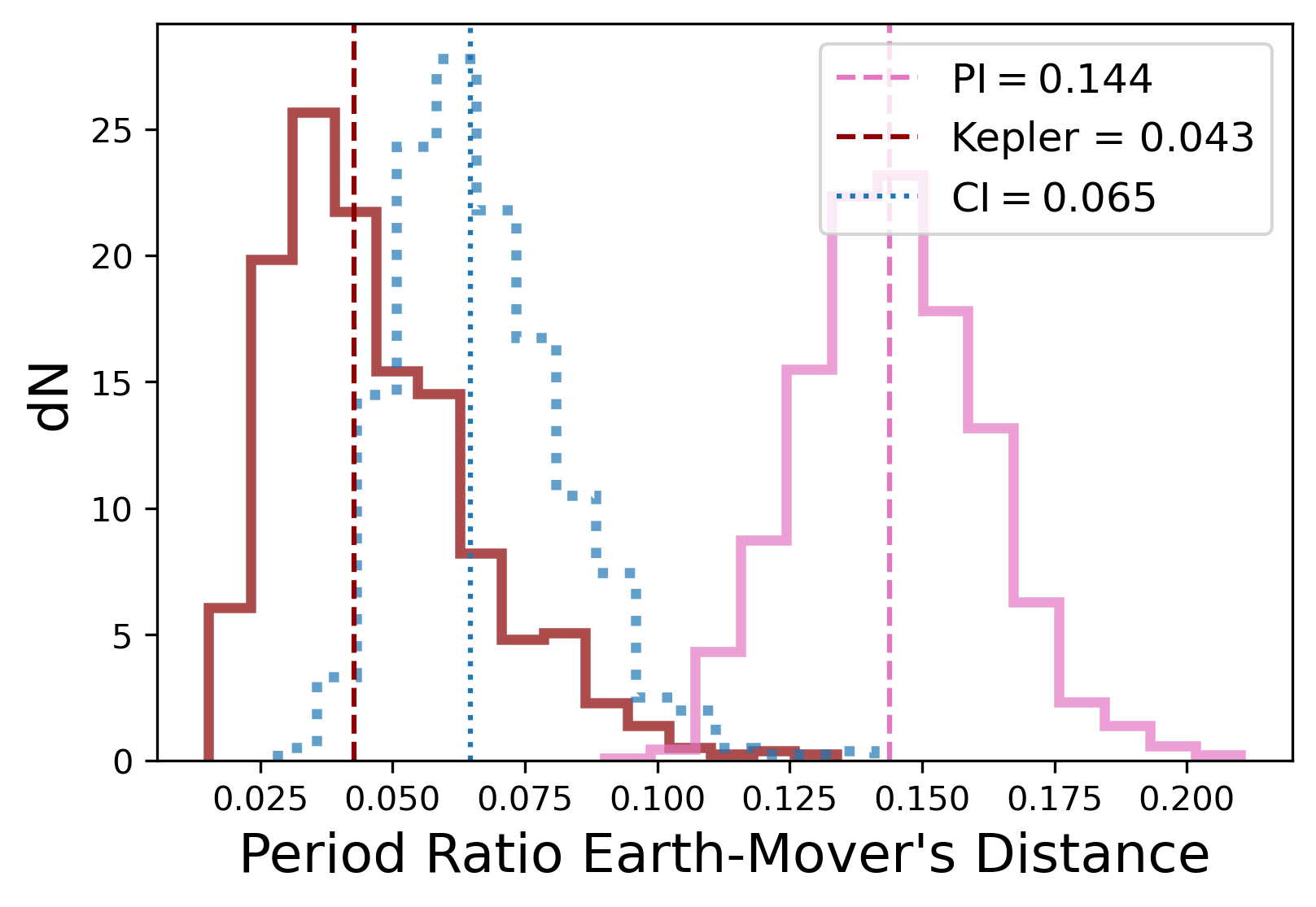}
    \includegraphics[width = 3.0in, height = 2.0in]{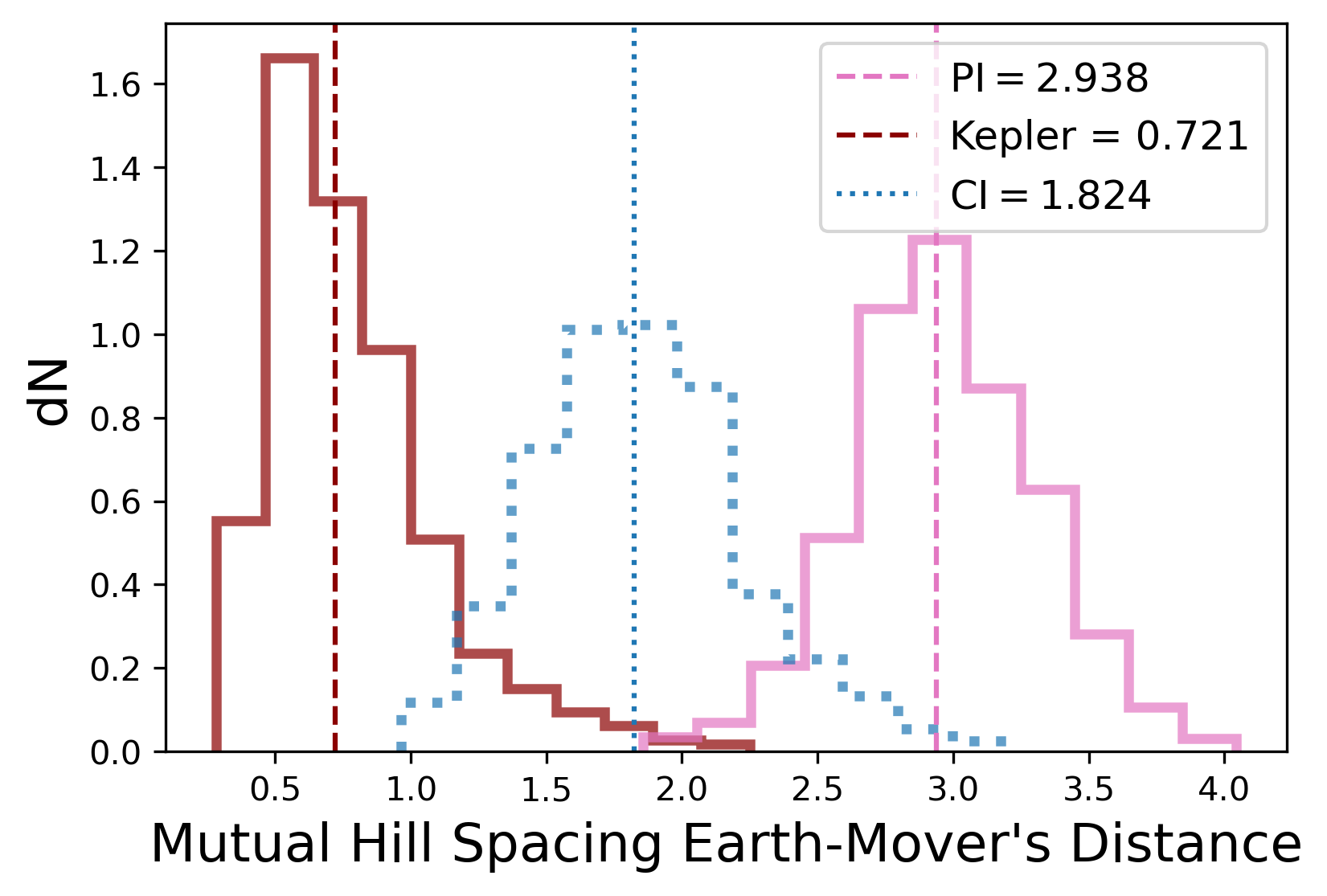}
    \includegraphics[width = 3.0in, height = 2.0in]{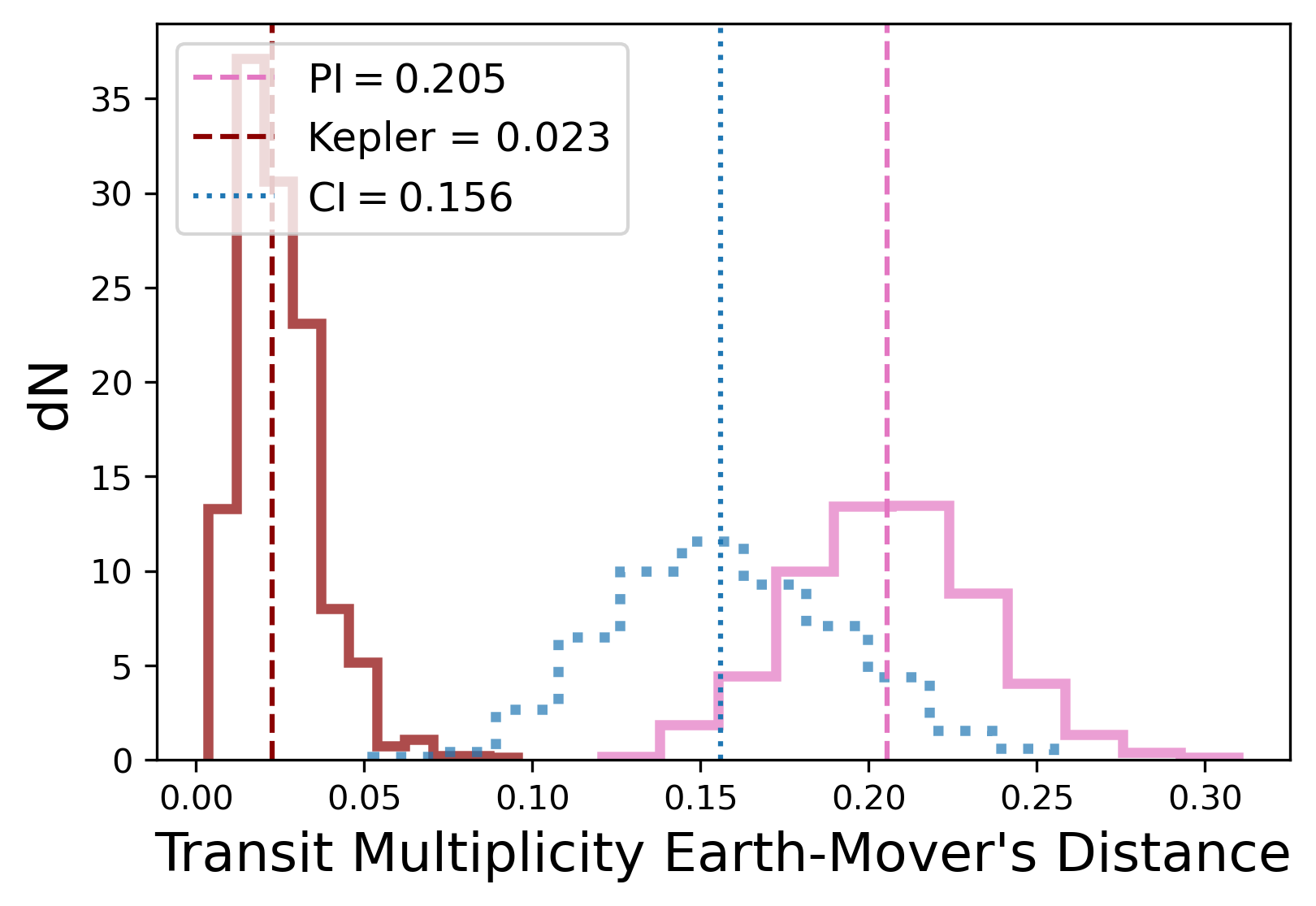}
    \includegraphics[width = 3.0in, height = 2.0in]{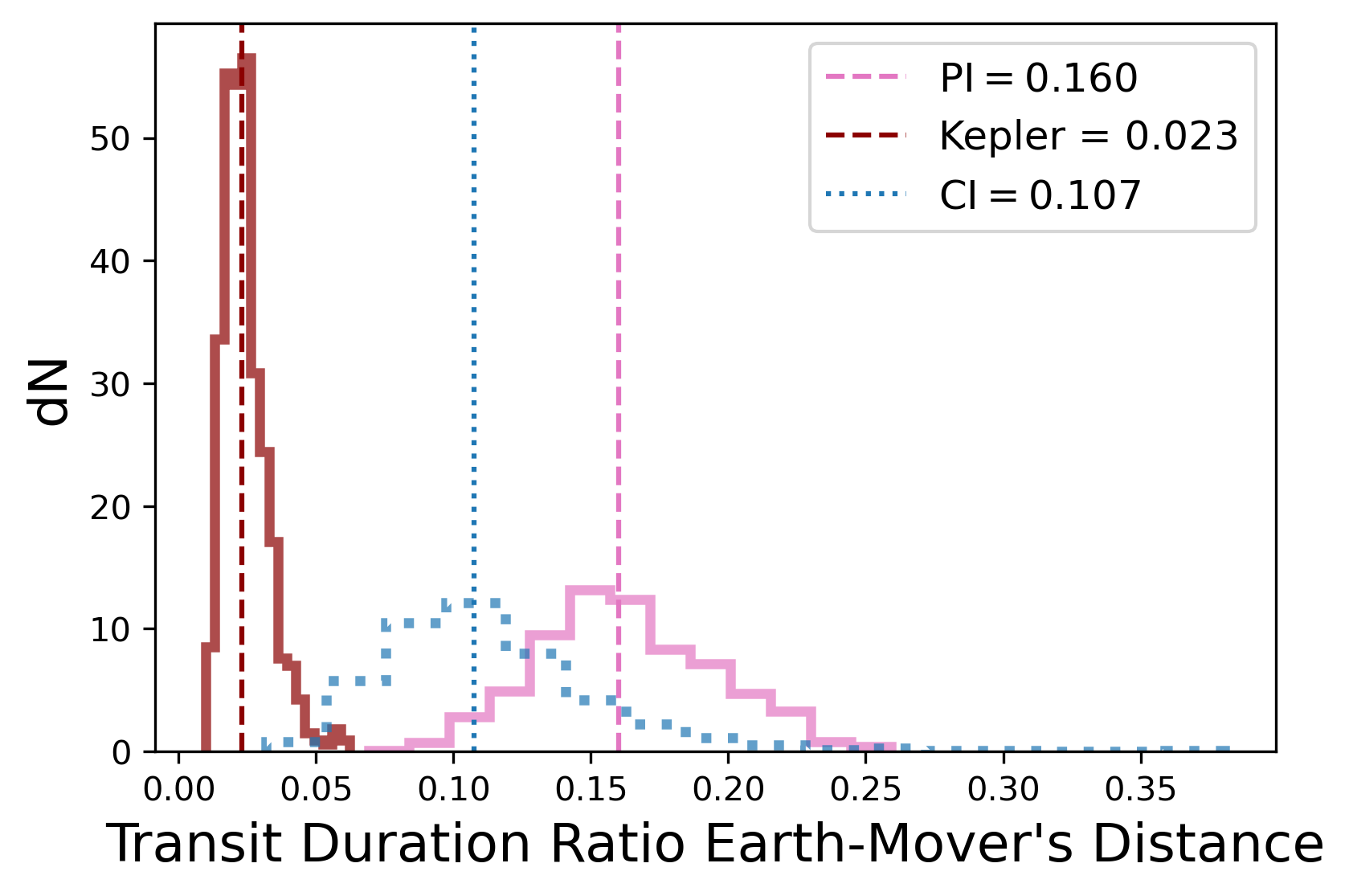}
    \caption{Earth Mover's distance distribution for the PI model (pink, solid) against the CI model (blue, dotted) and \emph{Kepler} EMD uncertainties (dark red, solid). 
    In all cases, the PI mass distributions peak at larger distances than the CI model, with some overlap between the two, as well as between PI and \emph{Kepler}.
    \changethree{However, CI performs better in every scenario, with lower overall EMDs and greater overlap with \emph{Kepler}.}
    In general, this leads us to the conclusion that the PI model presents a poorer fit to the \emph{Kepler} catalog data.}
    \label{fig:lamb_emd}
\end{figure*}

We use Figure~\ref{fig:lamb_emd} to quantify the quality of the match between the PI observables and the \emph{Kepler} catalog observables.
In the top left, we see the PI period ratio EMD distribution prefers larger distances relative to CI, with a slight overlap in the \changethree{tails of the PI and \emph{Kepler} distributions}.
Mutual Hill spacing is much the same for PI, with the EMD distribution peaking at larger distances than CI with \changethree{marginal PI-\emph{Kepler} agreement.}
\changethree{This trend is broken with transit multiplicity (bottom left): the PI distribution overlaps considerably with the CI distribution, but not at all with \emph{Kepler}.}
\changethree{The PI and \emph{Kepler} transit duration ratio EMD distributions once again do not match.}
The PI model requires much more fine-tuning to provide a similarly sufficient fit to the \emph{Kepler} catalog data as CI, and even then we are unable to capture many observable aspects, regardless of reweighting. \changethree{We provide a summary of the EMD statistics in Table \ref{tab:summary_stats_table}.}


\subsection{Implications of Formation by Pebble Isolation} \label{subsec:pebbleiso_implications}

The embryos formed from the PI mass are more massive than the CI planets, with similarly coplanar and circular orbits (see Figures~\ref{fig:lamb_mass_vs_a} and~\ref{fig:lamb_underlying_vs_observed}).
In comparison to the \emph{Kepler} catalog (Figure~\ref{fig:lamb_observables}), PI systems show a good mix of dynamically hot and cold configurations: they exhibit a transit duration ratio distribution skewed positive, and tight mutual Hill spacings, expected from high mass systems with high multiplicity.
However, the fraction of single-planet systems falls within $1\sigma$ of the \emph{Kepler} catalog, providing a good match.
In each of the other observables, \change{we are unable to qualitatively match the \emph{Kepler} catalog distributions with the PI model}. 
In addition to confounding factors such as giant planets and subsequent system-wide evolution (discussed in further detail in Section~\ref{sec:flowiso_implications}), the most likely factor in this inability to match the \emph{Kepler} catalog is an under-explored range of initial spacings. 
Figure~\ref{fig:hillrange} shows that the total mass of systems formed by the PI mass is quite large, even for wide initial spacings.
The range explored in this work ($6.99 < \Delta<12$) encompasses a total mass range of only $30-50~\mathrm{M_\oplus}$, well above the \citetalias{MacDonald2020} threshold for dynamically cold systems and also missing many of the most massive systems.
Like with the MFI1 model, initial spacings cannot exceed $\Delta_0 = 12$, as beyond this point embryos do not undergo late stage collisions. 
At any spacing closer than $\Delta_0=6.99$, the embryos collide rapidly and form giant planets interior to $1$ au, which is outside the scope of this work.
\citet{LambrechtsJohansen2014} suggest that this isolation mass is difficult to apply to planets in the inner disk, as the embryos begin too massive. 
The PI mass, when varying initial embryo spacing in a limited range, produces planets that are too massive to properly match \emph{Kepler} catalog period ratio and mutual Hill spacing distributions, and is thus not totally representative of the true observed system diversity.

\section{Conclusion and Future Work}
\label{sec:conclusion}

\subsection{\change{Summary}}

\begin{table*}
    \centering \ra{1.3}
    \begin{tabular}{*8l}    \toprule
    Observable Metric & \emph{Kepler} & CI & FI1 & FI2 & MFI1 & MFI2 & PI \\\midrule
        Period Ratio EMD & 0.043 & 0.065 & 0.171 & 0.162 & 0.123 & 0.142 & 0.144 \\
        Mutual Hill Spacing EMD & 0.706 & 1.815 & 3.077 & 3.974 & 2.003 & 8.073 & 2.938\\
        Transit Multiplicity EMD & 0.024 & 0.158 & 0.264 & 0.305 & 0.299 & 0.057 & 0.205\\
        Transit Duration Ratio EMD & 0.023 & 0.105 & 0.096 & 0.094 & 0.088 & 0.088 & 0.160\\\bottomrule
     \hline
    \end{tabular}
    \caption{Median Earth Mover's Distance for observable metrics between simulation suites and \emph{Kepler} catalog. The difference between the median EMD for a suite and the median EMD for the Kepler catalog distribution roughly shows how close the distributions are.}
    \label{tab:summary_stats_table}
\end{table*}

\change{Previous work \citep{Dawson2015, Dawson2016, MacDonald2020} modeled the giant impact stage of planet formation, forming their planetary embryos in situ and comparing them to observations.
Combined, they were able to identify the least complex model parameters that led to systems which matched observations, finding that initializing embryos with $\Delta_0=3$ and mild eccentricities and inclinations led to systems that were consistent with starting simulations prior to the giant impact phase.
Additionally, these works incorporated gas damping, exploring various degrees of gas depletion in the disk and finding that assuming \change{that the gas disk was partially depleted at the beginning of the simulated giant impact stage led} to better matches between the final systems and observations.
They also found that assuming a slope of $\alpha=1.5$ for the solid surface density distribution resulted in much more consistent systems than other slope assumptions.
Last, they found that relaxing the assumption of a fixed, underlying distribution of initial $\Sigma_0$ and assuming a continuum of $\Sigma_0$ ultimately matched observables better, including the so-called ``Kepler dichotomy,'' the radius valley, intra-system similarities, the scatter in the mass-radius relationship, and the resulting number and type of mean motion resonances.
More recently, \citet{Sandhaus2025} found that the inclusion of dynamically active exterior giant companions during the giant impact phase of the inner systems could account for the most dynamically hot systems that \citetalias{MacDonald2020} lacked, ultimately finding that up to 10\% of systems could contain exterior giant companions without disrupting the other observables but accounting for the more eccentric planets $e>0.3$ that previous simulation suites failed to produce.}

\change{While these works were able to sufficiently explain observations, each assumed that the initial embryos had reached their isolation masses by way of classic core accretion.
Here, we relax that assumption and explore three separate pebble accretion isolation mass models, investigating how they affect the final systems they form.
We simulate the giant impact phase of planetary formation, adopting the initial conditions found by \citetalias{MacDonald2020} to form systems that best match observations.}

We explored the effects of a continuum of initial formation conditions on the final properties of mock observed systems, dependent on the three different pebble accretion isolation mass models. 
We began our simulations within a gas disk structured like the MMSN, depleted by a factor of 100. 
We assumed the embryos in each simulation formed self-consistently within an undepleted MMSN disk prior to the beginning of the simulation, reaching the maximum isolation mass for that location in the disk given the chosen initial conditions.
To simulate this diffuse gas stage, we applied the gas damping force to these embryos for the first 1~Myr of simulation.
After 1~Myr had passed, the damping force was turned off, representing the complete dissolution of the gas disk.
The embryos then undergo unhindered collisional evolution for 29 Myr.

Finally, we simulated \emph{Kepler} observing the final systems \change{to create a mock catalog of ``observed'' systems.}


\subsection{\change{Comparing Pebble Accretion Models}}

For the flow isolation (FI) mass \change{models}, we varied maximum Stokes number for accreted pebbles in the inner protoplanetary disk given two different rates of solid mass accretion onto the host star.

In agreement with \citetalias{MacDonald2020}, we found that the more solid mass is present in the inner disk, the more likely a final system will end up dynamically cold: with high multiplicity, small spacings, and mostly coplanar orbits.
The more massive an embryo, the weaker the gas damping force it will feel, and as a result the more massive systems undergo a majority of their giant impacts before the dissipation of the gas disk.
Since most planets in these systems are formed \change{in the presence of gas}, the excitation of their eccentricities and inclinations by collisions is damped, leading to tightly spaced, coplanar systems on nearly circular orbits.
Dynamically hot systems occur when embryos remain relatively isolated from one another until disk dissipation.
Once the disk dissipates and the embryos undergo the majority of their giant impacts, they do so unhindered by gas damping, producing systems with large eccentricities and mutual inclinations, and thus fewer transiting planets.

\change{Without prior constraints on} the true underlying distribution of maximum Stokes numbers in protoplanetary disks, 
we reweight the simulations based on their initial $St$ for the FI1 and FI2 models so as to amplify the number of low mass, dynamically hot systems.
However, \change{even this reweighting} was not sufficient to reduce the \change{occurrence rate} of dynamically cold systems relative to dynamically hot systems \change{nor to more closely match the distribution of planet mass}.

\change{When comparing the underlying output of FI1 and FI2 directly to CI, both models produce a planetary mass distribution that is nearly identical to CI, with FI2 showing just under 10\% more planets in the lowest mass bin $0<M_p<1.0~M_\oplus$. 
FI2 shows slightly more eccentricities at $e\sim0.05$ but otherwise has an identical tail.
It also exhibits a larger fraction of mutual inclinations in the range $1^\circ<i_{mut}<4^\circ$ relative to both FI1 and CI.}

\change{Post mock-observation, FI1, FI2 and CI have qualitatively identical mass distributions.
In eccentricity, however, the fraction of nearly circular eccentricities in both FI suites decreases, with a slight enhancement at slightly larger values.
Beyond $e=0.1$, the FI distribution tails fall off much more rapidly than CI.
The fraction of nonzero mutual inclinations increases for both FI models once mock observed, whereas the opposite is true for CI.}

We found that for the low solid accretion rate case ($\dot{M}=10^{-9}~\mathrm{M_\odot~yr^{-1}}$, FI1), our continuum of Stokes numbers produced distributions of period ratio, mutual Hill spacing, and transit duration ratio that qualitatively match the \emph{Kepler} catalog distributions of the same parameters.
However, for this range of $St$, the model produced a deficit of single planet systems and too many multis (Figure~\ref{fig:flowiso_observables}, bottom left panel).

When we assumed formation occurred with a higher rate of accretion of solid material onto the host star ($\dot{M}=4\times10^{-8}~\mathrm{M_\odot~yr^{-1}}$, FI2), period ratio and transit duration ratio each matched the \emph{Kepler} catalog. 
The systems however had too high of multiplicity relative to the \emph{Kepler} catalog, in addition to overly-wide spacings between the planets (Figure~\ref{fig:flowiso_observables}, bottom left and top right panels respectively).
The maximum Stokes number for FI2 was $St_{max}=10^{-3}$, an order of magnitude lower than FI1's $St_{max}=10^{-2}$.
As a result, the most massive FI2 embryos were a factor of two less massive than the upper limit for FI1, which led to wider spacings between bodies and more bodies in the initial simulation.
In turn, the final FI2 planet spacings are generally wider than those for FI1.


Unlike the FI mass, the migration feedback isolation (MFI) mass distribution does not change as a function of $\Delta_0$. 
Thus, simulation-to-simulation within a given suite, the embryos generally began with similar masses, but varied in orbital phase, spacing, and initial inclination. We explored two separate prescriptions for MFI.
For MFI1, we varied the initial mutual Hill spacing between embryos to explore a range of initial total masses within the inner disk, under the assumption that these embryos formed within a disk with a gas profile commensurate with the MMSN, $\Sigma_{g,1}=1700~\mathrm{g~cm^{-2}}$.
For MFI2, we still varied initial mutual Hill spacing, but we relaxed the assumption of formation in an MMSN-like disk and instead depleted the gas by a factor of 100 ($\Sigma_{g,1}=17~\mathrm{g~cm^{-2}}$). \change{This depletion factor $d$=100 follows the results of \citetalias{MacDonald2020} who found it resulted in systems that most closely matched observations under the CI model.}

\change{MFI1 and MFI2 both show an enhancement in the fraction of planets in the $0<M_p<1.0~M_\oplus$ bin.
However, MFI2, on account of formation in a disk depleted by $d=100$, shows a 20\% relative enhancement in sub-Earth mass planets relatve to MFI1, and a $\sim40\%$ enhancement relative to CI.
MFI1 has a similar distribution to CI, but with a larger fraction at nearly coplanar eccentricities.
MFI2 planets generally exhibit slightly larger eccentricities, although they are mostly less than $e=0.1$.
Last, MFI1 show 10\% more planet pairs at mutual inclinations $i_{mut}<1^\circ$ relative to CI, whereas MFI2 has 40\% fewer, with a roughly 10\% enhancement relative to both MFI1 and CI in mutual inclinations $1^\circ<i_{mut}<5^\circ$.}

\change{Once mock observed, MFI1 shows a qualitative match to CI mass and eccentricity distributions, but 25\% fewer mutual inclinations $<1^\circ$.
MFI2 maintains its dramatic differences with the other two suites, with an exacerbated fraction of planets $<2.5~M_\oplus$ (as well as lacking the tail exhibited by MFI1 and CI beyond $M_p=3.0~M_\oplus$).
Its enhancement in eccentricities $0.05<e<0.085$ is maintained relative to MFI1 and MFI2, but beyond $e=0.1$ it lacks the tail toward the highest eccentricities.
in mutual inclination, very few coplanar MFI2 systems are mock observed. 
Most have $1^\circ<i_{mut}<5^\circ$, consistent with the underlying distribution, but it also shows a larger fraction of pairs with mutual inclinations beyond $5^\circ$ relative to MFI1 and CI.}

\change{For MFI1,} the distributions of period ratio and mutual Hill spacing provided the closest match to the \emph{Kepler} catalog when compared to the other pebble accretion isolation models presented in this work, with a sufficient match to transit duration ratio.
However, MFI1 produced too few single transiting planet systems relative to the \emph{Kepler} catalog (Figure~\ref{fig:migration_observables}, bottom left panel).

The MFI2 simulations resulted in the production of planets with spacings wider than the \emph{Kepler} catalog.  
The MFI2 model produced planet pairs with smaller transit duration ratios relative to the \emph{Kepler} catalog, implying comparatively larger mutual inclinations.


\change{For the pebble isolation mass model (PI), } we explored the impact of varying initial mutual Hill spacing. 
The PI mass generated planets with large masses, many of which were over the upper limit of $30~\mathrm{M_\oplus}$ and were removed from consideration in this work.
The surviving systems, when compared to the other pebble accretion models, fail to produce enough detectable planets $<2~\mathrm{M_\oplus}$, skewing the observable distributions accordingly.
The PI model \change{primarily} produced systems that were dynamically hot, with low transit multiplicity and wide spacings for all values of the initial mutual Hill spacing that we explored.

\change{When directly comparing to CI, the underlying eccentricity and mutual inclination distributions are effectively identical, with these similarities maintained post mock-observation (with the exception of slightly fewer nearly circular orbits relative to CI).
The largest difference comes in the comparison of PI and CI's mass distributions: PI produced planets that are skewed very heavily toward large masses, peaking at $M_p \sim5.0~M_\oplus$ with little to no planets within the bins $<2.5~M_\oplus$. This difference is maintained through mock observation.}


\subsection{\change{Matching Observations with Pebble Accretion}}

 For each model, we compared the observed distributions of transit multiplicity, orbital period ratio, mutual hill spacing, and period-normalized transit duration ratios to a subset of planetary systems discovered by \emph{Kepler}, focusing on systems with super-Earth/sub-Neptune planets with orbits less than 365 days.

\change{\textit{Flow Isolation, FI1}:} The observed distributions under the FI1 model provided an adequate fit to the \emph{Kepler} catalog \change{in terms of} period ratio, mutual Hill spacing, and transit duration ratio \change{distributions}, but failed to match the \change{multiplicity distribution. While it was able to produce}
both dynamically hot and cold planetary systems, it resulted in fewer single planet transiting systems than in the \emph{Kepler} catalog.

\change{\textit{Flow Isolation, FI2}:} The observed distributions of period ratio and transit duration ratio under the FI2 model are similar to \change{those from} the \emph{Kepler} catalog.  
Like FI1, FI2 produced both dynamically hot and cold systems, with a preference for multi-planet systems with moderate mutual inclinations. It produced too few single transiting planet systems relative to the \emph{Kepler} catalog. \change{In addition, the initial} embryos \change{formed by this model} were less massive than those produced by FI1, leading to wider final spacings.

\change{To explore how sensitive these results are to our initial assumed distribution of $St$,} we reweight these simulations based on their initial $St$ so as to amplify the number of low mass, dynamically hot systems.
However, \change{even the most extreme reweighting could not sufficiently balance the relative fractions of dynamically hot and dynamically cold systems}.
Thus, while the FI1 and FI2 models provided mostly adequate matches to the key properties of multi-planet systems in the \emph{Kepler} catalog, they failed to produce the observed abundance of systems with a single detected planet.  

\change{\textit{Migration Feedback Isolation, MFI1}:} The MFI1 model failed to produce enough low multiplicity, dynamically hot systems to match the \emph{Kepler} catalog. 
However, \change{it successfully matched} the \emph{Kepler} catalog in \change{distributions of} period ratio, mutual Hill spacing, and transit duration ratio. 
\change{Ultimately, this model resulted in the best match to these three observables between all five models under our other model assumptions, e.g. gas damping.}

\change{\textit{Migration Feedback Isolation, MFI2}:} The distribution of transit multiplicity \change{produced by} the MFI2 model was similar to that of the \emph{Kepler} catalog.
However, the planets detected under the MFI2 model tended to be smaller, more widely spaced, and moderately mutually inclined, leading to a poor match to the \emph{Kepler} catalog's \change{distributions of} period ratio, mutual Hill spacing, and transit duration ratio.  

In matching the \emph{Kepler} catalog multiplicity distribution whilst producing low mass, high mutual inclination planets, the MFI2 model shows that we should consider several observable properties when comparing theoretical results to \emph{Kepler} data.
Even after reweighting the simulations to upweight systems likely to achieve wider final spacings, the MFI1 model produced preferentially dynamically cold systems and overpredicted the abundance of systems with a single detected planet present in the \emph{Kepler} catalog.
The MFI2 model exhibited the opposite \change{behavior}.
It produced a larger fraction of dynamically hot systems relative to the \emph{Kepler} catalog, failing to capture the  population of nearly coplanar systems with multiple detected planets. 


\change{\textit{Pebble Isolation}:} The PI model produced a reasonable ratio of dynamically hot and cold systems.  
The observed distributions \changethree{of period ratio and mutual Hill spacing showed a marginal match to the \emph{Kepler} catalog in EMD-space, but showed no quantitative match in multiplicity or period-normalized duration ratio.}
\change{Overall we expect that these distributions were skewed towards larger separations} as a result of \change{large} final planet masses. 

PI planet pairs \change{ultimately} exhibited large period ratios and mutual Hill spacings. 
The PI model produced more coplanar systems than the other models under consideration, failing to quantitatively match the \emph{Kepler} catalog in transit duration ratio.
\changethree{PI resulted in more single transiting planets than the other models under consideration, but still failed to quantitatively match in regards to the \emph{Kepler} EMD distribution. }

However, even after reweighting toward tighter initial spacings, we could not fully match the observable distributions from \emph{Kepler}. 
\changethree{The PI model appears to match \emph{Kepler} in transit multiplicity and duration ratio, but in truth struggles to support this quantitatively.}




\subsection{\change{Future Directions}}

While each of the models considered in this study struggled to match the distribution of planetary systems observed by \emph{Kepler}, we can use these results to inform plans for future studies. For example, some models would benefit from additional dynamical excitation of eccentricities and inclinations. 
Previous work has shown that outer giants can have a strong effect on the dynamics and formation of an inner planetary system \citep{Huang2017, Bitsch2023, Sandhaus2025, Livesey2025}.
Other works have also shown that outer giant planets have a moderate occurrence rate in systems which also harbor inner super-Earths \citep{ZhuWu2018, Bryan2019, Rosenthal2022}.
Future research should include giant planets in the outer disk of our models.
In some cases, outer giants \changethree{with distributions of orbital properties that match those seen in radial velocity surveys (see \citealt{Huang2017, Sandhaus2025})} can excite planets in the inner disk, increasing their mutual inclinations and causing them to no longer transit their host star.
In this scenario, the fraction of dynamically hot systems in the FI and MFI models would likely increase, leading to the predicted distribution of planetary system architectures better matching those of the \emph{Kepler} catalog.  

\changethree{However, other configurations of outer giant planets can have different effects on the planets formed. For example, when inner planets form in the presence of the Solar System giants, studies show that the formed planets exhibit smaller eccentricities and mutual inclinations leading to dynamically colder systems \citep[e.g., ][]{ChildsQuintana2019_MNRAS, Sandhaus2025}.  Therefore, when trying to match the diversity of known exoplanets, it is important to consider how a particular configuration of outer giant planets may affect planet formation, while incorporating that configuration's true occurrence rate. As the number of observed systems that host small inner planets and outer giants grows, we can further explore super-Earth occurrence rates as a function of a variety of outer giant planet properties (e.g., eccentricity, inclination, multiplicity, mutual Hill spacing), and then incorporate this information into future simulations.}

In our prescription of the MFI and PI simulations, embryo masses grow assuming that the reservoirs of pebbles per embryo are independent of one another and thus all embryos would be able to reach their isolation masses. In reality, an embryo that reaches its isolation mass will influence the availability of solids for other growing embryos. Therefore, the embryos we model may be more massive than they would be were we to model a variable pebble flow.  We can gain insights into the potential effects of such scenarios by exploring results from previous works. 
One could also model formation in the gas disk by pebble accretion directly, rather than assuming each embryo has self-consistently reached isolation mass \citep{Izidoro2021, Izidoro2022}.

\citetalias{MacDonald2020} explored a continuum of initial solid surface densities of the protoplanetary disks that their embryos form within, finding that less massive embryos tend to complete a larger fraction of their collisions after the full dissipation of the gas disk, thereby resulting in more dynamically hot (high eccentricity and mutual inclination, low multiplicity) systems. They also performed simulations in which initial embryo mass has a steeper slope as a function of semi-major axis, representative of a system in which outer embryos are allowed to reach their isolation masses, but inner embryos are depleted. Relative to their baseline simulations, those simulations with steeper solid surface density profiles resulted in planetary systems with fewer single planet systems and more systems with multiple detected transiting planets, due to initial embryos being spaced $\Delta_0 = 10$ apart rather than $\Delta_0 = 3$. With tighter initial spacings, the multiplicity of these systems would decrease. Thus, if we were to conduct simulations with initial embryo masses at a fraction of the MFI or PI embryo masses considered in this study that increases with radial separation (e.g., not assuming a constant, independent flow of pebbles), then we could predict a higher fraction of dynamically hot systems than in our baseline simulations of the MFI or PI models. This would result in more planetary systems with lower multiplicities and lower mass planets. In the future, additional simulations that adopt an alternative prescription for initializing the embryos could investigate the impact of these alternate models on the final properties of the resultant planetary systems.

Other factors that may play a role in governing system diversity could also be explored.
Our current treatment of the gas disk is simplistic. 
We hold disk scale height and radial temperature profiles constant across simulations, when in reality morphology and temperature varies greatly from disk to disk. 
Future work could explore the sensitivity of isolation mass models to these initial disk conditions.
We could also improve in our representation of disk dissipation.
Instead of treating it as a step function, future studies could allow the depletion factor $d$ to decrease with time (e.g. \citet{Izidoro2021, Izidoro2022}), which will allow for more planet-planet excitation.
A radial dependence could also be added to the depletion factor, accounting for pressure bumps and gaps within the gas disk to allow for variations in embryo damping in a single system.

The amount of gas remaining in the protoplanetary disk at the beginning of embryo interaction is not well understood. 
We chose a depletion factor of $d=100$ in this work to mirror an environment in which the gas disk is nearing the end of its lifetime.
However, the truth case is surely more complex, with embryos in various disks reaching their isolation masses at different times depending on composition, disk mass, amount of gas, etc..
Thus, future work could explore the effects of a continuum of disk depletion factors on a set of initial embryos, reweighting the final results by depletion factor.

It is important to note that many assumptions we make in this work are carried over from previous works \citep{Dawson2015, Dawson2016, MacDonald2020} in an attempt to more faithfully compare the outcome of changing the isolation mass model. 
Since none of the isolation mass models we explore here can effectively match observations by tuning only one parameter, future work could perform an extensive study of the initial conditions of formation by exploring a grid of initial spacings, solid surface density distributions and slopes $\alpha$, and the amount \changethree{and/or distribution} of gas, among the parameters explored in this work. \changethree{Additionally, one could consider a more complex model with other factors, such as disk turbulence, giant planets, or stellar evolution.}

Photoevaporation is an extremely important process affecting planetary radius, atmospheric composition, and mass in the inner disk \citep{OwenWu2017}.
It is suspected to be a strong contributor to the ``radius gap'', a paucity of planets at $\mathrm{R_p}\sim1.7~\mathrm{R_\oplus}$ \citep{Fulton2017}, for planets with small orbital periods.
In this work, we use the probabilistic mass-radius relationship presented in \citet{ChenKipping2017} for translating planet masses into radii, as required to model the detection completeness using \emph{Kepler} data. 
Future work could consider more recent mass-radius estimates of the empirical mass-radius relationship.

Additionally, future research could consider photoevaporation to more accurately compare our simulated planetary systems to those in the \emph{Kepler} catalog.
A more accurate treatment of planetary atmospheres could allow us to compare our simulated systems with samples such as those in \citet{Vach2024}, \citet{Fernandes2025}, helping to place constraints on the time evolution of sub-Neptune and super-Earth occurrence rates around FGK stars.
This work would help to inform observations made by future missions, such as PLATO \citep{Rauer2025}.

\medskip
\section*{Acknowledgments}
L.B would like to thank Treazure DeBreo, Jiayin Dong, Rachel Fernandes, Kaden Kelly-Pojar, Christopher Lam, and Eve Lee for useful discussions and comments.
The Center for Exoplanets and Habitable Worlds is supported by the Pennsylvania State University, the Eberly College of Science, and the Pennsylvania Space Grant Consortium.  
The authors were supported in part by NASA Exoplanet Research Program grant No. 80NSSC24K0150.
The authors thank the referee for helpful comments and suggestions which improved the manuscript.
The authors of this work recognize the Penn State Institute for Computational and Data Sciences (RRID: SCR\textunderscore025154) for providing access to computational research infrastructure within the Roar Core Facility (RRID: SCR\textunderscore026424).
This content is solely the responsibility of the authors and does not necessarily represent the views of the Institute for Computational and Data Sciences.

\section*{Data Availability}
The data underlying this article are available on GitHub at \url{https://github.com/lucasbrefka/Pebble_Iso_Data/tree/main}.

\section*{Software} 
pandas \citep{mckinney-proc-scipy-2010}, IPython \citep{PER-GRA:2007}, matplotlib \citep{Hunter:2007}, scipy \citep{scipy}, numpy \citep{oliphant-2006-guide}, Jupyter \citep{Kluyver:2016aa}, Rebound \citep{ReinLiu2012}, Reboundx \citep{Tamayo2020} forecaster \citep{ChenKipping2017}

\bibliography{planetformrefs}
\bibliographystyle{mnras}

\bsp	
\label{lastpage}
\end{document}